\documentclass[fleqn,usenatbib]{mnras}

\usepackage{newtxtext,newtxmath}

\usepackage[T1]{fontenc}

\DeclareRobustCommand{\VAN}[3]{#2}
\let\VANthebibliography\thebibliography
\def\thebibliography{\DeclareRobustCommand{\VAN}[3]{##3}\VANthebibliography}

\usepackage{graphicx}	
\usepackage{amsmath}	
\usepackage{breqn}

\newcommand{\oiii}{[$\ion{O}{iii}$]}
\newcommand{\sii}{[$\ion{S}{ii}$]}
\newcommand{\nii}{[$\ion{N}{ii}$]}

\title[BASS XXXI: Outflow rate scaling relations in low-z AGN]{BASS XXXI: Outflow scaling relations in low redshift X-ray AGN host galaxies with MUSE}

\author[D. Kakkad et al.]{
D. Kakkad$^{1,2,3}$\thanks{E-mail: dkakkad@stsci.edu},
E. Sani$^{1}$, A. F. Rojas$^{4}$, Nicolas D. Mallmann$^{1,5}$, S. Veilleux$^{6}$, Franz E. Bauer$^{7,8,9}$, \newauthor
F. Ricci$^{10,11}$, R. Mushotsky$^{6}$, 
M. Koss$^{12}$, C. Ricci$^{13,14}$, E. Treister$^{7}$, George C. Privon$^{15,16}$,
N. Nguyen$^{1,17}$, \newauthor R. B\"{a}r$^{18}$,  F. Harrison$^{19}$, 
K. Oh$^{20,21}$, M. Powell$^{22}$,
R. Riffel$^{5}$, D. Stern$^{23}$, B. Trakhtenbrot$^{24}$, C. M. Urry$^{25}$  
\\
$^{1}$European Southern Observatory, Alonso de Cordova 3107, Vitacura, Casilla 19001, Santiago de Chile, Chile\\
$^{2}$Department of Physics, University of Oxford, Denys Wilkinson Building, Keble Road, Oxford, OX1 3RH, UK\\
$^{3}$Space Telescope Science Institute, 3700 San Martin Drive, Baltimore, MD 21218, USA\\
$^{4}$Centro de Astronom\'{i}a (CITEVA), Universidad de Antofagasta, Avenida Angamos 601, Antofagasta, Chile\\
$^{5}$Departamento de Astronomia, Instituto de F\'{i}sica, Universidade Federal do Rio Grande do Sul, CP 15051, 91501-970, Porto Alegre, RS, Brazil\\
$^{6}$Department of Astronomy and Joint Space-Science Institute, University of Maryland, College Park, MD 20742, USA\\
$^{7}$Instituto de Astrof\'{i}sica and Centro de Astroingenier\'{i}a, Facultad de F\'{i}sica, Pontificia Universidad Cat\'{o}lica de Chile, Casilla 306, Santiago 22, Chile\\ 
$^{8}$Millennium Institute of Astrophysics (MAS), Nuncio Monse\~{n}or S\'{o}tero Sanz 100, Providencia, Santiago, Chile\\
$^{9}$Space Science Institute, 4750 Walnut Street, Suite 205, Boulder, Colorado 80301\\
$^{10}$Dipartimento di Fisica e Astronomia, Università di Bologna, via Gobetti 93/2, 40129 Bologna, Italy\\
$^{11}$INAF - Osservatorio di Astrofisica e Scienza dello Spazio di Bologna, via Gobetti 93/3, 40129 Bologna, Italy\\
$^{12}$Eureka Scientific, 2452 Delmer Street Suite 100, Oakland, CA 94602-3017, USA\\
$^{13}$N\'{u}cleo de Astronom\'{i}a de la Facultad de Ingenier\'{i}a, Universidad Diego Portales, Av. E\'{j}ercito Libertador 441, Santiago 22, Chile\\
$^{14}$Kavli Institute for Astronomy and Astrophysics, Peking University, Beijing 100871, People's Republic of China\\
$^{15}$National Radio Astronomy Observatory, 520 Edgemont Rd, Charlottesville, VA, 22903, USA\\
$^{16}$Department of Astronomy, University of Florida, 211 Bryant Space Science Center, Gainesville, FL 32611, USA\\
$^{17}$Departamento de Astronom\'{i}a, Universidad de Chile, Camino el Observatorio 1515, Las Condes, Santiago, Casilla 36-D, Chile\\
$^{18}$Institute for Particle Physics and Astrophysics, Department of Physics, ETH Zurich, Wolfgang-Pauli-Strasse 27, CH-8093 Zurich, Switzerland\\
$^{19}$Cahill Center for Astronomy and Astrophysics, California Institute of Technology, Pasadena, CA 91125, USA\\
$^{20}$Korea Astronomy and Space Science Institute, Daedeokdae-ro 776, Yuseong-gu, Daejeon 34055, Republic of Korea\\
$^{21}$Department of Astronomy, Kyoto University, Kitashirakawa-Oiwake-cho, Sakyo-ku, Kyoto 606-8502, Japan\\
$^{22}$Kavli Institute of Particle Astrophysics and Cosmology, Stanford University, 452 Lomita Mall, Stanford, CA 94043, USA\\
$^{23}$Jet Propulsion Laboratory, California Institute of Technology, 4800 Oak Grove Drive, MS 169-224, Pasadena, CA 91109, USA\\
$^{24}$School of Physics and Astronomy, Tel Aviv University, Tel Aviv 69978, Israel\\
$^{25}$Yale Center for Astronomy \& Astrophysics, Physics Department, PO Box 208120, New Haven, CT 06520-8120, USA\\
}

\date{Accepted XXX. Received YYY; in original form ZZZ}

\pubyear{2020}

\begin{document}
\label{firstpage}
\pagerange{\pageref{firstpage}--\pageref{lastpage}}
\maketitle

\begin{abstract}
Ionised gas kinematics provide crucial evidence of the impact that active galactic nuclei (AGN) have in regulating star formation in their host galaxies. Although the presence of outflows in AGN host galaxies has been firmly established, the calculation of outflow properties such as mass outflow rates and kinetic energy remains challenging. We present the \oiii$\lambda$5007 ionised gas outflow properties of 22 z$<$0.1 X-ray AGN, derived from the BAT AGN Spectroscopic Survey using MUSE/VLT. With an average spatial resolution of 1\arcsec (0.1--1.2 kpc), the observations resolve the ionised gas clouds down to sub-kiloparsec scales. Resolved maps show that the \oiii ~velocity dispersion is, on average, higher in regions ionised by the AGN, compared to star formation. We calculate the instantaneous outflow rates in individual MUSE spaxels by constructing resolved mass outflow rate maps, incorporating variable outflow density and velocity. We compare the instantaneous values with time-averaged outflow rates by placing mock fibres and slits on the MUSE field-of-view, a method often used in the literature. The instantaneous outflow rates (0.2--275 $M_{\odot}$ yr$^{-1}$) tend to be 2 orders of magnitude higher than the time-averaged outflow rates (0.001--40 $M_{\odot}$ yr$^{-1}$). The outflow rates correlate with the AGN bolometric luminosity ($L_{\rm bol}\sim$ 10$^{42.71}$--10$^{45.62}$ erg/s) but we find no correlations with black hole mass (10$^{6.1}$--10$^{8.9}$ M$_{\odot}$), Eddington ratio (0.002--1.1) and radio luminosity (10$^{21}$--10$^{26}$ W/Hz). We find the median coupling between the kinetic energy and $L_{\rm bol}$  to be 1\%, consistent with the theoretical predictions for an AGN-driven outflow.
\end{abstract}




\section{Introduction} \label{sect1}


Understanding the role of Active Galactic Nuclei (AGN) in the galaxy evolution process is one of the major challenges in extra-galactic astronomy today. AGN are believed to be supermassive black holes in the centre of most massive galaxies that are powered by the accretion of gas \citep{soltan82, rees84, yu02, fabian12}. The net energy emitted by the AGN over its lifetime can greatly exceed the binding energy of the host galaxy \citep[e.g.,][]{begelman06}. The tremendous amount of energy could couple with the surrounding gas and dust in the interstellar medium (ISM), eventually influencing the host galaxy properties. Such a process, called AGN feedback, is often invoked in state-of-the-art cosmological simulations and analytical models to reproduce observed properties such as the galaxy luminosity function at the high mass end \citep[e.g.,][]{benson03,matteo05, hopkins10, gaspari11, faucher12, genel14, vogelsberger14, crain15, schaller15, sijacki15, hopkins16, torrey20}. These simulations predict that even a small fraction ($\sim$1--5\%) of the AGN energy that couples with the surrounding ISM is sufficient to regulate the growth of the black hole and the star formation in the host galaxy \citep[e.g.,][]{zubovas12,costa14, king_pounds15}. 

AGN feedback can exist in several forms such as radiation, thermal or non-thermal (cosmic rays) pressure-driven winds, jet-mode feedback and via magnetic forces on accretion disk scales. AGN feedback can explain several observed properties such as the presence of high velocity (>1000 km s$^{-1}$) multi-phase gas outflows in low and high redshift galaxies and observations of bubbles or cavities in X-ray observations of galaxy clusters \citep[e.g.,][]{blanton11, fabian12, sanders14, feruglio15, laha21}. High velocity outflows from AGN host galaxies have been reported in numerous studies in the literature (see \citet{veilleux20} for a review and the references therein) using optical spectroscopy \citep[e.g.,][]{greene11, mcelroy15, sun17, durre18, manzano-king19, perna20, santoro20, trindale-falcao21}, near-infrared spectroscopy \citep[e.g.,][]{kakkad16, bischetti17, zakamska16, diniz19, riffel20a,riffel20b} and sub-mm spectroscopy \citep[e.g.,][]{michiyama18, zschaechner18, impellizzeri19, audibert19, garcia-bernete21}. One of the key quantities that is not well understood through these observations is how efficiently does the outflow couple with the ISM \citep[e.g.,][]{harrison18}. The coupling efficiency i.e. the ratio between the kinetic power of the outflow ($\dot{E}_{\rm kin}$) and the bolometric luminosity of the AGN ($L_{\rm bol}$) or the star formation rate (SFR) of the host galaxy, is critical to quantify the true impact of AGN feedback on host galaxies - the higher the efficiency, the easier it is for these outflows to heat the gas or propagate the outflows to the galaxy outskirts. An accurate measurement of mass outflow rate and kinetic energy is therefore necessary to estimate the true coupling efficiency, which can also be used as constraints in cosmological simulations. 

The calculation of mass outflow rates, especially in the ionised gas phase, have often come from measurements using integrated fibre or long-slit spectra, where several assumptions are invoked in the outflow modelling. These assumptions, briefly described here, result in ``time-averaged global mass outflow rate" with large systematic uncertainties. First, due to the limitations of the current instruments even on large telescopes, an accurate modelling of the outflow geometry is not possible. This is especially true for high redshift galaxies (z$\sim$2) where, with currently available adaptive optics (AO) technology, one can at best achieve a spatial resolution of $\sim$2 kpc where the bulk of the outflow might reside \citep[e.g.,][]{brusa16, daviesR20}. Therefore, the outflow geometry is either assumed to be a uniformly filled conical, bi-conical or spherical thin shells \citep[e.g.,][]{veilleux01, fischer13, riffel13, ishibashi15, thompson15, bae16, husemann19, mingozzi19}. Second, if the data is obtained from fibre and single-slit spectroscopy, the size of the outflow is largely unconstrained. For long-slit observations, as an example, the outflow size depends on whether the slit is oriented along the outflow direction. This can be mitigated by using integral field spectroscopy (IFS) which is being increasingly used for extra-galactic studies \citep[e.g.,][]{rupke13,liu13, harrison14, maiolino17, astor19, husemann19, rupke21}, although there could still be projection effects with the IFS data. Third, accurate determination of electron density and electron temperature is required for the ionised mass outflow rate calculations. Electron density is usually derived from emission lines that arise out of two closely spaced ``meta-stable'' energy levels such as \sii$\lambda\lambda$6716,6731 (\sii ~doublet hereafter). Density measured from the \sii ~doublet is sensitive to values between $\sim$10--5000 cm$^{-3}$, typical in the Narrow Line Region (NLR) of AGN host galaxies \citep[e.g.,][]{osterbrock06, perna17, baron19, davies20}. The \sii ~doublet is significantly weaker than the lines used to trace ionised outflows such as the \oiii$\lambda$5007 and H$\alpha$. In high redshift galaxies, it is extremely challenging to detect these doublet lines, despite hours of observations on a single target. Therefore, nominal density values are often assumed in mass outflow rate calculations, resulting in systematic uncertainties of up to 2--3 orders of magnitude. Furthermore, the density structure within the outflowing medium is often non-uniform, when resolved in low redshift galaxies \citep[e.g.,][]{kakkad18}. Therefore assuming a constant density within the outflowing medium often leads to inaccurate outflow rate and kinetic energy values. Collectively, these assumptions result in a systematic uncertainty of approximately 3--4 orders of magnitude \citep[e.g.,][]{harrison18}. This implies that the quoted values of coupling efficiency in the literature have a wide range, with the actual efficiency still an unknown in most of the studies.

One of the ways to overcome the limitations of the previous studies is to use the IFS data sets to construct resolved outflow rate maps to get instantaneous outflow rates within the individual gas clouds. In other words, what was previously calculated for individual galaxies from integrated spectra (values averaged over the lifetime of the outflow) can now be calculated for every pixel or PSF (point spread function) element within the IFS field-of-view (FoV). The advantage of such a method is that an assumption on the outflow geometry is not strictly required and the spatial variation in the electron densities and outflow velocities are easily incorporated. This would result in outflow rate values with significantly smaller uncertainties than the methods previously used in the literature. The limitation, however, is that this method can only be applied for targets and/or pixels that show sufficiently high signal-to-noise (S/N) in the emission lines, especially the \sii$\lambda\lambda$6716, 6731 doublet, so as to resolve the wings in the individual emission lines. Furthermore, with the current instrumentation and their sensitivity, such a study can be performed only for low redshift targets. In addition to the sensitivity, IFS data are still limited by the spatial resolution, therefore assumptions need to be made for the ISM or outflow conditions within a single PSF element.

Previous studies targeting resolved mass outflow rates in low redshift AGN host galaxies have been limited to a few galaxies ($<$5). In \citet{venturi18}, an ionized gas outflow rate map was derived for NGC 1365 from H$\alpha$ emission within a biconical outflow. The low S/N in each spaxel was mitigated by co-adding spectra from multiple spaxels and forming a grid along the outflowing cone. The NGC 1365 outflow rate map exhibited both radial and angular variations as a result of inhomogeneous outflowing media. The outflow rate also decreased with distance from the AGN location, suggesting an energy exchange or momentum loss to the ISM as the outflow propagates across the host galaxy. Using long slit spectroscopy and \oiii ~imaging  in conjunction with emission line diagnostics and photoionisation models, \citet{revalski18} \citep[see also][]{revalski21} reported that instead of a radially decreasing function, the outflow rate in Mrk 573 has a peak of $\sim$3 $M_{\odot}$ yr$^{-1}$ at a distance of 210 pc from the AGN location before starting to decrease again, implying a strong variable outflow with time. In summary, spatially resolved data provide a clearer picture of how the mass propagates within an AGN-driven outflow compared to time-averaged outflow rates with assumed geometries.

In this paper, we derive resolved ionised gas mass outflow rate maps using optical IFS data for a sample of 22 low redshift X-ray selected AGN host galaxies derived from the BAT AGN Spectroscopic Survey \citep[BASS\footnote{www.bass-survey.com},][]{koss17}. We calculate the total outflow rate within these host galaxies and estimate the instantaneous coupling between the outflow properties such as the velocity, mass outflow rate and kinetic energy, and the properties of the AGN such as the bolometric luminosity, black hole mass, Eddington ratio and radio power. These relations are compared with the scaling relations previously reported in the literature \citep[e.g.,][]{carniani16,fiore17, fluetsch19}. Using IFS data, we additionally investigate how these scaling relations depend on the size and shape of the aperture (fibre and slit) used to extract the integrated spectrum.

This paper is arranged as follows: Section \ref{sect2} describes the sample used. In Sect. \ref{sect3}, we report the observations and data reduction procedure, followed by the analysis procedures in Sect. \ref{sect4}. The results of the analysis is presented in Sect. \ref{sect5}. We discuss the implications of these results in Sect. \ref{sect6} and the summary and conclusions are presented in Sect. \ref{sect7}. We adopt the standard $\Lambda$CDM cosmological parameters throughout the paper: H$_{\rm 0}$ = 70 km s$^{-1}$, $\Omega_{\rm m}$ = 0.3 and  $\Omega_{\Lambda}$ = 0.7. North is up and East is to left in all the maps presented in this paper. 

\section{Sample and observations} \label{sect2}
We aim to investigate the presence or absence of correlations between the outflow properties such as its kinetic power and AGN properties such as the black hole mass, bolometric luminosity and the Eddington ratio. This requires a sample that shows a wide range in these AGN properties. As described in Sect. \ref{sect1}, the sample should also consist of low redshift galaxies (z$<$0.1) such that the outflowing gas is spatially resolved to be able to construct the mass outflow rate maps resolved down to sub-kiloparsec scales. Lastly, we also restricted the AGN selection based on the X-ray emission between 14-195 keV, as selection contamination from sources other than the AGN, such as X-ray binaries or starbursts, is negligible in this band. 

\begin{figure}
\centering
\includegraphics[width=0.45\textwidth]{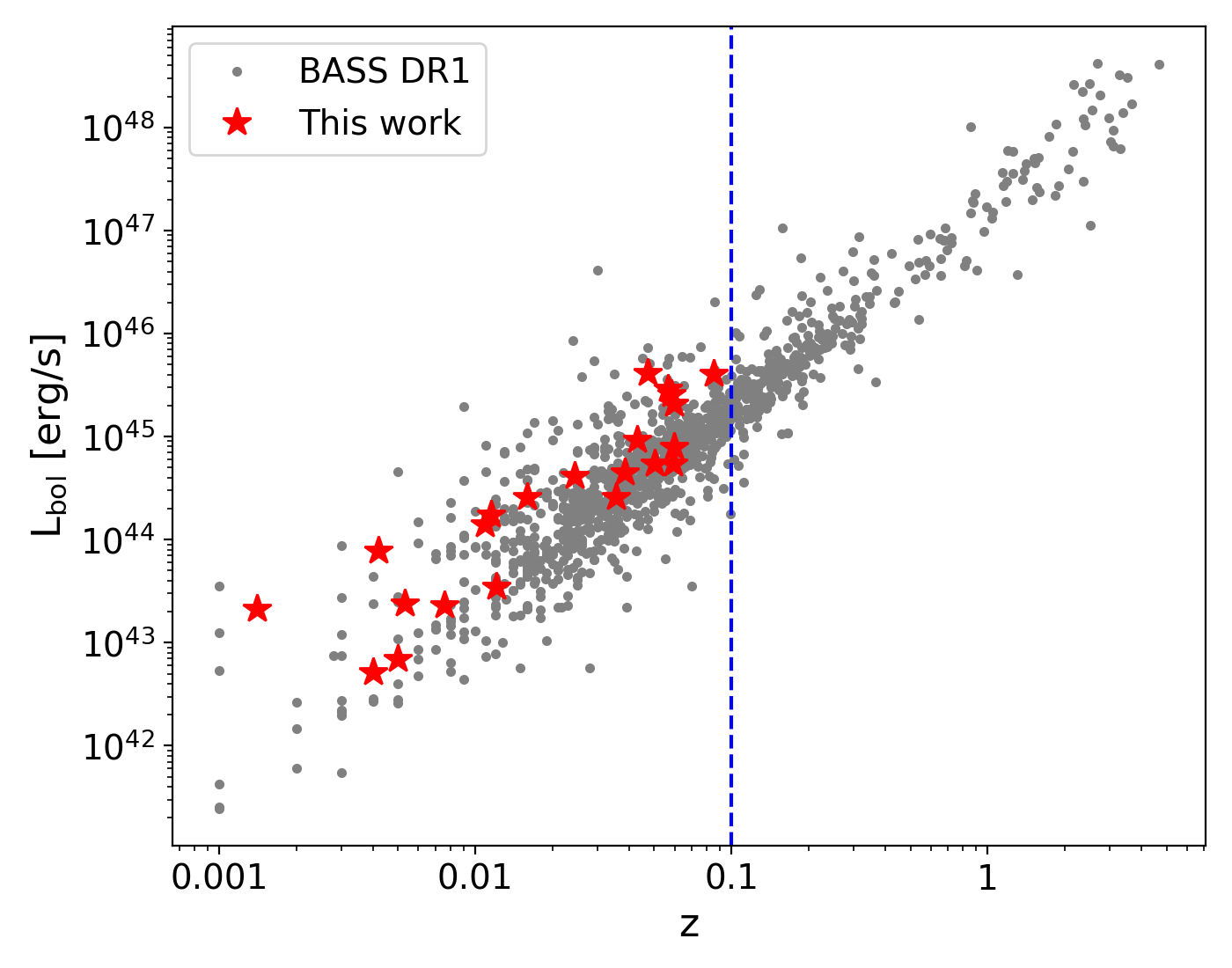}
\caption{The location of the 22 targets presented in this paper in the $L_{\rm bol}$ versus redshift plane. The grey data points show the parent BASS sample, while the red stars show the MUSE targets presented in this paper. The vertical dashed blue line shows the z = 0.1 line, which is chosen as the redshift cut for our targets to resolve the ionised gas down to sub-kiloparsec scales with MUSE. }
\label{fig:Lbol-z}
\end{figure}

\begin{figure*}
\centering
\includegraphics[scale=0.5]{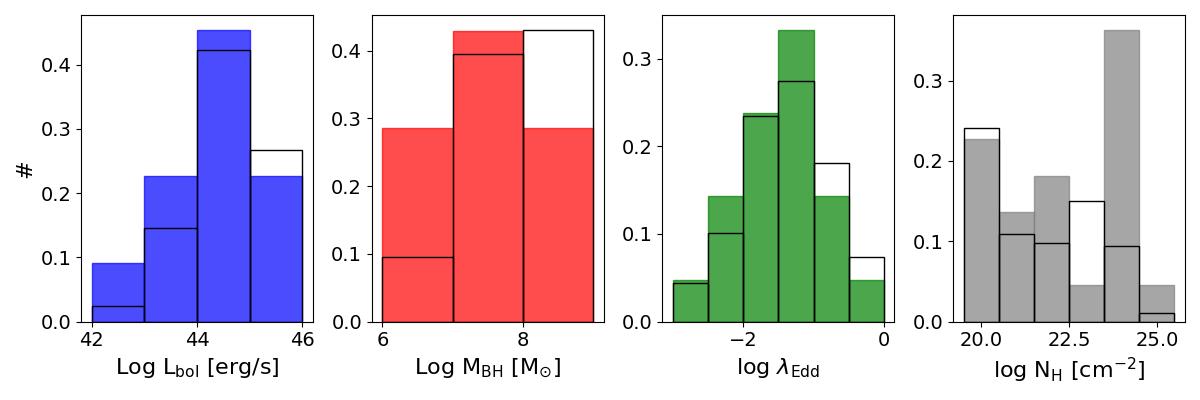}
\caption{From left to right: Distributions (fraction of targets in the respective bins) of the bolometric luminosity, black hole mass, Eddington ratio and the column density. The coloured histograms show the distributions for the BASS-MUSE sample, while the black outline shows the respective distribution for the parent BASS sample at z$<$0.1. The BASS-MUSE targets sample a similar range of parameters as the parent sample. Each of these parameters span more than three orders of magnitude, a useful characteristic to understand the presence of correlations of these parameters with the outflow properties.}
\label{fig:BASS-MUSE-properties}
\end{figure*}

\begin{figure}
\centering
\includegraphics[scale=0.12]{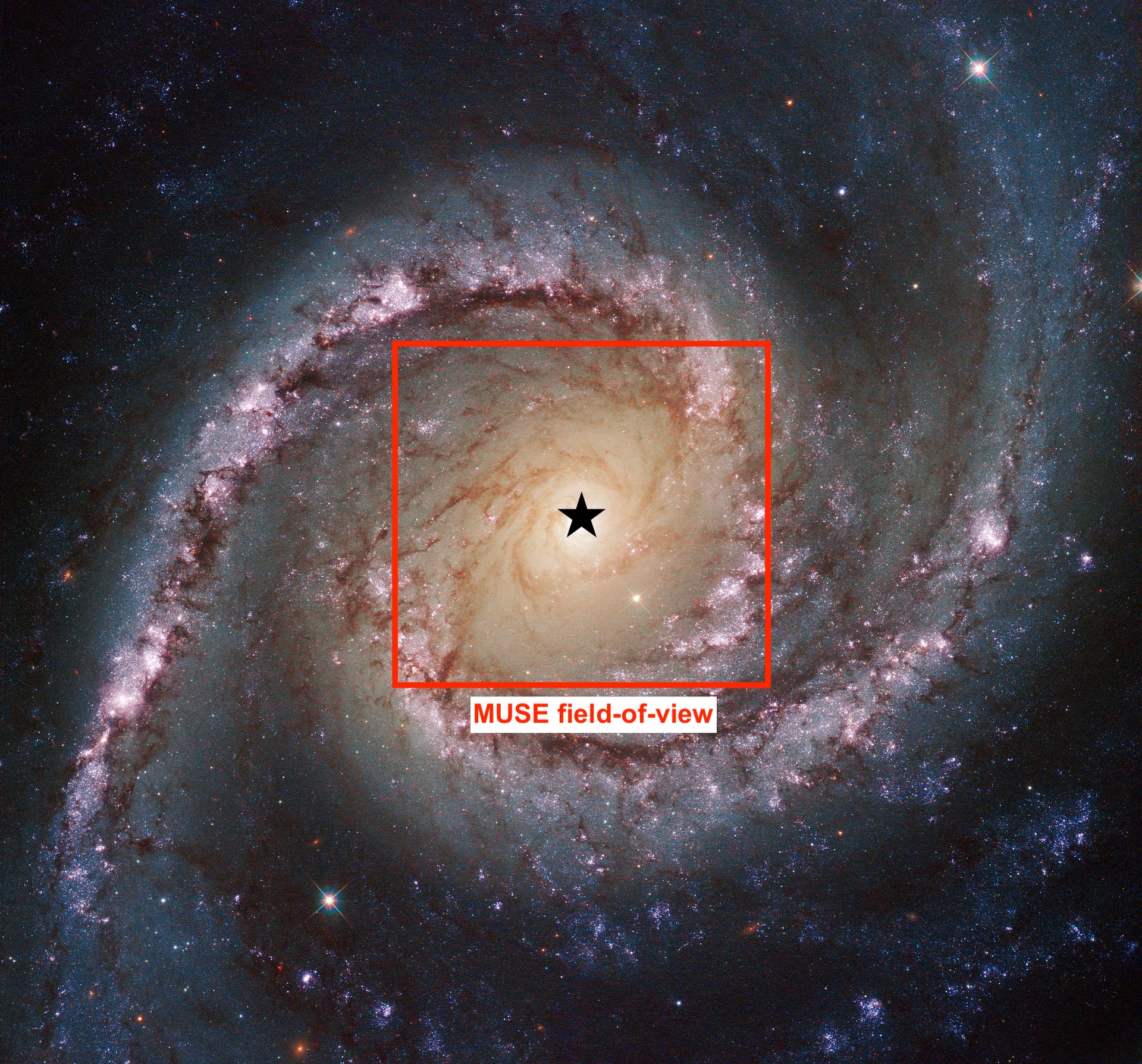}
\caption{The image shows the 2.7$\times$2.5 arcmin$^{2}$ HST WFC3 image of NGC 1566, one of the galaxies in the BASS-MUSE sample. The black star shows the AGN location and the red square shows the 1$\times$1 arcmin$^{2}$ MUSE FoV, which covers a distance up to 3 kpc from the AGN location. Image Credit: ESA Hubble and NASA.}
\label{fig:NGC1566_MUSE_cutout}
\end{figure}

The Burst Alert Telescope \citep[BAT, ][]{barthelmy05} on board Neil Gehrels Swift Observatory \citep{gehrels04} provides high-sensitivity all sky hard X-ray survey to sources such as AGN. The BASS survey \citep{koss17} is a project dedicated to spectroscopic follow-up of $\gtrsim$1000 BAT AGN with an aim to uniformly determine AGN properties via continuum, emission and absorption line measurements in multiple wavebands. Consequently, the BASS survey has a wealth of multi-wavelength ancillary data which has allowed the calculation of quantities such as the intrinsic X-ray luminosity and column density \citep{ricci17}, black hole mass, bolometric luminosity and Eddington ratio, \citep[see][]{koss17, lamperti17, oh17}, radio luminosity \citep[e.g.,][]{smith20}, stellar masses via Spectral Energy Distribution (SED) fitting, and other host galaxy properties such as gas fractions and dust properties via sub-mm follow-up \citep[e.g.,][]{koss21}. The high data quality and uniform measurement of black hole and host galaxy properties makes the BASS survey an ideal sample for the purpose of this study. 

The low redshift X-ray AGN sample presented in this paper is derived from the parent BASS sample from the first data release \citep[BASS-DR1,][]{koss17}. We first searched the ESO archive for prior optical IFS observations of the BASS sample with the Multi Unit Spectroscopic Explorer (MUSE) instrument \citep{bacon10} at the Very Large Telescope (VLT), which are publicly available. The archival search yielded 52 BASS galaxies with MUSE observations (as of January 2019). Out of these galaxies, we discarded targets at z$>$0.1, as those targets do not have sufficient spatial resolution to perform the resolved mass outflow rate study. The redshift cut is motivated by the fact that we would like to resolve outflows down to sub-kiloparsec scales in the host galaxies of the selected AGN. Furthermore, targets that have low signal in the \oiii$\lambda$5007 (\oiii ~hereafter) emission have also been removed from the analysis. The final BASS sub-sample (BASS-MUSE sample hereafter) selected for the ionised outflow analysis consists of 22 X-ray AGN with a mean redshift of 0.035 and mean X-ray luminosity, log $\mathbf{L_{2-10 ~keV}}$ of 43.16 erg s$^{-1}$ \footnote{ESO programme observation IDs: 60.A-9100(K), 60.A-9339(A), 094.B-298, 094.B-0321, 094.B-0345, 095.B-0015, 095.B-0482, 095.B-0532, 095.B-0934, 096.D-0263, 096.B-0309, 097.B-0080, 097.D-0408, 099.B-0137, 0100.B-0116}. Several targets presented in this paper are a part of other targeted surveys of AGN and/or star forming galaxies whose MUSE  or other optical IFS data have been previously presented in the literature \citep[e.g.,][]{dopita15, thomas17, powell18, treister18, venturi18, balmaverde19, erroz-ferrer19, mingozzi19, den-brok20, lopez-coba20, balmaverde21}. Whenever possible, we will compare our results with the already published data in the literature to check for consistency in the derived maps and quantities. 

Figure \ref{fig:Lbol-z} shows the bolometric luminosity versus the redshift of the BASS-MUSE sample (red stars), with the parent BASS sample as background grey data points. The bolometric luminosity has been derived from the intrinsic X-ray luminosity, which is described in detail in the BASS DR1 publication \citep[see][]{koss17}. The plot highlights the low redshift nature of the BASS-MUSE sample and the wide range covered in bolometric luminosity. This shows that the selected sample is representative of the parent BASS population at low redshift. The wide range in AGN properties is also clear from the histograms in Fig. \ref{fig:BASS-MUSE-properties}. Calculations of these properties are described in details in the data release papers \citep[e.g.,][]{koss17, ricci17}. The bolometric luminosity, $L_{\rm bol}$, is in the range 10$^{42.7}$--10$^{45.6}$ erg s$^{-1}$, the black hole mass, $M_{\rm BH}$, is in the range 10$^{6.2}$ -- 10$^{8.9}$ M$_{\odot}$ and the Eddington ratio is 0.002--1.1 (defined as $\lambda_{\rm Edd} = L_{\rm bol}/L_{\rm Edd}$, where $L_{\rm Edd} = 1.5\times10^{38} M_{\rm BH}$ erg s$^{-1}$, for solar composition gas). The sample consists of a nearly equal distribution of Type-1 (10) and Type-2 (or Type 1.9, 12) AGN, based on the presence or absence of broad lines in the optical spectra \citep{osterbrock81}. Similarly, the sample consists of both obscured and unobscured sources with the X-ray column densities spanning a full range between log $N_{\rm H}$ $<$20 to $\sim$25.4 cm$^{-2}$. The values of these quantities for individual targets are reported in Table \ref{table:sample}.

\begin{table*}
\centering
\caption{Properties of the sample used in this paper. See \citet{koss17} for a detailed description on the determination of these properties. (1) {\it Swift}-BAT 70 month hard X-ray survey ID; (2) Common name of the target; (3) \& (4) Optical coordinates of the target; (5) Redshift estimated from the \oiii$\lambda$5007 line; (6) AGN Type determined using the classification in \citet{osterbrock81}; (7) Bolometric luminosity of the AGN, determined from the intrinsic {\it Swift}-BAT hard X-ray luminosity (14-195 keV); (8) Black hole mass derived from broad emission lines in the optical and near-infrared wavelengths \citep[e.g.,][]{lamperti17}; (9) X-ray column density \citep{ricci17}; (10) Rest-frame radio luminosity obtained from 1.4 GHz flux density from \citet{veron10}; (11) Resolution of the MUSE images based on the seeing (corrected by airmass) at the time of observations (see Sect. \ref{sect3}).}
\label{table:sample}
\begin{tabular}{ccccccccccc}
\hline
BAT ID & Target Name & RA & DEC & z & Type & Log $L_{\rm bol}$ & Log $M_{\rm BH}$ & $N_{\rm H}$ & Log $L_{\rm 1.4GHz}$ & Seeing\\
&  & J200 & J200 & & & erg s$^{-1}$ & M$_{\odot}$ & cm$^{-2}$ & W Hz$^{-1}$ & arcsec\\
(1) & (2) & (3) & (4) & (5) & (6) & (7) & (8) & (9) & (10) & (11)\\
\hline\hline
57 & 3C033 & 01:08:52 & +13:20:14 & 0.059 & 2 & 45.32 & 8.83 & 23.76 & 26.0 & 1.2\\
58 & NGC 424 & 01:11:27 & -38:05:00 & 0.011 & 1.9 & 44.15 & 7.49 & 24.33 & 21.8 & 1.2\\
62 & IC 1657 & 01:14:07 & -32:39:03 & 0.012 & 2 & 43.54 & 7.68 & 23.4 & -- & 1.3\\
127 & HE 0224-2834 & 02:26:25 & -28:20:58 & 0.059 & 1 & 44.90 & 7.98 & 20.57 & -- & 1.4\\
134 & NGC 985 & 02:34:37 & -08:47:17 & 0.043 & 1 & 44.97 & 7.97 & 20.92 & 22.4 & 1.2\\
184 & NGC 1365 & 03:33:36 & -36:08:26 & 0.005 & 1 & 43.38 & 6.18 & 22.21 & 22.2 & 0.9\\
197 & HE 0351+0240 & 03:54:09 & +02:49:30 & 0.036 & 1 & 44.41 & 7.46 & 20 & -- & 0.8\\
213 & HE 0412-0803 & 04:14:52 & -07:55:39 & 0.037 & 1 & 44.65 & 8.09 & 20 & -- & 0.9\\
216 & NGC 1566 & 04:20:00 & -54:56:16& 0.005 & 1 & 42.84 & 6.10 & 20 & -- & 1.1\\
471 & NGC 2992 & 09:45:42 & -14:19:34 & 0.007 & 1.9 & 43.36 & 7.97 & 21.72 & 22.3 & 1.0\\
501 & HE1029-1401 & 10:31:54 & -14:16:51 & 0.086 & 1 & 45.61 & 8.67 & 20 & -- & 0.8\\
653 & NGC 4941 & 13:04:13 & -05:33:05 & 0.004 & 2 & 42.71 & 7.00 & 23.72 & 20.6 & 0.8\\
703 & Mrk 463 & 13:56:02 & +18:22:18 & 0.05 & 1.9 & 44.74 & 6.63 & 23.57 & 24.2 & 0.8\\
711 & Circinus & 14:13:09 & -65:20:20 & 0.001 & 2 & 43.33 & 6.23 & 24.4 & -- & 1.5\\
731 & NGC 5643 & 14:32:40 & -44:10:28 & 0.004 & 2 & 43.89 & 6.42 & 25.4 & -- & 0.8\\
783 & NGC 5995 & 15:48:24 & -13:45:27 & 0.025 & 1.9 & 44.62 & 7.85 & 21.97 & 22.5 & 0.7\\
817 & 2MASX J16311554+2352577 & 16:31:15 & +23:52:57 & 0.059 & 1.9 & 44.74 & 8.11 & 21.7 & -- & 0.9\\
1051 & 3C403 & 19:52:15 & +02:30:24 & 0.059 & 2 & 45.41 & 8.83 & 23.69 & 25.6 & 0.8\\
1092 & IC 5063 & 20:52:02 & -57:04:07 & 0.011 & 2 & 44.24 & 8.24 & 23.56 & -- & 1.1\\
1151 & 3C445 & 22:23:49 & -02:06:13 & 0.056 & 1 & 45.46 & -- & 23.54 & 25.6 & 1.4\\
1182 & NGC 7469 & 23:03:15 & +08:52:25 & 0.016 & 1 & 44.41 & 6.96 & 20.53 & 23.0 & 0.9\\
1183 & Mrk 926 & 23:04:43 & -08:41:08 & 0.046 & 1 & 45.62 & 7.99 & 20 & 23.1 & 0.8\\
\hline
\end{tabular}
\end{table*}

\section{Observations and data reduction} \label{sect3}

The archival MUSE data were reduced using the standard ESO MUSE pipeline \citep[e.g.,][]{weilbacher14, weilbacher20}, which performs bias correction, flat fielding, wavelength and astrometry calibrations. The science exposures were interleaved with separate sky exposures which were used for sky-subtraction. Although the pipeline sky subtraction provided satisfactory results in most cases, we used ZAP \citep[Zurich Atmospheric Purge, see][]{soto16} in the cases where the pipeline output left significant sky residuals. The spatial resolution of the observations was determined using the unresolved H$\beta$ and H$\alpha$ broad line region (BLR) emission, or isolated stars within the field of view or the airmass-corrected DIMM seeing during observations. In case of multiple observations of the same target, the data cubes were combined using the pipeline. The average spatial resolution (seeing) for each target is reported in Table \ref{table:sample}. The total on-source exposure times are in the range 2400--7200s.

The final data cubes consist of a FoV of $\sim1 \times 1$ arcmin$^{2}$ approximately centred on the AGN in most cases with a spatial sampling of 0.2\arcsec $\times$ 0.2\arcsec. The data cubes have a spectral coverage of $\sim$480--930 nm and the spectral resolution ranges from 1750--3750 in this wavelength range. Figure \ref{fig:NGC1566_MUSE_cutout} shows the 1$\times$1 arcmin$^{2}$ MUSE FoV superposed on a 2.7$\times$2.5 arcmin$^{2}$ image of NGC 1566, one of the targets analysed in this paper, taken with the Wide Field Camera 3 (WFC3) onboard the Hubble Space Telescope (HST). The MUSE field covers a physical distance of up to $\sim$3 kpc from the AGN location. With an average spatial resolution of $\sim$1.0\arcsec, the observations are able to resolve gas kinematics down to $\sim$100 pc scales for the galaxies presented in this paper.

\section{Analysis} \label{sect4}
\subsection{Spectral modelling} \label{sect4.1}

\begin{figure*}
\centering
\includegraphics[scale=0.5]{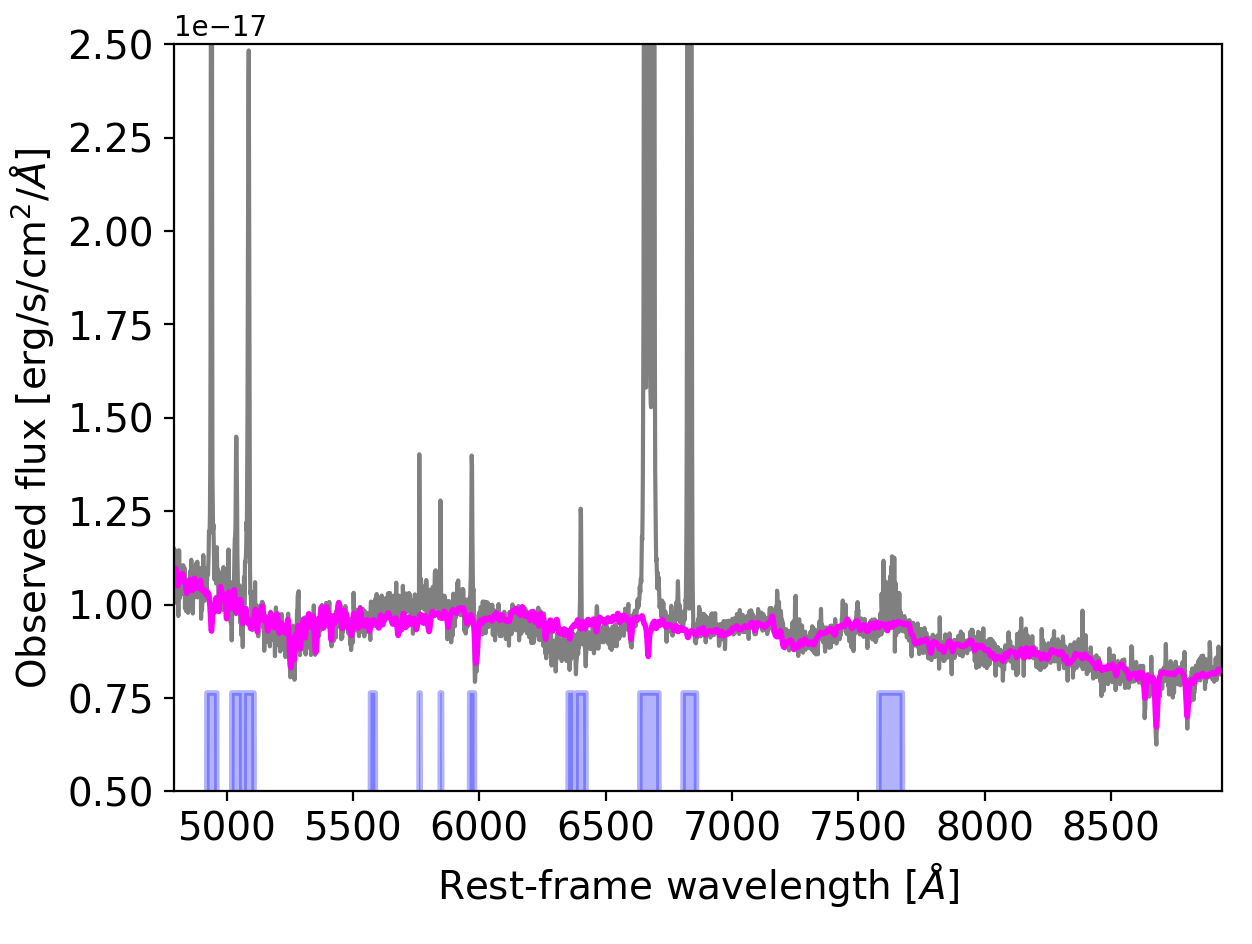}\\
\includegraphics[scale=0.5]{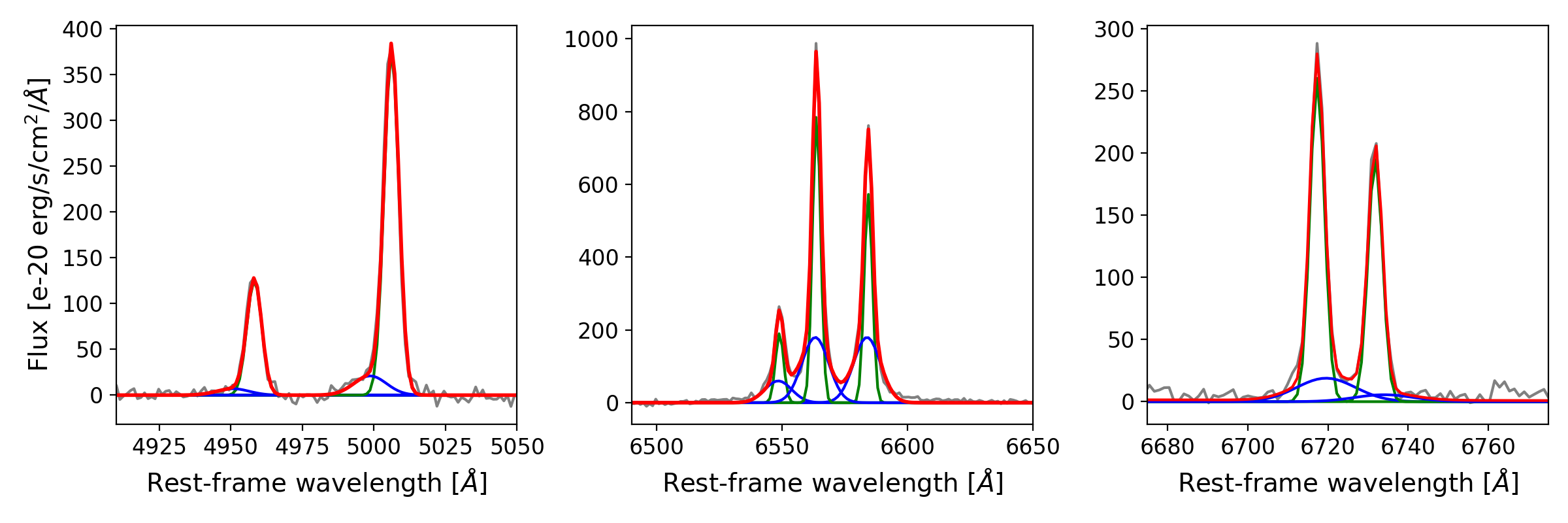}\\
\caption{Top panel shows an example of stellar continuum fit for one of the pixels in the MUSE spectra of NGC 7469. The grey curve shows the raw data and the magenta overlaid curve shows the stellar continuum fit. The blue bars on the bottom of the plot show the spectral channels masked during the continuum fitting procedure. The bottom panels show the examples of the multi-Gaussian emission line models after the stellar continuum subtraction for [OIII]$\lambda\lambda$4959,5007 (lower left panel), [NII]$\lambda\lambda$6549,6585 \& H$\alpha$ (lower middle panel) and \sii$\lambda\lambda$6716,6731 (lower right panel). The grey curve show the spectra after continuum subtraction, the green curve shows the narrow Gaussian component, the blue curve shows the broad Gaussian component and the red curve shows the overall emission line model.}
\label{fig:spectral_modelling}
\end{figure*}

We first model the stellar continuum emission using the $\mathtt{LZIFU}$ code, which has been routinely used for optical IFS spectroscopy \citep[e.g.,][]{ho16, kreckel18}. $\mathtt{LZIFU}$ adopts the penalised pixel fitting routine \citep[PPXF, ][]{cappellari04, cappellari17} to fit the stellar continuum using a series of input spectral templates from stars or modelled simple stellar populations (SSPs) convolved with parametrised velocity distribution. In this paper, we used MILES stellar template libraries \citep{sanchez-blazquez06}, which includes templates from $\sim$1000 stars with a wide range in stellar ages and metallicities  The regions in the spectra with strong skylines, emission lines such as H$\beta$, \oiii, \nii, H$\alpha$ and \sii ~and/or NaD doublet contamination in case of adaptive optics (AO) assisted observations were masked during the continuum fitting procedure. Further details of the continuum fitting procedure are given in \citet{ho16}. An example of a stellar continuum fit in one of the MUSE pixels of NGC 7469 is shown in the top panel in Fig. \ref{fig:spectral_modelling}. 

The modelled stellar continuum emission was subtracted from all the pixels across the MUSE FoV. We use the resulting stellar continuum-subtracted cube to fit key emission lines such as H$\beta$, \oiii$\lambda\lambda$4959, 5007, \nii$\lambda\lambda$6549,6585, H$\alpha$ and \sii$\lambda\lambda$6716, 6731. We will use the \oiii$\lambda$5007 line to trace the ionised gas outflows across the field of view. The aim of this step is to derive morphological and kinematic maps of the ionised gas in the AGN host galaxies. The emission lines were modelled with multiple Gaussian functions using the \texttt{scipy.curve-fit} package in python \citep{scipy20}. We started fitting the emission lines with a single Gaussian and additional Gaussian functions were added to minimise the $\chi^{2}$. In the case of multiple Gaussian fitting, the component with relatively lower width is termed the narrow component and the one with larger width is termed the broad component, without imposing a strict upper or lower limit on the individual Gaussian components. We use the broad Gaussian as a tracer of the outflowing component of the ionised gas and the narrow component as the systemic component tracing the kinematics of the disk. Given the quality and the moderate spectral resolution of the MUSE data, a maximum of two Gaussians were required for the forbidden transitions (e.g., \oiii, \nii ~and \sii) and three Gaussians for the Balmer lines, H$\alpha$ and H$\beta$. The third component in the case of Balmer lines was required to reproduce the emission from the fast moving clouds from the BLR. Hereafter, we will specifically term this third component the "BLR component", to avoid confusion with the broad component tracing the outflow. Furthermore, we impose constraints on the line centroids and widths of each of the Gaussian components of the emission lines. For example, the line centroids of the narrow (broad) component of H$\beta$, \oiii$\lambda\lambda$4959, 5007, \nii$\lambda\lambda$6549,6585, H$\alpha$ and \sii$\lambda\lambda$6716, 6731 were tied to each other, based on their expected positions in the rest-frame spectra. The width of each of the narrow (broad) component for all the lines were coupled with each other. Lastly, the line fluxes were kept free for all lines, except the line ratios \oiii$\lambda$5007:\oiii$\lambda$4959 and \nii$\lambda$6585:\nii$\lambda$6549 were fixed at 3, based on theoretical values \citep{storey00, dimitrijevic07}. 

To make sure that the fitting parameters are reliable, the multi-Gaussian fitting was only performed for spaxels with a S/N $\gtrsim$10 (the ratio between the peak of the lines and the rms noise in the continuum regions) in all the emission lines (H$\beta$, \oiii, \nii, H$\alpha$ and \sii). To increase the S/N, especially in the case of faint emission lines such as the \sii ~doublet, we perform a running median of nine neighbouring pixels during the line fitting procedure. This procedure does not affect the spatial resolution of the observations ($\geq$0.8 arcsec) as the mean is calculated for a region within the seeing of the observations. We do not use voronoi binning for our purpose, as the S/N in the lines obtained from the running median are sufficient at the location of these outflows (where additional broader Gaussian components are required in spaxels closer to the AGN location) and faint or negligible in the galaxy outskirts. Furthermore, the emission line widths (e.g., $w_{80}$) were corrected for the limited spectral resolution of MUSE at the location of \oiii ~lines \citep[$\sim$150 km/s,][]{bacon17}. An example of the emission line modelling for one of the MUSE pixels of NGC 7469 is shown in the bottom panels of Fig. \ref{fig:spectral_modelling}. 

We note that several works in the literature have also used a cut in the non-parametric velocity dispersion (width containing 80\% of the emission line flux), $w_{80}$ to define the presence or absence of outflows \citep[e.g.,][]{mullaney13, harrison14, perna17, sun17, rojas20}. However, in low redshift targets such as the ones presented in this paper, even a relatively low width can represent outflowing gas. This is because the emission lines in some of the regions require an additional broader Gaussian (with width lower than the adopted cuts of 600 km s$^{-1}$) to reproduce the overall line profile, while a single Gaussian fit results in significant residuals. Furthermore, compared to high redshift galaxies, most of these low redshift targets have low Eddington ratios, therefore the observed outflow velocities are expected to be lower. Therefore, using a particular cut in the velocity dispersion to define the presence or absence of outflows in a high spatially resolution data might eliminate regions which show such asymmetry in emission line profiles. Furthermore, the calculation of outflow flux from non-parametric definitions also depend on this user-defined velocity cut, which also might lead to an underestimation of the outflow flux if the cut is too high. We will therefore follow the parametric definition as described earlier i.e. assuming that the broad Gaussian component is tracing the outflow. We will make use of the \oiii ~$w_{80}$ values in Sect. \ref{sect5.1} to compare the overall width of the \oiii ~line in the regions ionised by AGN and star formation to understand the impact of AGN on \oiii ~kinematics. 

\subsection{Derivation of mass outflow rates} \label{sect4.2}

As a result of our assumption that the broad Gaussian component is the outflow component, the parameters of the broad Gaussian of the \oiii ~line are considered for the calculation of outflow properties. Conventionally, mass outflow rates are computed from integrated spectra, which often invoke assumptions in the parameters such as a particular outflow geometry with uniform electron density and velocity. This results in an outflow rate value which is time-averaged over the lifetime of the outflowing gas. We will now derive these outflow properties for every pixel as an individual aperture, to compute local-instantaneous outflow rate across each pixel \citep[see e.g.,][]{veilleux17}. For this purpose, we first calculate the outflow mass in the warm ionised gas phase for every pixel from the flux of the outflowing component of \oiii ~emission line (broad Gaussian), assuming Case B recombination in a fully ionised medium with electron temperature of 10$^{4}$ K \citep[for further details on the derivation of these equations, see][]{rupke05, genzel11, carniani15, cresci17, veilleux20}:

\begin{multline}
M_{\rm out} = 0.8\cdot 10^{8}{\rm M_{\odot}} \left(
\frac{1}{10^{\rm [O/H]-[O/H]_{\odot}}} \right)
\left( \frac{L_{\rm [OIII]}}{10^{44} ~erg/s} 
\right) \\
\left( \frac{<n_{\rm e}>}{500 ~{\rm cm^{-3}}}\right)^{-1}
\label{eq:outflow_mass}
,\end{multline}

\noindent
where $M_{\rm out}$ is the outflow mass, $L_{\rm [OIII]}$ is the luminosity of the broad \oiii ~component, [O/H] is the metallicity (assumed solar metallicity here) and $n_{\rm e}$ is the electron density. Most often in the literature, the value of the electron density is assumed to be $\sim$200 cm$^{-3}$, distributed uniformly within the outflowing medium. The high S/N in the \sii ~doublet and spatially resolved data allows us to compute the electron density within the outflowing medium for every pixel. The electron density is measured using the flux ratio of the \sii ~doublet \citep[e.g.,][]{osterbrock06, sanders16, kaasinen17, kakkad18, rose18} using the following equation:

\begin{equation}
n_e = \frac{627.1 R - 909.2}{0.4315 - R},
\label{eq:electron_density}
\end{equation}

\noindent
where R is the flux ratio of the outflowing component of the \sii ~doublet: $f$(\sii$\lambda$6731)/$f$(\sii$\lambda$6716). See \citet{osterbrock06} for further details on the derivation and other assumptions used in the derivation of Eq. \ref{eq:electron_density}. The mass outflow rate within each pixel is then computed as:

\begin{equation}
\dot{M}_{\rm out} = \frac{M_{\rm out} v_{\rm out}}{\Delta \rm R}
\label{eq:outflow_rate}
\end{equation}

\noindent 
where $M_{\rm out}$ is plugged in from Eq. \ref{eq:outflow_mass}, the width (FWHM) of the broad Gaussian component of \oiii ~is taken as a proxy for the outflow velocity, $v_{\rm out}$ and $\Delta$R is the pixel size, 0.2\arcsec. It is assumed in Eq. \ref{eq:outflow_rate} that the physical properties of the outflow such as the outflow velocity, density and temperature remain constant within the 0.2\arcsec ~pixel. The total mass outflow rate within the host galaxy is obtained by summing the outflow rates obtained within these individual pixels across the FoV. We will call the outflow rate obtained using Eq. \ref{eq:outflow_rate} as the instantaneous outflow rate. We note that, in the cases where the outflow is resolved, the Eq. \ref{eq:outflow_rate} does not require an additional factor for geometry of the outflows, as the mass outflow rate is computed for each pixel. 

Studies across the literature make use of spectra which are obtained from varying apertures such as a fibre or slit. It has been unclear how such differences in the apertures lead to deviations in the derivation of outflow properties for a large sample of galaxies such as the ones considered in this paper. Therefore, one of the aims of this paper is to compare the outflow scaling relations obtained from the resolved data set with the methods used in the literature. We extract integrated spectra from a circular aperture of 3\arcsec, similar to optical data from the latest Sloan Digital Sky Survey \citep[SDSS, see][]{bundy14, ahumada20}, and from a 1.5\arcsec$\times$10\arcsec ~slit, similar to the instruments at the VLT such as XSHOOTER \citep{vernet11}. Both the circular aperture and the rectangular slit were centred on the AGN location. Emission line fluxes, widths and the mass outflow rates are calculated for both apertures to characterise the impact of the extraction aperture on the line parameters. Multiple slit orientations were also chosen and the values presented in this paper correspond to the orientation where a maximum difference is seen with respect to the circular aperture. 

\begin{figure}
\centering
\includegraphics[scale=0.5]{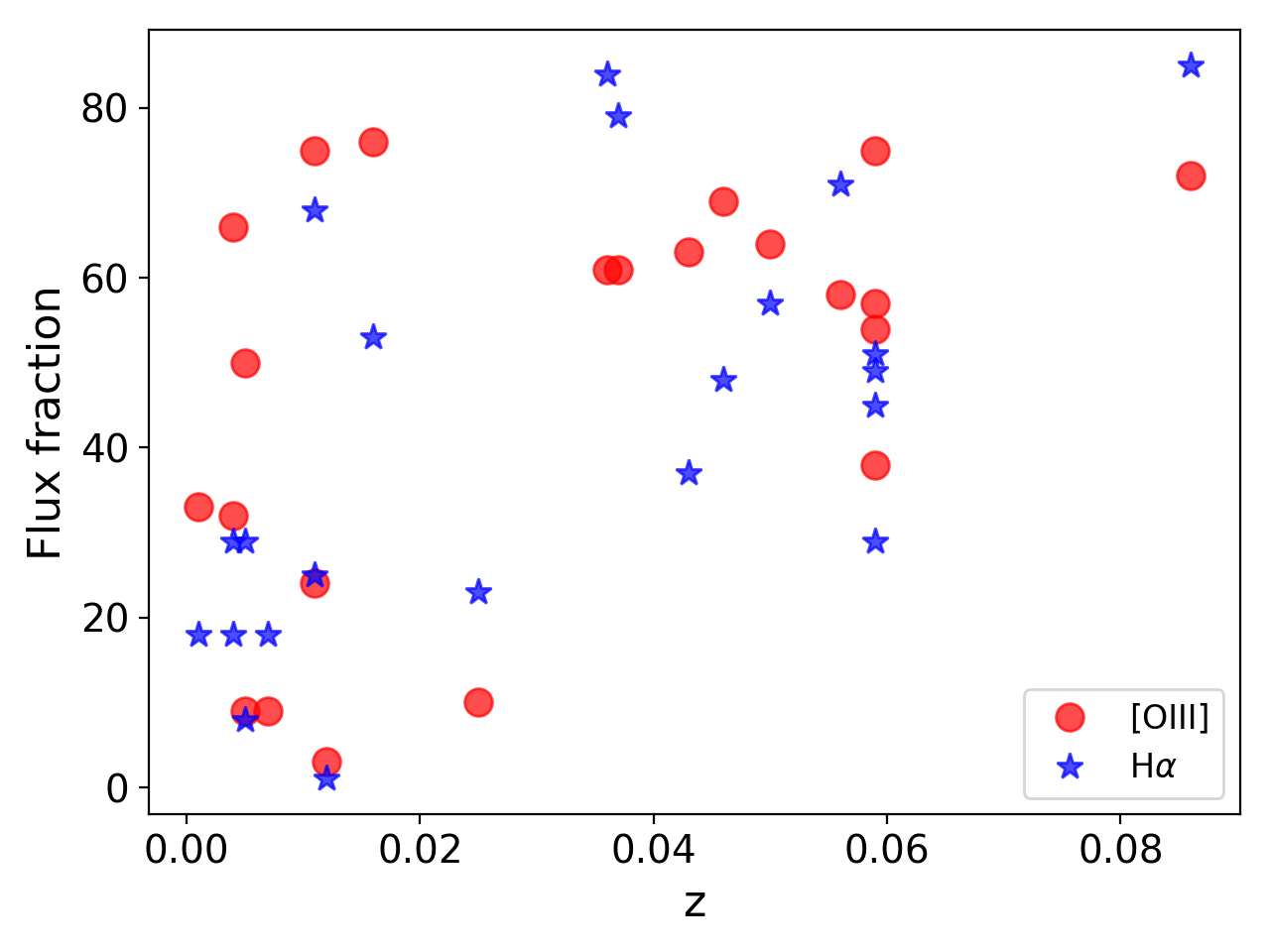}
\caption{Fraction of \oiii ~(red circles) and H$\alpha$ (blue stars) flux within 3\arcsec ~fibre aperture compared to total flux within the MUSE FoV versus the redshift of the targets. The plot shows a wide range in the flux fraction for the entire range of redshift explored in this paper.}
\label{fig:flux_frac_redshift}
\end{figure}

Figure \ref{fig:flux_frac_redshift} already highlights the impact of varying aperture on emission line fluxes, which would consequently also affect the $L_{\rm [OIII]}$ parameter in Eq. \ref{eq:outflow_mass}. Figure \ref{fig:flux_frac_redshift} shows the fraction of \oiii ~and H$\alpha$ flux within a 3\arcsec ~aperture compared to the total \oiii ~and H$\alpha$ flux obtained from the entire MUSE FoV. The \oiii ~and H$\alpha$ radial flux gradient plots for individual galaxies are moved to the Appendix \ref{sect:appendix}. The choice of 3\arcsec ~aperture is motivated by the SDSS fibre aperture often used in the literature. The H$\alpha$ fraction is also shown to highlight the differences in the flux distribution of \oiii ~and H$\alpha$ in the individual galaxies. For targets at redshift $<$ 0.03, the fraction of flux varies between 1--80 \%, while for targets between redshift of 0.03 and 0.1, the fraction is in the range 20--80\%. Such a large range in the fluxes makes it challenging to incorporate aperture corrections with fibre or slit spectroscopy, but can be mitigated with the help of an IFS. In most of the cases, the flux drops with increasing radius, except in some galaxies, such as NGC 1365 and IC 1657, where the gradient is either flat or shows bumps due to the presence of ionisation cones or HII regions.

The stellar continuum and the emission line fitting procedure were kept the same for the integrated spectra as described earlier in this section for the pixel-by-pixel analysis. The errors on the \oiii ~line parameters in the integrated spectra are obtained by repeating the fitting procedure 100 times after adding rms noise from an emission line-free region in the spectra. The errors are the standard deviation of the different values obtained from the repeated fitting procedure. While calculating the outflow properties from the integrated spectra, the outflow mass equation remains the same as in Eq. \ref{eq:outflow_mass}. However, we must invoke the standard assumptions on the outflow geometry which is filled with a uniformly dense ionised gas. We assume a bi-conical outflow geometry, which is commonly adopted in the literature. In this case, Eq. \ref{eq:outflow_rate} takes the form $\dot{M}_{\rm out}$ = $\zeta \cdot$ ($M_{\rm out} v_{\rm out}$/R), where the width (FWHM) of the broad component is the outflow velocity and R is the outflow radius, which we determined from the broad component flux map. As a result, the outflow rate computed here is averaged over the wind lifetime R/$v_{\rm out}$. We note that the instantaneous outflow rate values will be a factor R/$\Delta$R higher than time-averaged outflow rate due to the equations used. The constant $\zeta$ depends on the outflow geometry used and in the case of a uniformly filled bi-conical outflow, $\zeta = 3$ \citep[e.g.,][]{lutz20, veilleux20}. Other outflow morphologies such as a thin-shell model have also been used in the literature, but the systematic error from the assumption of different geometries are within an order of magnitude \citep[e.g.,][]{kakkad20}. 

In summary, we calculate the \oiii ~width and mass outflow rates for spectra obtained from both circular (which we will refer to as ``fibre'' hereafter) and rectangular apertures (referred as ``slit'' hereafter). We also compute the total mass outflow rate obtained from the resolved mass outflow rate maps (also referred to as instantaneous outflow rates) and compare to the values obtained from the integrated spectra (also refereed to as time-averaged outflow rates). The velocity and outflow rate values from the different methods described in this section are reported in Table \ref{table:outflow_properties}.

\begin{figure*}
\centering
\includegraphics[width=0.99\textwidth]{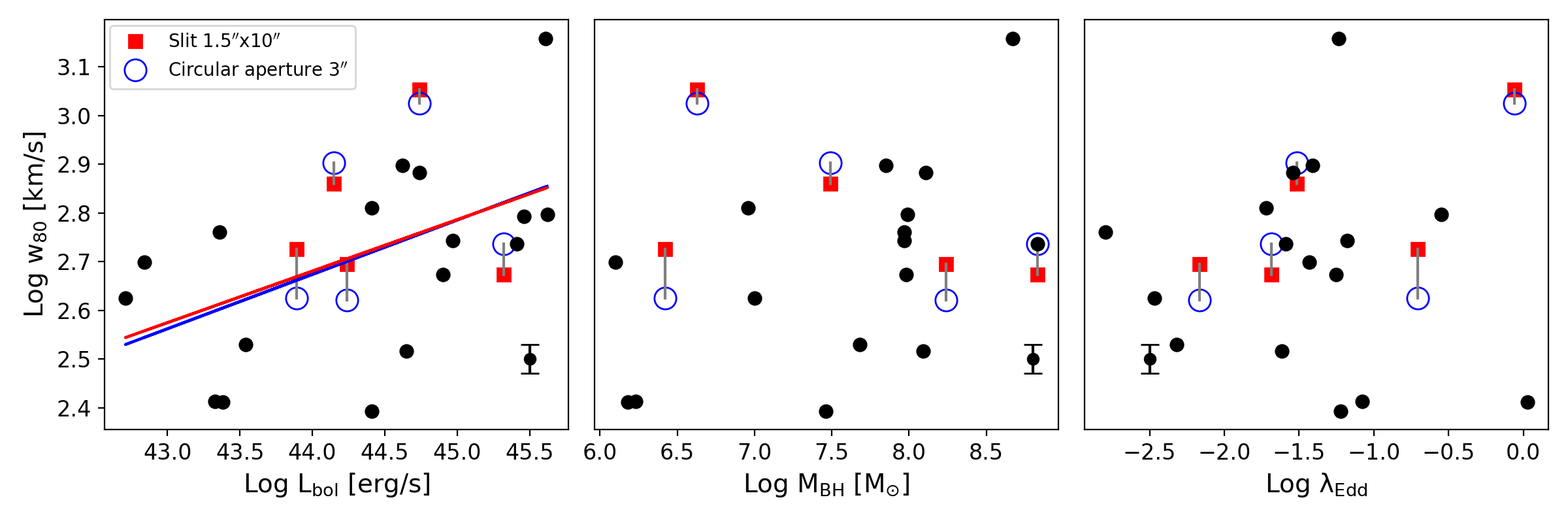}
\caption{These plots show the relation between the non-parametric velocity dispersion, $w_{80}$ (corrected for spectral resolution) obtained from the 3\arcsec ~fibre and 1.5\arcsec$\times$10\arcsec ~rectangular slit versus the AGN properties, namely L$_{\rm bol}$ (left panel), M$_{\rm BH}$ (middle panel) and $\lambda_{\rm Edd}$ (right panel). The $w_{80}$ for majority of these galaxies show consistent measurements between the different methods of spectral extraction (shown as solid black circles). The exception is for 4 galaxies where the difference in $w_{80}$ values is at most 75 km s$^{-1}$ (shown as open blue circles for fibre aperture and solid red squares for slit aperture, both connected by a grey line). The blue and red curves show the best-fit linear relations for the fibre and slit spectra respectively. The best-fit relation is only shown in the cases where the p-value for a non-correlation is $<$0.05. The two curves show that there is a negligible impact of the aperture size and shape on the global $w_{80}$ scaling relations. The black symbol with the error bar in the lower right (or lower left) region of the plots show the typical uncertainty in the $w_{80}$ values.}
\label{fig:w80_integrated}
\end{figure*}

\begin{figure*}
\centering
\includegraphics[scale=0.355]{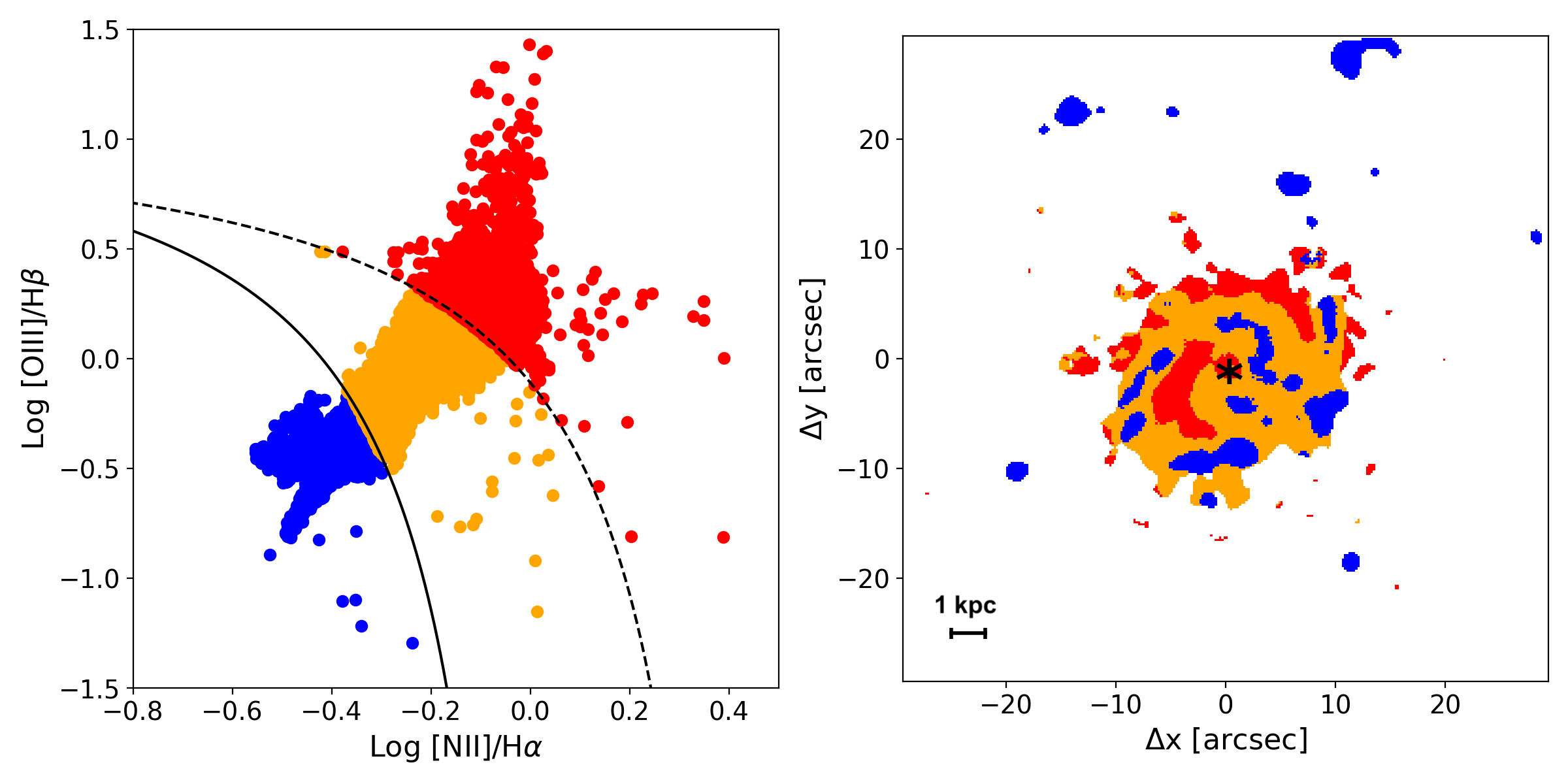}\\
\includegraphics[scale=0.35]{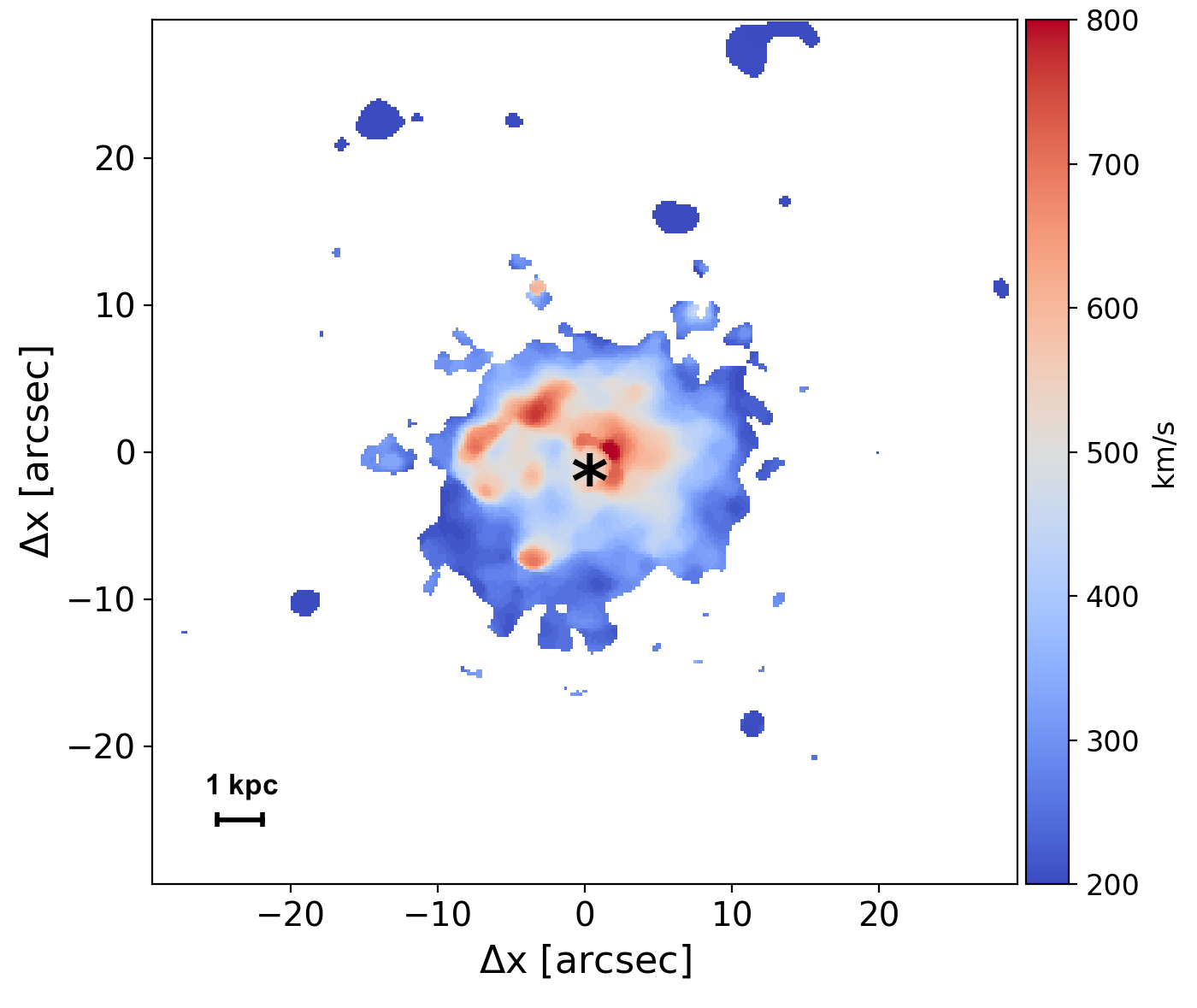}
\includegraphics[scale=0.35]{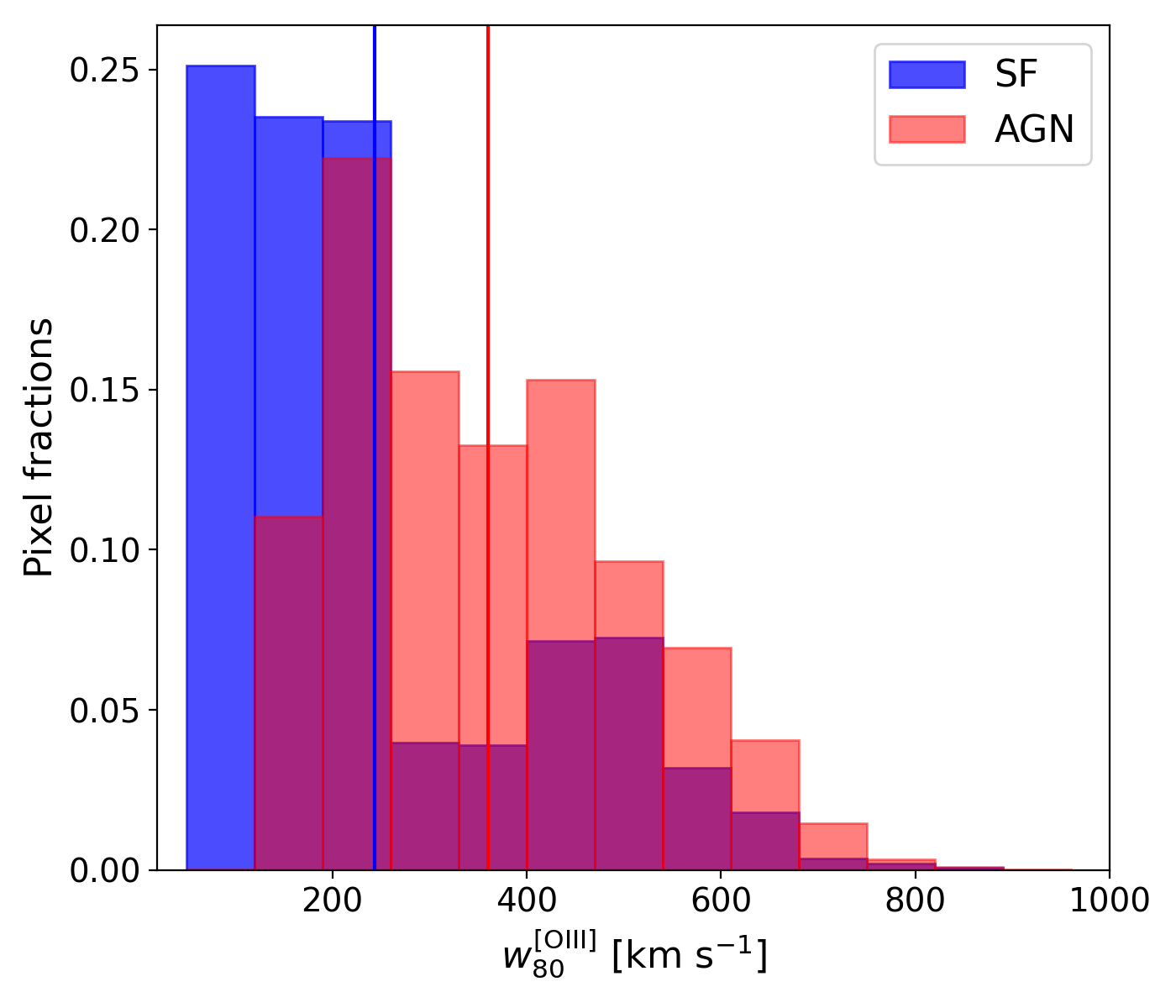}
\caption{The top panels show the location of spaxels in the \nii ~BPT diagram (i.e. \oiii/H$\beta$ versus \nii/H$\alpha$ plot, top left panel) and the corresponding BPT map (top right panel) of NGC 7469. Regions dominated by AGN, star formation, and composite ionisation are denoted in red, blue, and orange colours, respectively. The dashed curve in the top left panel corresponds to the extreme starburst line from \citet{kewley01} and the solid black curve obtained from \citet{kauffmann03} shows the demarcation between ionisation due to star forming and composite processes. The AGN \& composite ionisation is centrally concentrated, while the ionisation due to star formation processes are dominant in the spiral arms. Only the regions with S/N$>$10 in all the emission lines are shown in this map. The lower left panel shows the \oiii ~$w_{80}$ map for this galaxy with the same S/N cut as in the top left panel. The lower right panel shows the \oiii ~$w_{80}$ distribution for spaxels ionised by the AGN (red) and spaxels ionised by star formation (blue) from the top and lower left panels. Composite regions of the BPT diagram are not shown in this distribution plot as only the regions ionised by the extreme two processes- star formation or AGN are considered. The $w_{80}$ distribution in AGN ionised regions are clearly skewed towards higher velocities compared to the star forming regions. The vertical blue and red lines show the mean value of the star forming and AGN $w_{80}$ distributions, respectively. The mean velocities of the two distributions differ by $\sim$120 km s$^{-1}$ (after taking into account the limited spectral resolving power of MUSE). The horizontal black bar in the maps represents the 1 kpc physical scale.}
\label{fig:resolved_BPT_NGC7469}
\end{figure*}

\begin{figure}
\centering
\includegraphics[scale=0.3]{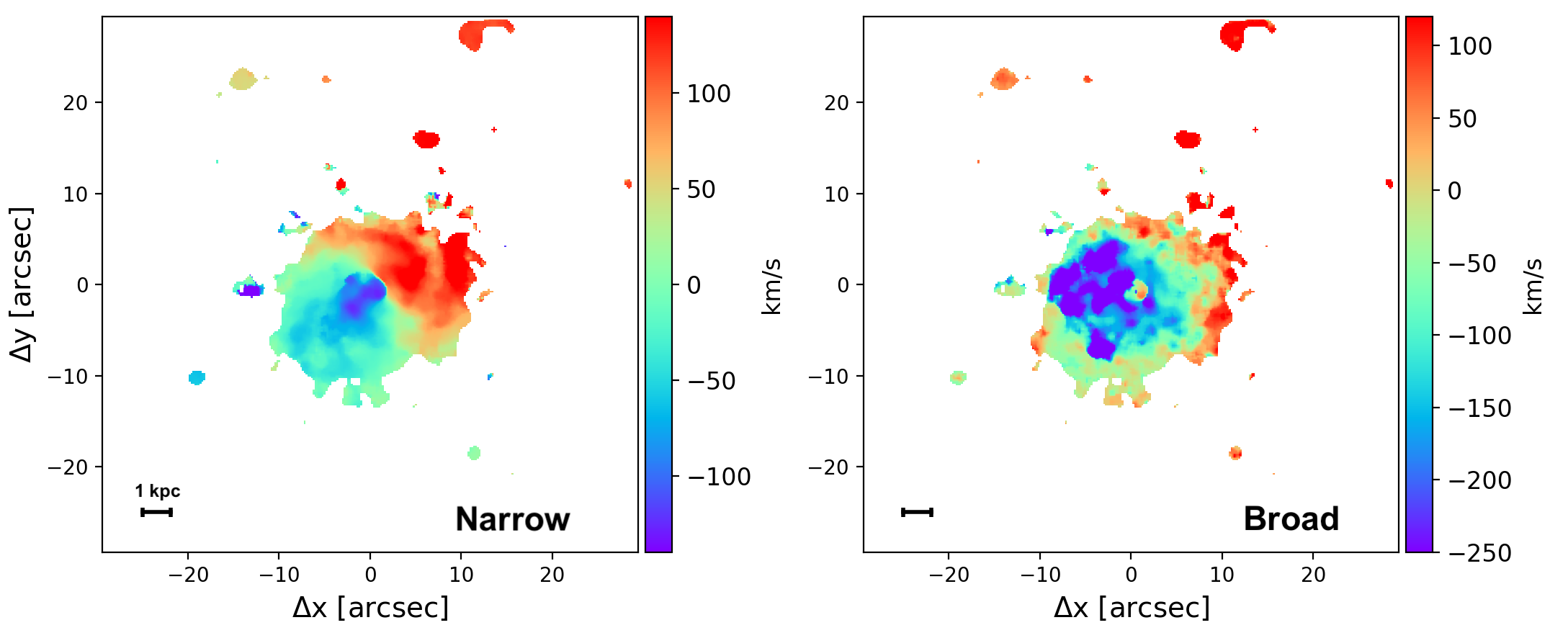}
\includegraphics[scale=0.3]{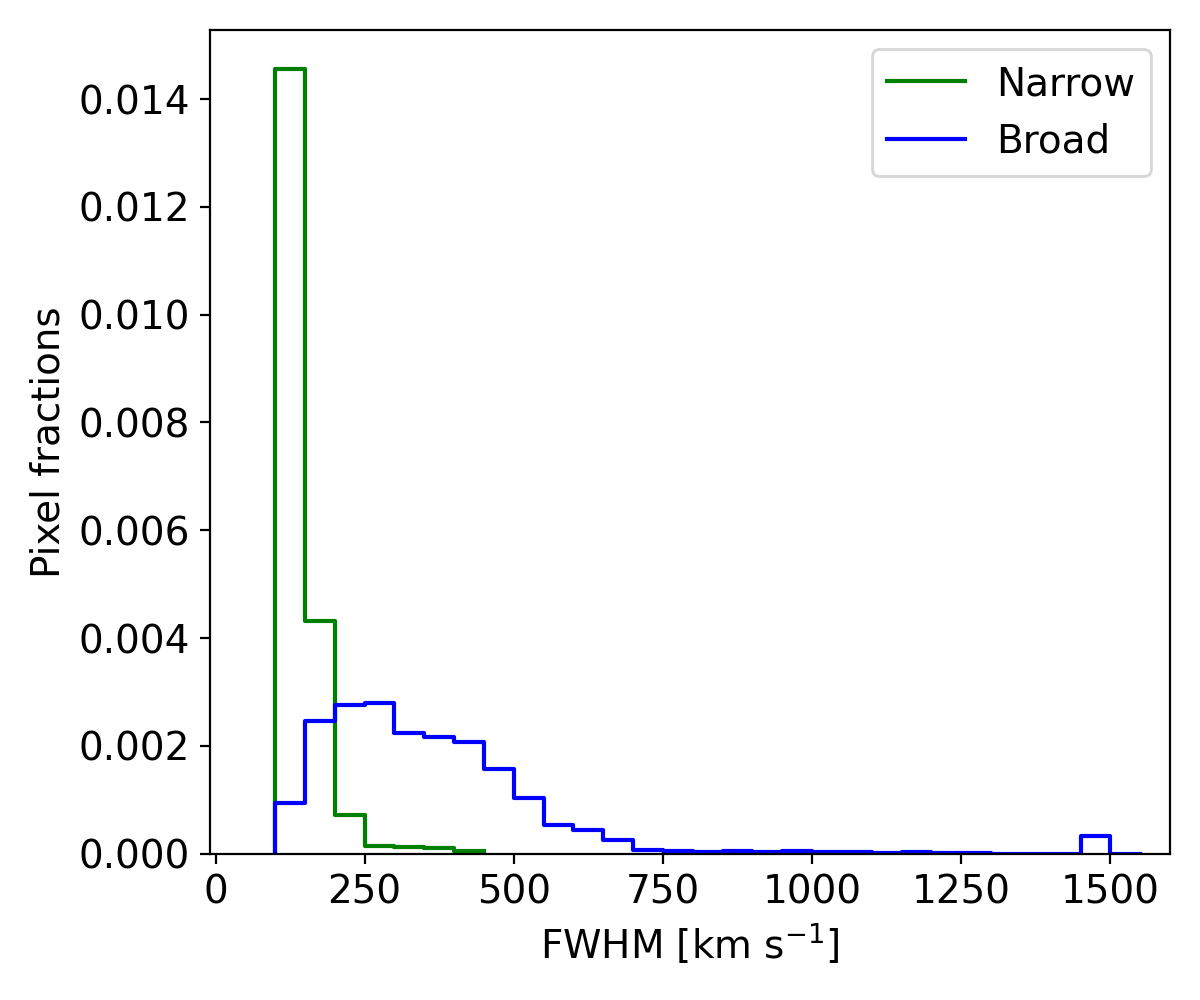}
\caption{The top panels show maps of the best-fit line centroids of the narrow (top left panel) and broad (top right panel) Gaussian components of the \oiii ~emission in NGC 7469 as an example. The centroid map of the narrow \oiii ~component shows a smooth variation from $\sim$-150 to +150 km s$^{-1}$, suggesting that the narrow component traces the systemic component of the host galaxy (or in other words, the non-outflowing component). The bottom panel shows the width (FWHM) distribution of the narrow (green) and the broad (blue) Gaussian component of the \oiii ~emission across the MUSE FoV where the S/N in the \oiii ~emission is $>$10. The broad component widths can reach values $>$700 km s$^{-1}$, which cannot be explained by bar streaming in NGC 7469.  This indicates a possible AGN origin of the central outflows in this galaxy \citep[e.g.,][]{luminari21}.}
\label{fig:narrow_broad_distribution}
\end{figure}

\section{Results} \label{sect5}

In the following section, we present the integrated and spatially resolved properties of the outflows such as the velocity, mass outflow rates and outflow scaling relations from the analysis methods described in Sect. \ref{sect4}. 

\subsection{Velocity distributions} \label{sect5.1}

Before describing the spatially resolved velocity distributions from the individual components of the \oiii ~line, we first analyse the overall \oiii ~profile in the integrated fibre and slit spectra extracted from the MUSE cubes. This will help us to understand if the choice of aperture shape and size introduces any differences in the overall \oiii ~line width, and consequently the kinematics of the ionised gas. The $w_{80}$ parameter introduced in Sect. \ref{sect4.1} is the ideal choice to quantify such changes in the observed total \oiii ~line width, as it is not dependent on the adopted model to reproduce the emission line profile. However, we note that results presented here using the non-parametric $w_{80}$ values also hold true if a parametric velocity measurement, such as $v_{\rm max}$ is used instead (defined as $\Delta\lambda$ + 2$\sigma_{\rm broad}$, where the former component is the difference in the centroids of the narrow and broad Gaussian. See \citet{rojas20} for more details). The relevant plots with $v_{\rm max}$ can be found in the Appendix \ref{sect:appendix}. We hereafter refer to the $w_{80}$ value as the \oiii ~width. Table \ref{table:outflow_properties} reports the $w_{80}$ values for fibre ($w_{80}^{\rm fibre}$) and slit ($w_{80}^{\rm slit}$) spectra for the targets presented in this paper. The majority of the galaxies (>80\%) show consistent \oiii ~widths for spectra obtained from fibre or slit, while five galaxies show differences which are within $\sim$75 km s$^{-1}$, which is within the spectral resolving power of MUSE at the location of the \oiii ~line. These small differences arise due to the slit orientation along a bi-conical outflow in these cases, while the fibre partially misses the bulk of the outflowing gas. Notably, these differences do not significantly affect the relations between velocity dispersion ($w_{80}$) and the AGN properties, as seen in Fig. \ref{fig:w80_integrated}, which shows the relation between the \oiii ~$w_{80}$ obtained from the fibre and slit apertures versus the bolometric luminosity ($L_{\rm bol}$, left panel), black hole mass ($M_{\rm BH}$, middle panel) and Eddington ratio ($\lambda_{\rm Edd}$, right panel) of the AGN. The $w_{80}$ measurements that are consistent between the two methods are shown as black solid circles in Fig. \ref{fig:w80_integrated}, while for the five galaxies that show differences, red squares represent $w_{80}$ measurement from the slit aperture and open blue circles represent measurement from the fibre aperture.

We also performed correlation tests to verify the presence or absence of correlations between the $w_{80}$ values and the AGN properties, namely $L_{\rm bol}$, $M_{\rm BH}$ and $L_{\rm Edd}$. The Pearson correlation coefficient, the p-value for testing non-correlation and the slope of these relations are reported in Table \ref{table:correlation_tests}. The tests show similar results for the $w_{80}$ obtained from fibre and slit spectra. The probability for non-correlation with $L_{\rm bol}$ is $\sim$2\%. This probability increases in the case of $M_{\rm BH}$ (9--13\%) and $\lambda_{\rm Edd}$ (67-76\%), where we conclude that there is a weak or no correlation with these two quantities. The best-fit linear relations, only shown where the p-value $<$0.05, also show nearly consistent results for the two integrated spectra, suggesting a negligible impact of the aperture shape and size on relation between the global velocity dispersion versus AGN properties. Similar correlation results are obtained when using $v_{\rm max}$ instead of $w_{80}$ (considering uncertainties, the correlations with $v_{\rm max}$ are more robust), as shown in Fig. \ref{fig:vmax_scalings} and Table \ref{table:correlation_tests}.

The possible presence of linear correlation between the $w_{80}$ parameter from the integrated spectrum and $L_{\rm bol}$ suggests a possible role of the AGN in regulating the kinematics of the ionised gas. We therefore investigated whether the evidence of such an AGN impact on the \oiii ~based ionised gas velocity is also present in the resolved data. Therefore, we first made use of the BPT diagnostic maps \citep[Baldwin, Phillips \& Terlevich:][]{baldwin81, veilleux87} to identify pixels or regions ionised by the AGN, star formation or composite sources \citep[e.g.,][]{dopita14, belfiore16, davies17, kakkad18}. This is shown in the top panels of Fig. \ref{fig:resolved_BPT_NGC7469}. We then compare the \oiii ~$w_{80}$ distribution (after correcting for the spectral resolution) in the pixels ionised by AGN and star formation, as shown in the bottom right panel of Fig. \ref{fig:resolved_BPT_NGC7469}. The BPT maps and $w_{80}$ distribution plots for the rest of the AGN sample presented in this paper are moved to the Appendix \ref{sect:appendix}. 

Figure \ref{fig:resolved_BPT_NGC7469} clearly shows that the pixels or regions ionised by the AGN are skewed towards higher \oiii ~$w_{80}$ values in NGC 7469. Star forming regions show a mean $w_{80}$ value of $\sim$240$\pm$150 km s$^{-1}$, while the AGN ionised regions show a higher fraction of $w_{80}$ values greater than 600 km s$^{-1}$, with a mean of 360$\pm$140 km s$^{-1}$. Similar distributions have previously been observed in the case of MaNGA galaxies in \citet{wylezalek20} where the AGN ionised pixels show a much higher velocity compared to star formation ionised pixels. The bottom right panel in Fig. \ref{fig:resolved_BPT_NGC7469} shows that even with the finer sampling of higher spatially resolved MUSE data (compared with the lower resolution and coarsely sampled MaNGA sample), the previous literature results hold true. We note that this does not mean that the spaxels with AGN ionisation will necessarily show higher \oiii ~$w_{80}$ value, as from figure \ref{fig:resolved_BPT_NGC7469} it is clear that even spaxels with AGN ionisation can show lower $w_{80}$ values. However, the AGN ionised regions will tend to show a collectively higher ionised gas velocity dispersion than the star forming pixels. In most of the cases, regions close to the AGN (which are clearly in the AGN ionisation region of the BPT diagram) show higher $w_{80}$ values. The results in Fig. \ref{fig:resolved_BPT_NGC7469} are therefore indicative of AGN activity having a greater impact on the kinematics of the ionised gas compared to star formation processes. The $w_{80}$ radial profiles (shown in the Appendix for individual galaxies) show a common trend that the values uniformly drop from the centre. However this is not always true as apparent from the $w_{80}$ distribution in NGC 7469 in Figs. \ref{fig:resolved_BPT_NGC7469} (bottom left panel) \& \ref{fig:plots_NGC7469} (panel d) and NGC 1365 in Fig. \ref{fig:plots_NGC1365}. The $w_{80}$ value increases from the AGN location, peaking at a distance of a few arcsec before dropping off at larger distances.

So far in this section we focused on the overall \oiii ~line profile to indicate that the AGN has an impact on the width ($w_{80}$) of the \oiii ~line. However, as mentioned earlier, we make use of the narrow and broad Gaussian decomposition to define a systemic gas (non-outflowing component) and the outflowing component respectively. Figure \ref{fig:narrow_broad_distribution} shows an example of the narrow and broad Gaussian width distributions in NGC 7469, which supports this assumption for the analysis presented in this paper. The top left panel of Fig. \ref{fig:narrow_broad_distribution} shows the centroid distribution of the narrow Gaussian component which shows a smooth rotation-like profile about an axis oriented approximately along the S-W direction. Such a smooth profile is not present in the centroid map of the broad component (top right panel in Fig. \ref{fig:narrow_broad_distribution}). Furthermore, the bottom panel in Fig. \ref{fig:narrow_broad_distribution} shows the width (FWHM) distribution of the individual narrow and broad Gaussian components across the MUSE FoV. The narrow component width is always less than 250 km s$^{-1}$, while the broad component reaches values of >600 km s$^{-1}$ in some pixels, suggesting that this component is indeed tracing the high velocity outflowing gas possibly driven by the AGN \citep[e.g.,][]{luminari21}. 

\subsection{Resolved mass outflow rate maps} \label{sect5.2}

In this and the next sections, we show the resolved ionised gas mass outflow rate maps derived from spaxel-by-spaxel analysis and compare these values with measurements from the integrated fibre or slit spectra. 

\begin{figure*}
\centering
\includegraphics[width=\textwidth]{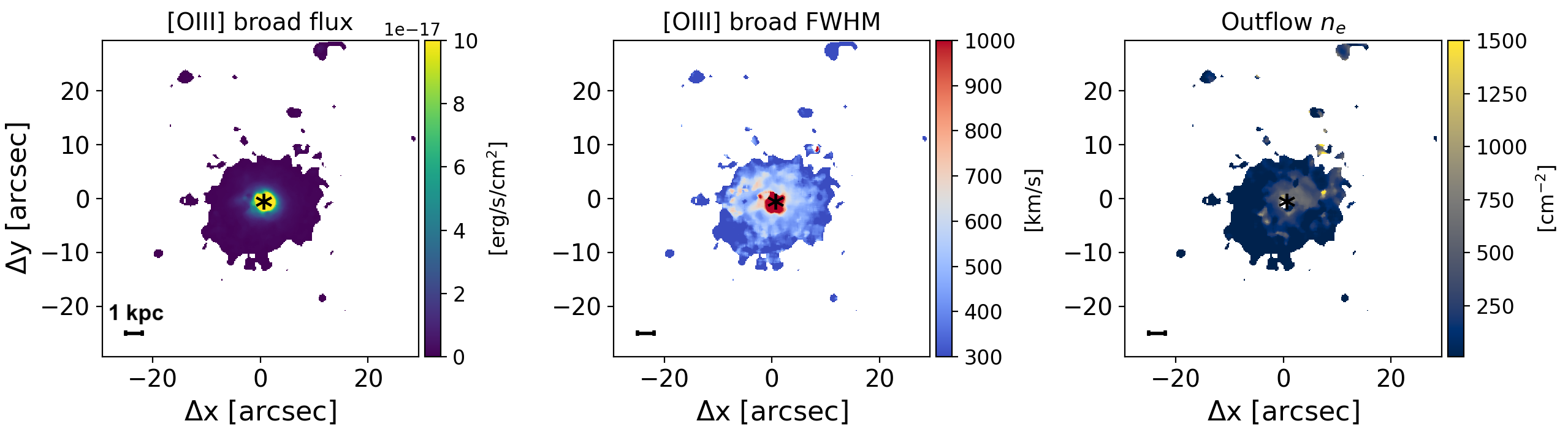}
\caption{The left panel shows the broad component \oiii ~flux map of NGC 7469, one of the galaxies in the BASS-MUSE sample, which indicates the presence of a strong outflow (high broad \oiii ~flux, marked by the yellow regions) close to the AGN location (black star). The middle panel shows the width (FWHM) map of the broad Gaussian component of \oiii, as a tracer of the outflow velocity. The outflow velocity is also highest close to the AGN and in clumps distributed across the FoV. The right panel shows the electron density map, obtained from the flux ratio of the broad Gaussian components of the \sii ~doublet (Eq. \ref{eq:electron_density}). The electron density also shows a non-uniform structure and is in the range $<$10-1500 cm$^{-3}$. These three maps are used to derive the ionised gas outflow rate map shown in Fig. \ref{fig:outflow_rate_map_NGC7469}. Further details about these maps are given in Sect. \ref{sect5.2}. The maps for the rest of the targets are moved to the appendix \ref{sect:appendix}. \label{fig:flux_velocity_density_NGC7469}}
\end{figure*}

Figure \ref{fig:flux_velocity_density_NGC7469} shows the outflow flux (left panel), velocity (FWHM, middle panel) and the electron density (right panel) distribution in NGC 7469, one of the galaxies in the BASS-MUSE sample as an example. These maps are obtained from the broad Gaussian component of the \oiii ~line. Specifically in the case of NGC 7469, the outflow flux is concentrated close to the AGN location, marked by the black star, and the flux significantly drops with increasing distance from the AGN (by a factor of $>$10). Similar to the flux distribution, the outflow velocity defined by the width (FWHM) of the broad Gaussian, is also maximum at the center ($\sim$1000 km s$^{-1}$) and falls to $\sim$200 km s$^{-1}$ towards the spiral arms. However, this drop is not uniform as the velocity and flux distributions both show clumpy profiles within the outflowing medium. We note that the outflow across the FoV might not be entirely due to the AGN, as the asymmetry in the \oiii ~line profile could result from residual turbulence from star formation or supernovae driven winds \citep[e.g.,][]{daviesR19, avery21, herrera-camus21}. The electron density map also shows a non-uniform distribution, similar to the ones observed previously in the literature with lower spatial resolution data \citep[e.g.,][]{kakkad18}.

Several galaxies in the BASS-MUSE sample also show a bi-conical ionisation morphology such as IC 1657, NGC 1365 \citep[e.g.,][]{venturi18}, NGC 2992 \citep[e.g.,][]{veilleux01} and Mrk 463 \citep{treister18}. Often the outflow flux is concentrated towards the tip of this ionisation cone and falls off as a function of distance from the AGN. The outflow velocity and density, on the other hand, are distributed in the form of clumps within this ionisation cone. This is expected as the gas might interact with the stars and the dust within the ISM leading to non-uniform distribution of these quantities. The flux, velocity and electron density maps for the rest of the sample are shown in the Appendix \ref{sect:appendix}. 

\begin{figure}
\centering
\includegraphics[scale=0.45]{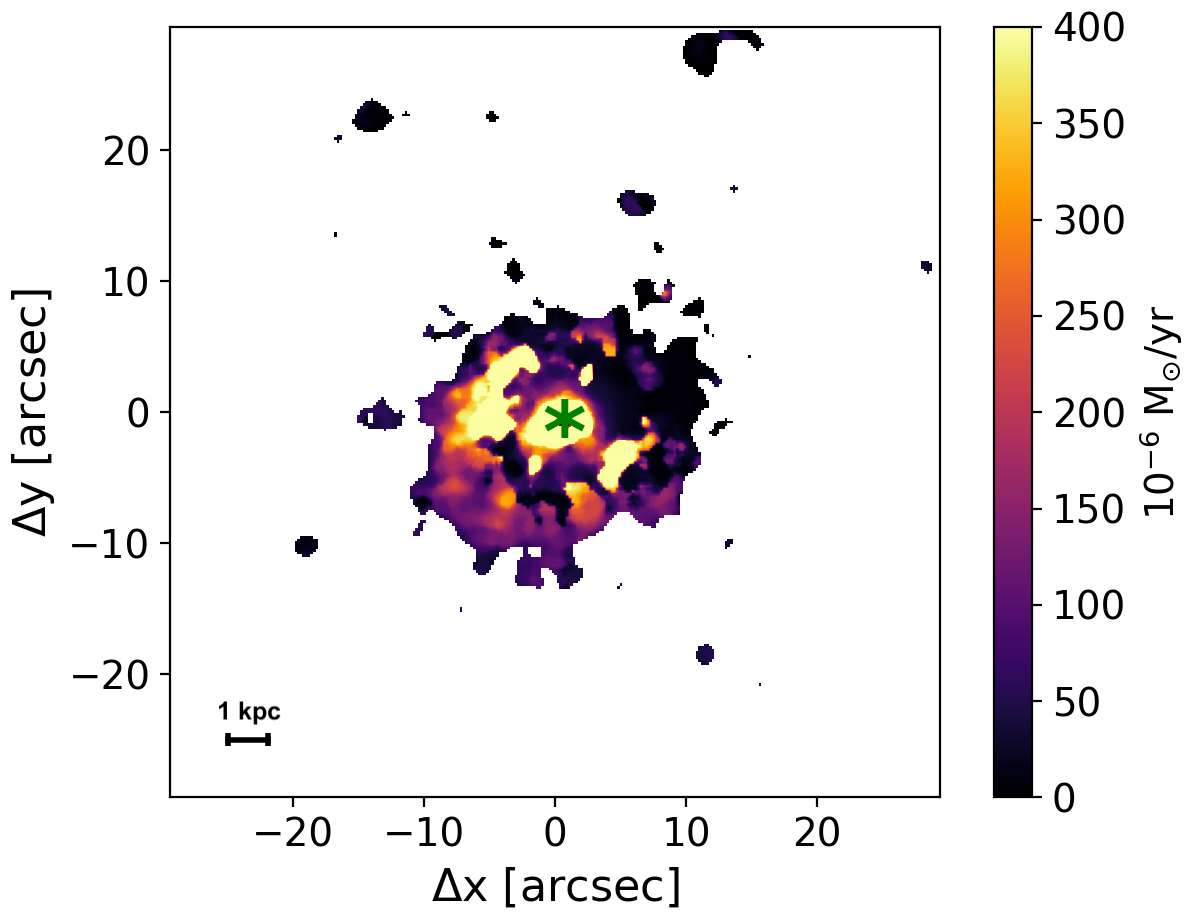}
\caption{The map shows the local instantaneous ionised gas mass outflow rate distribution in NGC 7469. The mass outflow rate shows a non-uniform distribution and is concentrated along an arc-like structure towards the SE, which is also seen in multiple galaxies in the BASS-MUSE sample. The observed distribution is a consequence of the variable density and outflow velocity, apparent from the middle and right panels in Fig. \ref{fig:flux_velocity_density_NGC7469}. The green star shows the location of the AGN. The outflow rate maps for the rest of the targets in this paper are moved to the appendix.  \label{fig:outflow_rate_map_NGC7469}}
\end{figure}

The outflow luminosity does not represent the distribution of the ionised gas mass or the mass outflow rate. This is because the outflow mass and the mass outflow rates are dependent on the three quantities -- flux, velocity and electron density -- as apparent from Eqs. \ref{eq:outflow_mass} and \ref{eq:outflow_rate}. The ionised gas mass outflow rate map of NGC 7469 is shown in Fig. \ref{fig:outflow_rate_map_NGC7469}. The mass outflow rate distribution is non-uniform across the FoV and is concentrated along an arc towards the SE direction. The arc-like distribution in the mass outflow rate is also observed in several other galaxies such as NGC 5995 and 3C403. Such outflow morphology could be indicative of expansion of the ionised gas in the form of spherical shells from the AGN location. Within the spiral arms, such as in the case of NGC 7469, the outflow mass is negligible compared to regions close to the AGN, suggesting that the regions close to the AGN dominate the outflow mass as well as the outflow rate budget. We further discuss the outflow rates close to the AGN location later in this section. 

The total instantaneous outflow rates ($\dot{M}_{\rm res}^{\rm total}$) derived from the resolved maps are in the range $\sim$0.2--275 M$_{\odot}$ yr$^{-1}$, which are nearly consistent with previously published values in the literature for some of the galaxies presented in this paper \citep[e.g., NGC 1365 and Mrk 463, ][]{venturi18, treister18}. Table \ref{table:outflow_properties} also reports the mass outflow rate within the central 3\arcsec ~of these outflow rate maps, outflow rate obtained from fibre ($\dot{M}_{\rm fibre}$) or slit ($\dot{M}_{\rm slit}$) spectra and the outflow rate if the electron density is assumed to be 200 cm$^{-3}$ ($\dot{M}_{\rm 200}$), a commonly adopted value in the literature \citep[e.g.,][]{fiore17, davies20}.

Table \ref{table:outflow_properties} highlights that the outflow rate values are highly dependent on the method of spectral extraction. The ratio between $\dot{M}_{\rm res}^{\rm total}$ and the time-averaged mass outflow rate obtained from fibre or slit spectra ($\dot{M}_{\rm fibre}$ or $\dot{M}_{\rm slit}$) ranges from 0.4--525 with a mean of $\sim$100. In other words, $\dot{M}_{\rm res}^{\rm total}$ is on average about two orders of magnitude higher than $\dot{M}_{\rm fibre}$ or $\dot{M}_{\rm slit}$. To understand the difference between the two methods, we also compared the outflow mass obtained using the pixel-by-pixel analysis and using the fibre and slit aperture integrated spectra. The ratio of the outflow masses obtained from the resolved data to that of fibre or slit aperture spectra is in the range 0.01--130, with a mean ratio of $\sim$20 i.e. summed local outflow masses are an order of magnitude higher than outflow masses calculated from integrated spectra. Therefore, the observed difference in the mass outflow rate values could be a consequence of the fact that in the resolved maps, the outflow mass is calculated from a larger area (entire galaxy) compared to integrated spectra where the mass is obtained from a limited aperture size.  Furthermore, these differences are also expected due to the different equations used in the computation of mass outflow rates, as described in Sect. \ref{sect4.2}. The summed instantaneous outflow rates are higher than the time-averaged integrated measurements by a factor of R/$\Delta$R. 

A further important insight is obtained by calculating the mass outflow rate from a 3\arcsec ~aperture on the resolved maps, $\dot{M}_{\rm res}^{\rm cent}$ (so as to match the area with the fibre extraction spectrum). We find that the mean ratio of $\dot{M}_{\rm res}^{\rm cent}$ to $\dot{M}_{\rm fibre}$ is $\sim$35. The average fraction of mass outflow rate concentrated in a 3\arcsec ~region centred on the AGN (i.e. $\dot{M}_{\rm res}^{\rm cent}$/$\dot{M}_{\rm res}^{\rm total}$) is 55\% i.e. roughly half of the total instantaneous outflow rate is concentrated in the central 3\arcsec. 

We do not make a distinction in the source of ionisation while calculating the mass outflow rates. Therefore, the entire FoV is used to calculate the local instantaneous mass outflow rates. This is also to ensure that all the outflowing gas is taken into account while calculating the mass outflow rates. As one of the main aims of this paper is to compare the local instantaneous mass outflow rate values (obtained from resolved maps) with that of the time averaged values (obtained from the integrated spectra) and that the scaling relations with the time-averaged values are not distinguished based on the source of ionisation, we do not make the distinction for consistency between the two measurements. Also, $>$90\% of the mass outflow rate comes from AGN-ionised regions and 16 out of the 22 galaxies presented in the paper have their field-of-view dominated by AGN ionisation. Therefore, our results would not change significantly if only AGN ionised regions are used.

\begin{figure}
\centering
\includegraphics[scale=0.45]{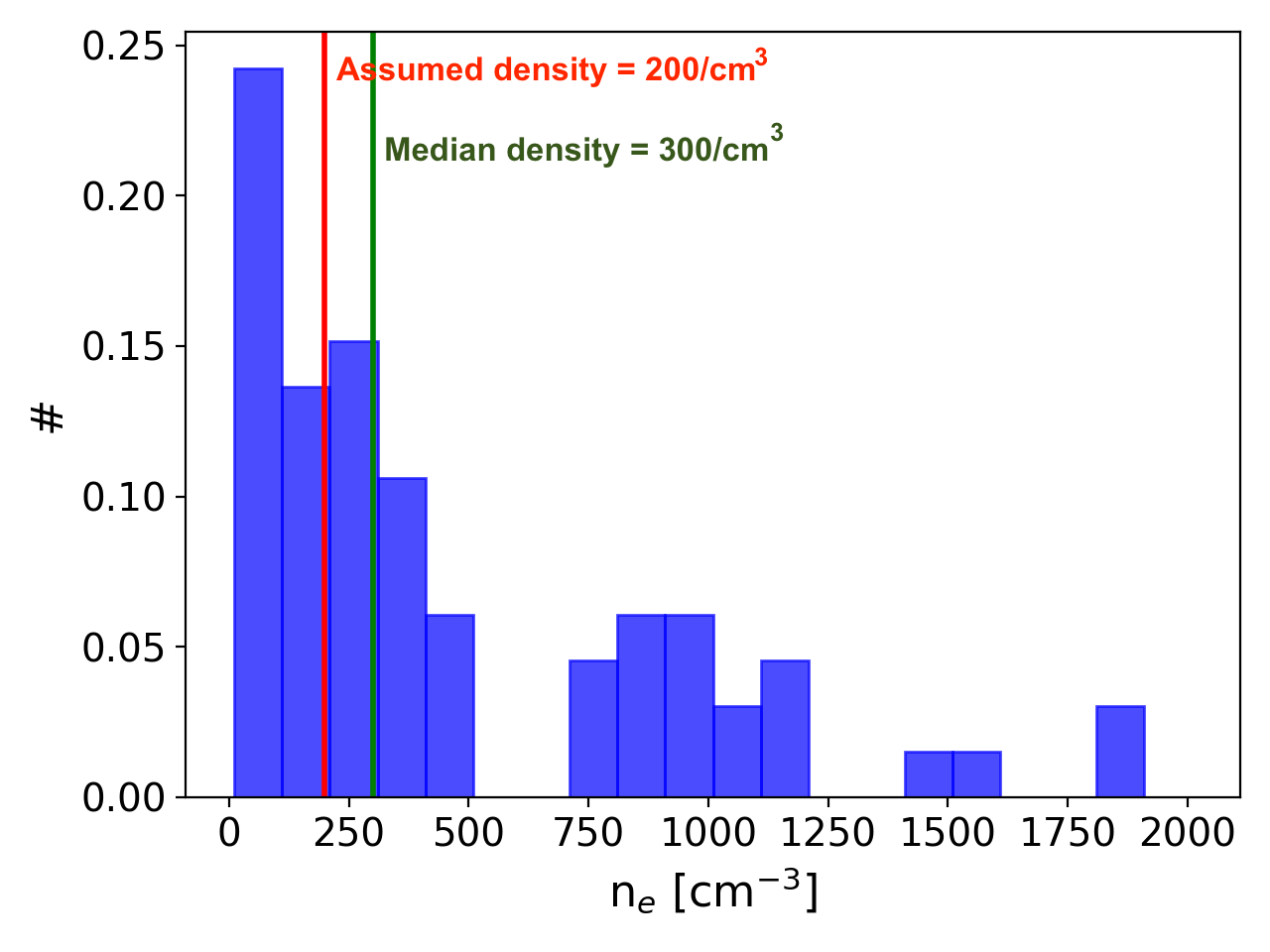}
\caption{Electron density distribution obtained from the \sii ~doublet ratios in the fibre and slit integrated spectra of all the BASS-MUSE targets. The y-axis shows the fraction of targets with the corresponding electron density. The commonly assumed electron density value of 200/cm$^{3}$ is similar to the median value of the distribution, 300/cm$^{3}$, we find for the targets presented in this paper. The electron density values span a wide range as they depend on the shape of the aperture (fibre or slit) and the orientation of the slit (along or away from the outflow). As the electron density values are obtained from integrated spectra with high S/N, the measurement errors are estimated at $\pm$200 cm$^{-3}$.}
\label{fig:ne_distribution}
\end{figure}

We also analyse the values of electron density derived from the outflowing component of \sii ~doublet for integrated fibre and slit spectra. We use Eq. \ref{eq:electron_density} to derive the electron density from the outflowing components of the \sii ~doublet emission lines. Figure \ref{fig:ne_distribution} shows the distribution of the outflow electron density obtained from the integrated spectra, which is in the range $<$10--1800 cm$^{-3}$. We find a median electron density value of $\sim$300 cm$^{-3}$, which is similar to the commonly assumed density value of 200 cm$^{-3}$ in the literature \citep[e.g.,][]{fiore17}. This density is much lower than other integrated spectra studies targeting low redshift X-ray AGN hosts \citep[e.g.,][]{perna17}. However, there is a large range in the calculated densities, which depend on the method of spectral extraction and the slit orientation.  

To summarise, the summed instantaneous mass outflow rates from resolved data show a higher outflow rate value than time-averaged outflow rates from integrated spectra. This is due to a combination of the equations used in the computation of these quantities and that a larger mass is incorporated with the larger FoV in the instantaneous case. The resolved outflow rate maps provide a means to incorporate variable outflow density, velocity and flux distribution to give a realistic picture of the distribution of the outflow mass across the host galaxies.

\begin{figure*}
\centering
\includegraphics[width=0.99\textwidth]{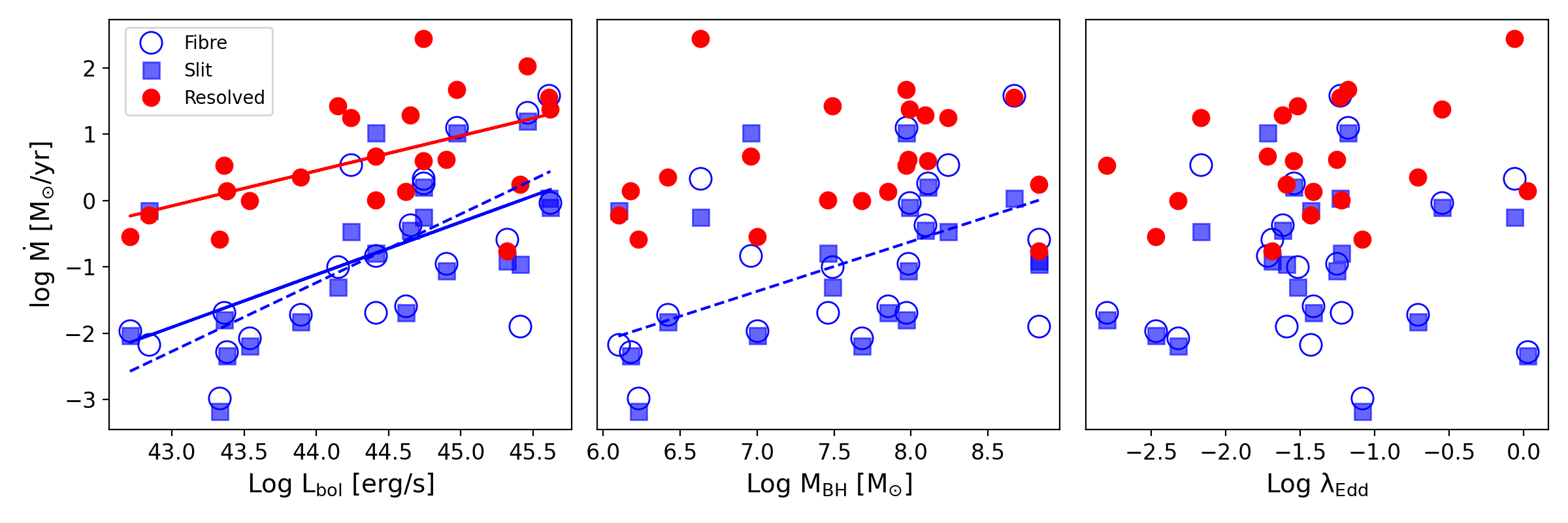}
\caption{The plots shows the relation between mass outflow rate, obtained from different methods, versus $L_{\rm bol}$ (left panel), $M_{\rm BH}$ (middle panel) and $\lambda_{\rm Edd}$ (right panel). The red data points show the summed instantaneous mass outflow rate obtained from co-adding contributions from all pixels in the mass outflow rate maps (labelled resolved) and the red curve shows the best-fit linear relation. The open and filled blue data points show the time-averaged mass outflow rates obtained from integrated fibre and slit spectra respectively. The dashed and solid lines show the best-fit linear relations in the case of fibre and slit apertures respectively. The best-fit relations are shown if the p-value for non-correlation is $<$0.05. The results of statistical correlation tests are reported in Table \ref{table:correlation_tests}. The mass outflow rates from all the methods show a high probability of correlation with $L_{\rm bol}$. The correlations are weak or non-existent with $M_{\rm BH}$ and $\lambda_{\rm Edd}$. For further details, see Sect. \ref{sect5}. \label{fig:outflow_scaling_aperture}}
\end{figure*}

\begin{figure}
\centering
\includegraphics[scale=0.4]{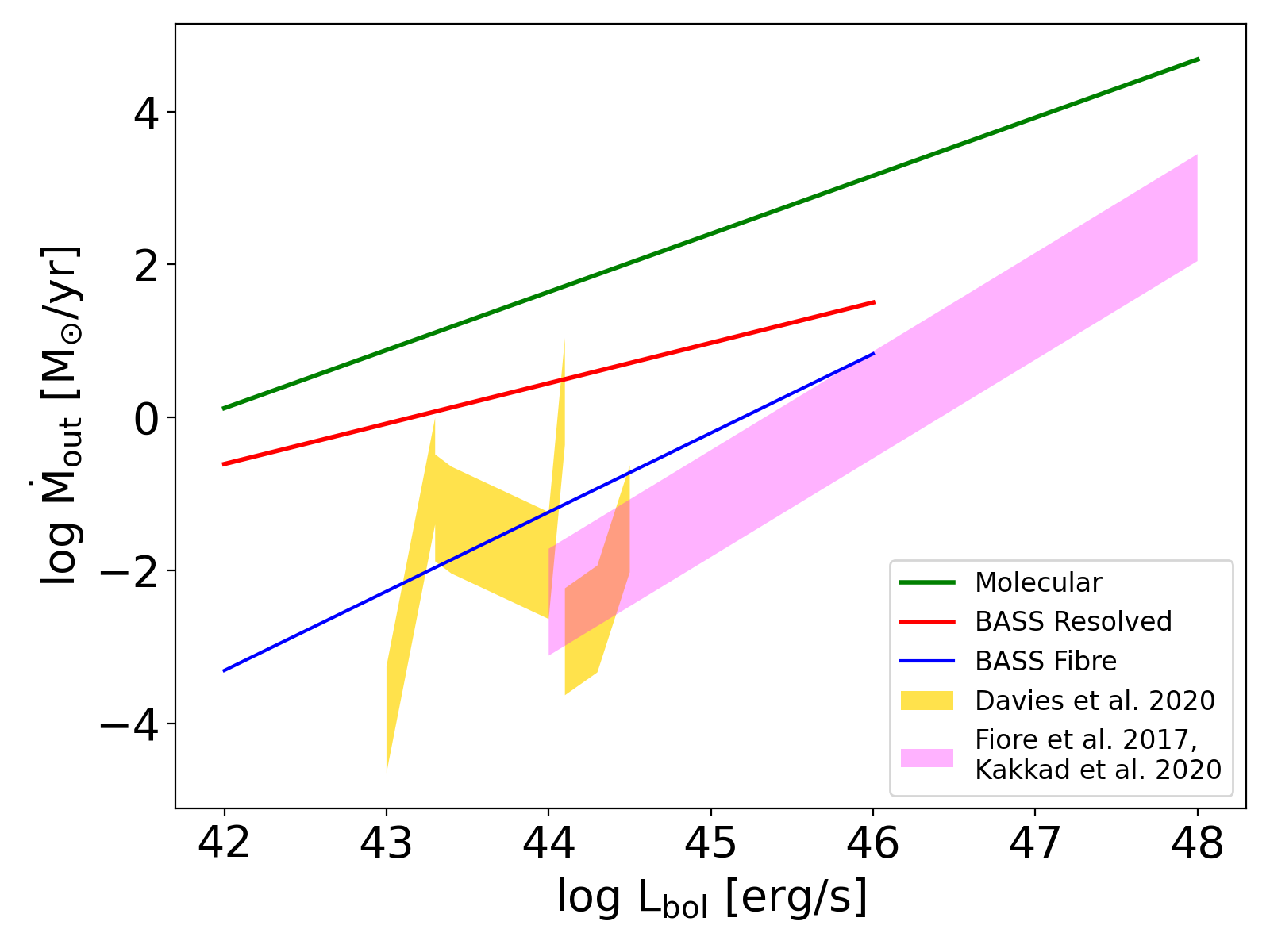}
\caption{Correlations between ionised, molecular gas mass outflow rate and the bolometric luminosity of the AGN for the targets presented in this paper and from the literature \citep[e.g.,][]{cicone14, carniani16, fiore17, davies20, fluetsch20, kakkad20}. Ionised outflow relations are represented in red, blue, yellow and magenta curves (or shaded regions) while the molecular outflow relation is shown in green. The shaded regions correspond to ionised gas mass outflow rates for electron density of 200-5000 cm$^{-3}$. The literature mass outflow rates are mostly obtained from integrated spectra, which match well with the fibre and slit integrated spectrum analysis for the targets in this paper (solid and dashed blue curves). The instantaneous outflow rates derived from the resolved data are $\sim$2 orders of magnitude higher than those derived from the integrated spectrum at the low bolometric luminosity end. However, molecular outflows still seem to represent the bulk of the outflowing gas in AGN host galaxies across all luminosities.  \label{fig:outflow_scaling_literature}}
\end{figure}

\begin{table*}
\caption{Outflow properties of the ionised gas derived from the MUSE data. (1) {\it Swift}-BAT 70 month hard X-ray survey ID; (2) Common name of the target; (3) \& (4) Non-parametric velocity dispersion, $w_{80}$ calculated from the fibre and slit spectra respectively. Typical 1$\sigma$ uncertainty in the $w_{80}$ value is $\pm$70 km/s. Columns (5)--(9) report the mass outflow rate values, $\dot{M}$ calculated using different methods. (5) $\dot{M}^{\rm total}_{\rm res}$: Total mass outflow rate from the resolved map; (6) $\dot{M}^{\rm cent}_{\rm res}$: Mass outflow rate in the central 3\arcsec of the resolved map, typical uncertainty in the outflow rate values from the resolved maps is 0.2 dex; (7) $\dot{M}_{\rm fibre}$ is the outflow rate from the integrated fibre spectra and (8) $\dot{M}_{\rm slit}$ from integrated slit spectra. In both (7) \& (8), the electron density was obtained using the flux ratio of the broad components of the \sii ~doublet. (9) $\dot{M}_{200}$ is the outflow rate from integrated fibre spectra, assuming an electron density of 200 cm$^{-3}$ commonly adopted in the literature. (10) reports the fraction of outflow mass within the central 3\arcsec of the resolved mass outflow rate maps ($\dot{M_{\rm res}^{\rm cent}}/\dot{M_{\rm res}^{\rm total}}$)}
\centering
\begin{tabular}{cccccccccc}
\hline
BASS ID & Target & $w_{80}^{\rm fibre}$ & $w_{80}^{\rm slit}$ & Log $\dot{M}_{\rm res}^{\rm total}$ & Log $\dot{M}_{\rm res}^{\rm cent}$ & Log $\dot{M}_{\rm fibre}$ & Log $\dot{M}_{\rm slit}$ & Log $\dot{M}_{200}$ & $f$\\
& & km s$^{-1}$ & km s$^{-1}$ &M$_{\odot}$/yr & M$_{\odot}$/yr & M$_{\odot}$/yr & M$_{\odot}$/yr & M$_{\odot}$/yr & \%\\
(1) & (2) & (3) & (4) & (5) & (6) & (7) & (8) & (9) & (10)\\
\hline\hline
57 & 3C033 & 565 & 495 & -0.76 & -1.26 & -0.59 & -0.91 & -0.54 & 32\\
58 & NGC 424 & 814 & 740 & 1.42 & 1.40 & -1.00 & -1.30 & -0.25 & 95\\
62 & IC 1657 & 370 & 370 & -0.01 & -1.28 & -2.08 & -2.19 & -2.24 & 5\\
127 & HE 0224-2834 & 495 & 495 & 0.62 & 0.44 & -0.95 & -1.06 & -0.91 & 66\\
134 & NGC 985 & 574 & 574 & 1.67 & 1.63 & 1.10 & 1.02 & -0.20 & 91\\
184 & NGC 1365 & 298 & 298 & 0.14 & -2.50 & -2.28 & -2.34 & -2.05 & <1\\
197 & HE 0351+0240 & 289 & 289 & 0.01 & -0.18 & -1.68 & -0.79 & -1.55 & 64\\
213 & HE 0412-0803 & 361 & 361 & 1.29 & 1.14 & -0.37 & -0.45 & -0.04 & 70\\
216 & NGC 1566 & 522 & 522 & -0.22 & -0.69 & -2.18 & -0.15 & -1.43 & 33\\
471 & NGC 2992 & 595 & 595 & 0.52 & -1.06 & -1.69 & -1.79 & -1.08  & 3\\
501 & HE1029-1401 & 1449 & 1449 & 1.55 & 1.53 & 1.58 & 0.04 & 0.28 & 95\\
653 & NGC 4941 & 448 & 448 & -0.55 & -1.11 & -1.96 & -2.04 & -1.27 & 28\\
703 & Mrk 463 & 1070 & 1141 & 2.44 & 2.37 & 0.33 & -0.25 & 0.68 & 85\\
711 & Circinus & 299 & 299 & -0.26 & -0.96 & -2.98 & -3.18 & -2.30 & 20\\
731 & NGC 5643 & 448 & 522 & 0.35 & 0.09 & -1.72 & -1.83 & -1.08 & 55\\
783 & NGC 5995 & 804 & 804 & 0.13 & -0.53 & -1.59 & -1.69 & -1.62 & 20\\
817 & 2MASX J1631+2352 & 778 & 778 & 0.60 & 0.55 & 0.26 & 0.20 & -1.04 & 89\\
1051 & 3C403 & 566 & 566 & 0.24 & -0.05 & -1.90 & -0.96 & -1.96 & 51\\
1092 & IC 5063 & 518 & 444 & 1.25 & 0.96 & 0.54 & -0.46 & -0.76 & 51\\
1151 & 3C445 & 638 & 638 & 2.03 & 1.99 & 1.32 & 1.19 & 0.02 & 91\\
1182 & NGC 7469 & 663 & 663 & 0.67 & 0.51 & -0.83 & 1.02 & -0.23 & 70\\
1183 & Mrk 926 & 644 & 644 & 1.38 & 1.33 & -0.03 & -0.11 & 0.16 & 89\\
\hline
\end{tabular}
\label{table:outflow_properties}
\end{table*}

\subsection{Mass outflow rate scaling relations} \label{sect5.3}
Several publications in the literature have investigated the presence or absence of correlations between the outflow properties and the AGN properties in the different ionised, molecular and neutral gas phases \citep[e.g.,][]{cicone14, cicone18, carniani15, fiore17, kakkad20, fluetsch20, rojas20}. Most of these are based on integrated spectra with fibre or slit spectroscopy or marginally resolved IFS data. Using the resolved MUSE data set, we investigate if the relations between outflow properties and the AGN properties change with the aperture or underlying methods of analysis.

Scaling relations between the \oiii ~velocity dispersion ($w_{80}$) and $L_{\rm bol}$, $M_{\rm BH}$ and $\lambda_{\rm Edd}$ were already presented in Sect. \ref{sect5.1}, where correlations (p-value$<$0.05) were only suggestive in the case of $L_{\rm bol}$. However, as we stressed previously, the mass outflow rate depends not only on velocity dispersion but also on the flux/luminosity and electron density. Thus, we revisit this theme, focusing on correlations between the mass outflow rates obtained from different methods and the AGN properties, namely $L_{\rm bol}$, $M_{\rm BH}$ and $\lambda_{\rm Edd}$. We compute the Pearson's r correlation coefficient implementation within \texttt{pymccorrelation} package \citep[seee][]{curran14, privon20}. 

Figure \ref{fig:outflow_scaling_aperture} shows the relations of the summed instantaneous mass outflow rates obtained from the resolved mass outflow rate map ($\dot{M}_{\rm res}^{\rm total}$, solid red circles), and time-averaged outflow rates from the integrated fibre spectra ($\dot{M}_{\rm fibre}$, open blue circles) and slit spectra ($\dot{M}_{\rm slit}$, solid blue squares) versus $L_{\rm bol}$ (left panel), $M_{\rm BH}$ (middle panel) and $\lambda_{\rm Edd}$ (right panel), respectively. The results from the Pearson correlation tests between mass outflow rates derived from the different methods and the AGN properties are reported in Table \ref{table:correlation_tests}.

\vspace{0.3cm}

\noindent
{\bf $\dot{\bf M}$ versus ${\bf L_{\rm bol}}$:} Similar to the case of the velocity dispersion, the mass outflow rates show a maximum probability of correlation with the bolometric luminosity. The null hypothesis for a non-correlation (p-value) with $L_{\rm bol}$ is $\leq$1\% for all methods of outflow rates computations (although considering uncertainties, the correlation between $\dot{M}^{\rm total}_{\rm res}$ and $L_{\rm bol}$ is relatively weak). The correlation with $L_{\rm bol}$ is strongest in the case of time-averaged mass outflow rates obtained from the fibre spectra. The outflow rates derived from summed resolved data have a higher intercept and flatter slope compared to the integrated spectra obtained from fibre or slit aperture (0.51$\pm$0.03 in the former case compared to 1.04$\pm$0.05 \& 0.79$\pm$0.05 in the latter). The p-value for the outflow rate derived from slit spectra versus $L_{\rm bol}$ is slightly higher at 0.003, compared to 2.0e-4 in the case of fibre spectra. We note here that in the case of the slit spectroscopy, the mass outflow rate values depend on the slit orientation. Different slit orientations were explored and Table \ref{table:outflow_properties} reports the values for slit spectroscopy which show the most extreme differences with respect to the circular aperture. However, despite these differences, the time-averaged mass outflow rates obtained from fibre and slit spectra roughly cover the same parameter space in Fig. \ref{fig:outflow_scaling_aperture}. We also note that the p-value reduces in the case of outflow rate values extracted from the central 3\arcsec aperture in the resolved maps ($\dot{M}_{\rm res}^{\rm cent}$), compared to the total summed resolved map ($\dot{M}_{\rm res}^{\rm total}$). This result might already suggest that the AGN influence on the ISM is the strongest within the inner one kiloparsec of host galaxies.  

\noindent
{\bf $\dot{\bf M}$ versus ${\bf M_{\rm BH}}$ and ${\lambda_{\rm \bf Edd}}$:} The mass outflow rates show  a weaker correlations (or no correlation) with the black hole mass and Eddington ratio, compared to the bolometric luminosity. However, we find a strong correlation between the time-averaged outflow rate value from fibre spectra versus $M_{\rm BH}$ with a p-value of 0.01, again highlighting that the presence or absence of correlation is highly dependent on the observation and analysis method. These correlations will be further discussed in the context of the driving mechanism of outflows in Sect. \ref{sect6}. 

\vspace{0.3cm}
We note that there is no significant change in the presence or absence of correlations if we assume a constant density value in the case of time-averaged mass outflow rate values, although the relative slopes might be different. This is apparent from the $\dot{M}_{200}$ relations in Table \ref{table:correlation_tests}. We also compare these correlations with the results previously published in the literature for ionised and molecular gas outflows, as shown in Fig. \ref{fig:outflow_scaling_literature} \citep[e.g.,][]{cicone14, cicone18, walter17, zchaechner18}. For simplicity, we only consider the relations with the bolometric luminosity. The ionised gas outflows are shown in red, blue, yellow and magenta curves/shaded regions, while the molecular outflow scaling relation is shown as green curve. The shaded regions correspond to an assumed electron density range of 200--5000 cm$^{-3}$ in ionised outflows. The mass outflow rate scaling relation from the fibre integrated spectra of the BASS-MUSE sample is consistent with the literature results, which are also obtained from integrated spectroscopy using fibre or slit. The summed instantaneous outflow rates obtained from resolved data are clearly $\sim$1--2 orders of magnitude higher than the time-averaged outflow rate values from the integrated spectra from the literature. The difference is more pronounced at lower bolometric luminosities. The molecular outflows still seem to dominate the outflowing gas content in low as well as high redshift galaxies and across a wide range of bolometric luminosities. However, similar analysis as the one presented in this paper, will be required to investigate the scaling relations and observational biases in the case of molecular outflows \citep[e.g., MODA, PUMA surveys][]{treister18, perna21}. This will require high spatial resolution and deeper observations with sub-mm facilities such as ALMA, VLA, PdBI and SMA \citep[see also][]{sun14, brusa18}. 

\begin{table}
\centering
\begin{tabular}{cccc}
\hline
Relation  & Correlation & p-value & Slope\\
& coefficient & & \\
\hline \hline
Relations with $L_{\rm bol}$ \\
$w_{80}^{\rm fibre}$ vs. L$_{\rm bol}$ & 0.52$^{+0.11}_{-0.13}$ & 0.02$^{+0.04}_{-0.01}$ & 0.102$\pm$0.001\\
$w_{80}^{\rm slit}$ vs. L$_{\rm bol}$ & 0.50$^{+0.09}_{-0.11}$ & 0.02$^{+0.07}_{-0.01}$ & 0.096$\pm$0.002\\
$v_{\rm max}^{\rm fibre}$ vs. $L_{\rm bol}$ & 0.58$^{+0.08}_{-0.10}$ & 0.003$^{+0.02}_{-0.002}$ & 0.128$\pm$0.002\\
$v_{\rm max}^{\rm slit}$ vs. $L_{\rm bol}$ & 0.58$^{+0.10}_{-0.11}$ & 0.004$^{+0.03}_{-0.003}$ & 0.126$\pm$0.002\\
$\dot{M}_{\rm res}^{\rm total}$ vs.  L$_{\rm bol}$ & 0.54$^{+0.17}_{-0.20}$ & 0.01$^{+0.10}_{-0.009}$ & 0.51$\pm$0.03\\
$\dot{M}_{\rm res}^{\rm cent}$ vs.  L$_{\rm bol}$ & 0.63$^{+0.13}_{-0.15}$ & $<$0.01 & 0.91$\pm$0.06\\
$\dot{M}_{\rm fibre}$ vs.  L$_{\rm bol}$ & 0.73$^{+0.08}_{-0.13}$ & $<$0.01 & 1.04$\pm$0.05\\
$\dot{M}_{\rm slit}$ vs.  L$_{\rm bol}$ & 0.62$^{+0.13}_{-0.14}$ & $<$0.01 & 0.79$\pm$0.05\\
$\dot{M}_{200}$ vs.  L$_{\rm bol}$ & 0.58$^{+0.15}_{-0.16}$ & 0.004$^{+0.04}_{-0.003}$ & 0.56$\pm$0.03\\
\\
Relations with $M_{\rm BH}$\\
$w_{80}^{\rm fibre}$ vs. M$_{\rm BH}$ & 0.37$^{+0.17}_{-0.22}$ & 0.10$^{+0.39}_{-0.09}$ & --\\
$w_{80}^{\rm slit}$ vs. M$_{\rm BH}$ & 0.32$^{+0.19}_{-0.23}$ & 0.15$^{+0.43}_{-0.13}$ & --\\
$v_{\rm max}^{\rm fibre}$ vs. $M_{\rm BH}$ & 0.36$^{+0.18}_{-0.21}$ & 0.11$^{+0.38}_{-0.10}$ & --\\
$v_{\rm max}^{\rm slit}$ vs. $M_{\rm BH}$ & 0.34$^{+0.18}_{-0.26}$ & 0.12$^{+0.45}_{-0.11}$ & --\\
$\dot{M}_{\rm res}^{\rm total}$ vs. M$_{\rm BH}$ & 0.18$^{+0.26}_{-0.25}$ & 0.34$^{+0.45}_{-0.30}$ & --\\
$\dot{M}_{\rm res}^{\rm cent}$ vs. M$_{\rm BH}$ & 0.29$^{+0.23}_{-0.25}$ & 0.18$^{+0.39}_{-0.17}$ & --\\
$\dot{M}_{\rm fibre}$ vs. M$_{\rm BH}$ & 0.56$^{+0.14}_{-0.20}$ & 0.009$^{+0.08}_{-0.008}$ & 0.75$\pm$0.08\\
$\dot{M}_{\rm slit}$ vs. M$_{\rm BH}$ & 0.34$^{+0.20}_{-0,23}$ & 0.13$^{+0.42}_{-0.12}$ & --\\
$\dot{M}_{200}$ vs. M$_{\rm BH}$ & 0.33$^{+0.20}_{-0.23}$ & 0.13$^{+0.47}_{-0.12}$ & --\\
\\
Relations with $\lambda_{\rm Edd}$\\
$w_{80}^{\rm fibre}$ vs. $\lambda_{\rm Edd}$ & 0.00$^{+0.38}_{-0.33}$ & 0.22$^{+0.56}_{-0.18}$ & --\\
$w_{80}^{\rm slit}$ vs. $\lambda_{\rm Edd}$ & 0.03$^{+0.42}_{-0.37}$ & 0.21$^{+0.56}_{-0.20}$ & --\\
$v_{\rm max}^{\rm fibre}$ vs. $\lambda_{\rm Edd}$ & 0.08$^{+0.26}_{-0.25}$ & 0.42$^{+0.37}_{-0.31}$ & --\\
$v_{\rm max}^{\rm slit}$ vs. $\lambda_{\rm Edd}$ & 0.08$^{+0.27}_{-0.27}$ & 0.40$^{+0.38}_{-0.31}$ & --\\
$\dot{M}_{\rm res}^{\rm total}$ vs. $\lambda_{\rm Edd}$ & 0.33$^{+0.20}_{-0.23}$ & 0.14$^{+0.43}_{-0.13}$ & --\\
$\dot{M}_{\rm res}^{\rm cent}$ vs. $\lambda_{\rm Edd}$ & 0.27$^{+0.35}_{-0.35}$ & 0.17$^{+0.50}_{-0.16}$ & --\\
$\dot{M}_{\rm fibre}$ vs. $\lambda_{\rm Edd}$ & 0.13$^{+0.19}_{-0.21}$ & 0.46$^{+0.35}_{-0.33}$ & --\\
$\dot{M}_{\rm slit}$ vs. $\lambda_{\rm Edd}$ & 0.20$^{+0.26}_{-0.21}$ & 0.35$^{+0.42}_{-0.31}$ & --\\
$\dot{M}_{200}$ vs. $\lambda_{\rm Edd}$ & 0.25$^{+0.24}_{-0.24}$ & 0.24$^{+0.45}_{-0.22}$ & --\\
\\
Relations with $L_{\rm 1.4 ~GHz}$\\
$w_{80}^{\rm fibre}$ vs. $L_{\rm 1.4 ~GHz}$ & 0.17$^{+0.21}_{-0.27}$ & 0.47$^{+0.35}_{-0.31}$ & --\\
$w_{80}^{\rm slit}$ vs. $L_{\rm 1.4 ~GHz}$ & 0.14$^{+0.24}_{-0.24}$ & 0.53$^{+0.32}_{-0.35}$ & --\\
$v_{\rm max}^{\rm fibre}$ vs. $L_{\rm 1.4 ~GHz}$ & 0.47$^{+0.23}_{-0.27}$ & 0.12$^{+0.36}_{-0.11}$ & \\
$v_{\rm max}^{\rm slit}$ vs. $L_{\rm 1.4 ~GHz}$ & 0.33$^{+0.32}_{-0.33}$ & 0.27$^{+0.46}_{-0.25}$ & \\
$\dot{M}_{\rm res}^{\rm total}$ vs. $L_{\rm 1.4 ~GHz}$ & 0.13$^{+0.29}_{-0.26}$ & 0.39$^{+0.37}_{-0.34}$ & --\\
$\dot{M}_{\rm res}^{\rm cent}$ vs. $L_{\rm 1.4 ~GHz}$ & 0.25$^{+0.30}_{-0.31}$ & 0.37$^{+0.41}_{-0.30}$ & --\\
$\dot{M}_{\rm fibre}$ vs. $L_{\rm 1.4 ~GHz}$ & 0.42$^{+0.26}_{-0.33}$ & 0.17$^{+0.50}_{-0.15}$ & --\\
$\dot{M}_{\rm slit}$ vs. $L_{\rm 1.4 ~GHz}$ & 0.43$^{+0.21}_{-0.25}$ & 0.16$^{+0.40}_{-0.14}$ & --\\
$\dot{M}_{200}$ vs. $L_{\rm 1.4 ~GHz}$ & 0.23$^{+0.31}_{-0.31}$ & 0.39$^{+0.41}_{-0.32}$ & --\\
\hline
\end{tabular}
\caption{The table reports the results from the correlation tests between the outflow properties and the AGN properties quantities presented in this paper. The outflow properties considered in these tests are: Non-parametric velocity dispersion obtained from integrated fibre spectrum (w$_{80}^{\rm fibre}$) and slit spectrum (w$_{80}^{\rm slit}$), parametric velocity computed from integrated fibre spectrum ($v_{\rm max}^{\rm fibre}$) and slit spectrum ($v_{\rm max}^{\rm slit}$), total mass outflow rate obtained from resolved maps ($\dot{M}_{\rm res}^{\rm total}$), mass outflow rate obtained from resolved maps within a 3\arcsec ~aperture centered on the AGN ($\dot{M}_{\rm res}^{\rm cent}$), mass outflow rate obtained from integrated fibre spectrum ($\dot{M}_{\rm fibre}$) and slit spectrum ($\dot{M}_{\rm slit}$) and mass outflow rate assuming an outflow electron density of 200 cm$^{-3}$. The AGN properties considered in these tests are the bolometric luminosity (L$_{\rm bol}$), black hole mass (M$_{\rm BH}$) and the Eddington ratio ($\lambda_{\rm Edd}$). We report the Pearson's r correlation coefficient, p-value which represents the null-hypothesis probability (for non-correlation) and the slope of the relations. The details about the implications and interpretation of these correlation tests are further discussed in Sects. \ref{sect4} and \ref{sect5}.}
\label{table:correlation_tests}
\end{table}

\section{Discussion} \label{sect6}

In this paper, we have presented the impact of observational bias and the methods of analysis on the observed correlations between the properties of ionised gas outflows and the AGN. We showed that the parameters of the linear correlations change depending on whether the individual data points are obtained from a spatially resolved or integrated fibre/slit spectroscopy or from assumed values while modelling the outflows. In this section, we discuss the implications of the results presented in Sect. \ref{sect5} and explore the possible reasons for the presence or absence of correlations. By comparing the scaling relations obtained from the resolved data and the literature (or integrated fibre/slit spectra), we will also provide a means to correct (or cross-calibrate) for observational biases in future studies. Lastly, we will discuss the nature of these outflows in the context of predictions from current state-of-the-art hydrodynamical simulations.

For $\sim$80\% of the galaxies, the \oiii ~$w_{80}$ parameter (a measure of ionised gas kinematics) derived from the integrated fibre or slit spectra show consistent values, suggesting a minor dependence of the $w_{80}$ parameter on observational method. The $w_{80}$ values correlate with the AGN bolometric luminosity, but not the black hole mass or Eddington ratio. Also, in the resolved data, the $w_{80}$ distribution is skewed towards larger values in regions ionised by the AGN compared to the star formation ionised regions. These results, therefore, suggest that the ionised gas kinematics is dependent on the power of the ionising radiation from the central source. However, the line broadening could also result from shocks \citep[e.g.,][]{rich14} induced in targets that are undergoing or recently underwent a merger \citep[e.g., Mrk 463][]{treister18}. However, since the fraction of mergers in our sample is relatively low, the AGN radiation seems to be the dominant mechanism behind the observed turbulence. We note that unlike observations at high redshift, where high velocity ($>$800 km s$^{-1}$) ionised gas is observed in a few kiloparsec scales, the low redshift X-ray AGN show high velocity gas restricted to sub-kiloparsec scales. 

The mass outflow rate maps show the distribution of instantaneous outflow rates through local ionised clouds. Although we observe a diverse morphology in the outflow rate maps themselves, some targets such as NGC 7469 (Fig. \ref{fig:outflow_rate_map_NGC7469}) notably show a semi-arc like structure which may suggest a spherical propagation of the outflow closer to the AGN location. Several numerical simulations invoke or predict a spherical or semi-spherical shockfront that results from an AGN outburst \citep[e.g.,][]{gabor14, nayakshin14, zubovas14}. A thin shell-like outflow morphology was also proposed in \citet{husemann19} for a nearby AGN, HE 1353-1917. The observed mass outflow rate maps are also consistent with recent results from resolved spectroscopy in the literature \citep[e.g.,][]{revalski21} where the mass outflow rate is shown to have a peak value at a distance of $\sim$500--1500 pc from the AGN location, before dropping off at larger distances.

\begin{figure*}
\centering
\includegraphics[scale=0.7]{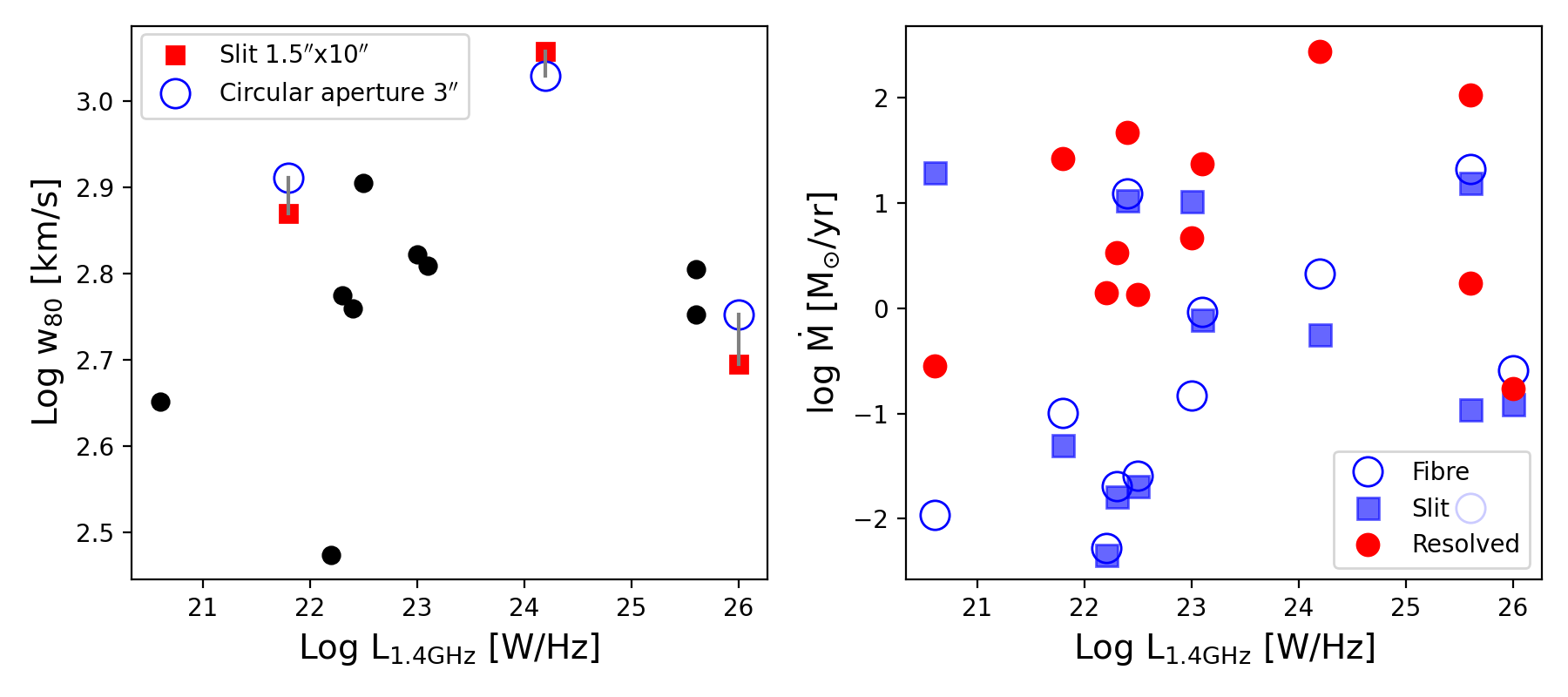}
\caption{The \oiii ~$w_{80}$ value versus the radio luminosity (1.4 GHz) of the targets presented in this paper (left panel). The right panel shows the relation between the mass outflow rates obtained from resolved and integrated apertures versus the radio luminosity (1.4 GHz). The colour coding in these plots are the same as Figs. \ref{fig:w80_integrated} and \ref{fig:outflow_scaling_aperture} \label{fig:scaling_radio}}
\end{figure*}

A comparison of the resolved mass outflow rate maps and the mass outflow rate calculated from fibre or slit spectra show that the values and their scaling relations with AGN properties can highly depend on the analysis method, as apparent from Fig. \ref{fig:outflow_scaling_aperture} and Tables \ref{table:outflow_properties} and \ref{table:correlation_tests}. The mass outflow rates show a correlation with the AGN bolometric luminosity, regardless of the calculations using integrated or spatially resolved data. The fact that we observe the correlations with the bolometric luminosity suggests that the AGN radiation may be, at least in part, responsible for driving these outflows. The probability of correlation increases (lower p-value in Table \ref{table:correlation_tests}) if the mass outflow rates are obtained from time-averaged values of the integrated fibre or slit spectra. Between the outflow rates obtained from the fibre and slit spectra, the correlations are stronger in the case of fibre spectra. This further indicates possible AGN driven nature of the ionised gas outflows over the lifetime of the outflow (R/$v_{\rm out}$). And finally, in the case of instantaneous mass outflow rates from resolved maps, the $\dot{M}-L_{\rm bol}$ correlation is stronger when the mass outflow rate values extracted from the central 3\arcsec aperture is considered, compared to the summed outflow rate integrated over the entire FoV. Therefore, the AGN radiation shows the strongest influence on the ionised gas in its vicinity compared to the gas in the galactic outskirts. This is also supported by the fact that we do not observe high velocity winds in the galaxy outskirts, unless the system is disturbed due to external factors such as mergers. 

As described in Sect. \ref{sect5}, the observed difference in the summed instantaneous mass outflow rate from the resolved maps and the time-averaged outflow rate from the integrated spectra could be due to a combination of higher outflow mass from the larger FoV of the resolved maps and the radius factor in the respective equations. One may use Fig. \ref{fig:outflow_scaling_aperture} to define an ``effective electron density'' for summed instantaneous outflows, $n_{e\rm }^{\rm eff}$. The effective electron density is the density value that can be used in the time-averaged outflow rate equation to derive the summed instantaneous outflow rate i.e. this value can be used to cross-calibrate between the instantaneous and time-averaged methods. We find that the $n_{e\rm }^{\rm eff}$ has a value of $\sim$10 cm$^{-3}$. Therefore, if the mass outflow rate calculations need to be made from integrated spectra using the biconical outflow model with uniform density of ionised gas, an electron density of $\sim$10 cm$^{-3}$ would provide outflow rate values consistent with those of resolved maps presented in this paper.

The correlation of ionised gas mass outflow rates and the bolometric luminosity has been reported in several works \citep[e.g., see a literature compilation in][]{fiore17}. Recent spatially resolved observations with STIS/HST show that the mass outflow rates and the kinetic energy seem to monotonically increase with the AGN luminosity for 6 nearby AGN \citep{revalski21}. However, we note that contrary results have also been reported in the literature, where no correlation is observed between the outflow rates and the bolometric luminosity \citep[e.g.,][]{baron19, davies20} for a similar range of bolometric luminosity also explored in this paper. One of the main difference in these works is that the electron density is calculated based on the ionisation parameter, the AGN luminosity and the distance of the outflowing gas from the AGN location, also called $logU$ method. The $logU$ method is sensitive to higher density values compared to the \sii ~doublet method used in this paper and as a result, \citet{baron19} and \citet{davies20} find electron densities that are $\sim$3--100 times larger than the ones reported in this paper. Furthermore, the bolometric luminosity ranges compared in these works are slightly lower than the ones displayed by the sample in this paper (This paper, 3 dex $\sim 10^{42.5}$--10$^{45.5}$ erg s$^{-1}$, \citet{davies20}: $<$2 dex $\sim$10$^{43}$--10$^{44.5}$ and \citet{baron19}: 2 dex $\sim$10$^{43.5}$--10$^{45.5}$). Similarly, \citet{rojas20} use a constant density value of 10$^{4.5}$ cm$^{-3}$ for their (time-averaged) mass outflow rate calculations for a sub-sample of BASS galaxies, based on integrated spectrum measurements. The observed differences in the outflow rate values in \citet{rojas20} and this paper can be attributed to these different methodologies. A robust comparison between these works, therefore, cannot be made within the scope of this paper and a future work will address the spatially resolved mass outflow rates considering electron density determination using the multiple methods. However, upon using the assumed electron densities in these works (which are based on the electron density derived from the \sii ~doublet values), the scatter in the $\dot{M}-L_{\rm bol}$ is reduced, which suggests that the scaling relations are indeed subject to the methods of analysis. We estimate a maximum error in the mass outflow rate of $\sim$0.5 dex and in the bolometric luminosity of 0.3 dex. The presence of correlation in this work will also be observed after taking into account the errors in the different quantities.


We find a weak or no correlation of any of the outflow quantities, namely the outflow velocity and mass outflow rate, with the black hole mass and the Eddington ratio. Therefore, the observed outflows for the sample used in this paper cannot be explained by radiation pressure driven winds from the accretion disk. Extra-galactic studies in the literature have also shed light on the impact of radio jets in driving outflows in multiple gas phases, via jet-mode feedback \citep[also called radio-mode or mechanical-mode of feedback, e.g.,][]{villar-martin14, nesvadba17, santoro18, molyneux19, jarvis21}. Several theoretical simulations support a scenario where radio jets transfer the energy into the ISM, which can also create a clumpy distribution of gas \citep[e.g.,][]{sutherland07, wagner12, cielo18, mukherjee18b}. Therefore, we also explore the presence of any correlation of the outflow properties with the radio luminosity in Fig. \ref{fig:scaling_radio}. We obtain the 1.4 GHz radio fluxes from \citet{veron10} catalogue, which are available for 12 out of the 22 AGN presented in this paper. The radio luminosity, $L_{\rm 1.4GHz}$, of our targets is in the range 10$^{21}$--10$^{26}$ W/Hz ($\sim$5 dex) and Table \ref{table:correlation_tests} reports the correlation test results of various outflow properties presented earlier, but with the radio luminosity. Similar to the Eddington ratio, we do not find a robust correlation with the radio luminosity. 

The presence of non-correlations of total mass outflow rate from resolved maps with both the Eddington ratio and the radio luminosity can be an interpretation of the fact that the observed outflows may be a combination of radiation pressure driven or thermal winds, radio jets and external influences such as shocks induced by mergers. Therefore, correlations with a single quantity may not exist. Relatively weaker processes such as star formation could also contribute to the scatter in the scaling relations \citep[e.g.,][]{dipompeo18}. This is evident from an in-depth analysis of individual targets that have been published in the literature. For instance, ionised and molecular gas observations of IC 5063 have shown evidence of jet-ISM interaction in this system \citep[e.g.,][]{kulkarni98, morganti15, tadhunter14, dasyra15, oosterloo17, mukherjee18a, venturi21}. Jet-ISM interaction has also been proposed in some of the 3C sources, part of the MURALES survey \citep[e.g.,][]{balmaverde19, balmaverde21}. Similarly, Mrk 463 is a late-stage merger and the observed outflow could be due to the accretion disk triggered by the merging activity \citep[see][]{treister18}. Thermal winds are believed to be driving the outflows in NGC 2992 in the biconical morphology, also observed in the MUSE data presented in this paper \citep[e.g.,][]{veilleux01}, while in the low luminosity AGN in NGC 1365, star formation is also inferred to be driving the outflows \citep[e.g.,][]{venturi18}. In summary, all these processes contribute to the observed mass outflow rate scaling relations with the bolometric luminosity. At redshift corresponding to cosmic noon (z$\sim$2--3), the black hole accretion activity is expected to be at its maximum \citep[e.g.,][]{silk12, madau14}. Therefore, the chances are that the high redshift surveys pick up a large fraction of targets where accretion disk driven winds are dominant, which might explain the observed scaling relations with the Eddington ratio at this epoch \citep[e.g.,][]{kakkad20}. 

We also note that other emission lines such as H$\beta$ and H$\alpha$ have also been used in the literature to derive the properties of ionised gas outflows. The H$\beta$ line is usually faint in the NLR and the H$\alpha$ is more susceptible to dust extinction effects compared to the \oiii ~line. Furthermore, H$\alpha$ could also be contaminated by the closely spaced \nii$\lambda\lambda$6549, 6585 lines due to blending in the case of high velocity outflows. These factors may partly explain the observed differences between the results presented in this paper and in \citet{ruschel-dutra21} who also report the ionised outflow properties of 30 low redshift AGN host galaxies using the H$\alpha$ line. For example, \citet{ruschel-dutra21} find mass outflow rates in the range Log $\dot{M}$ = -3.91--2.38 with a median value of -2.1, while the local instantaneous mass outflow rates presented in this paper are in the range Log $\dot{M}$ = -0.76--2.44, with a median value of 0.56, nearly two orders of magnitude more than that in \citet{ruschel-dutra21}. Furthermore, only two sources in \citet{ruschel-dutra21} show a coupling efficiency $>$1\%, in contrast to the methods used in this paper where some sources show much higher coupling efficiencies (discussed below). The dependence of outflow properties on different diagnostic lines will be explored in a future work. 

\begin{figure}
\centering
\includegraphics[scale=0.4]{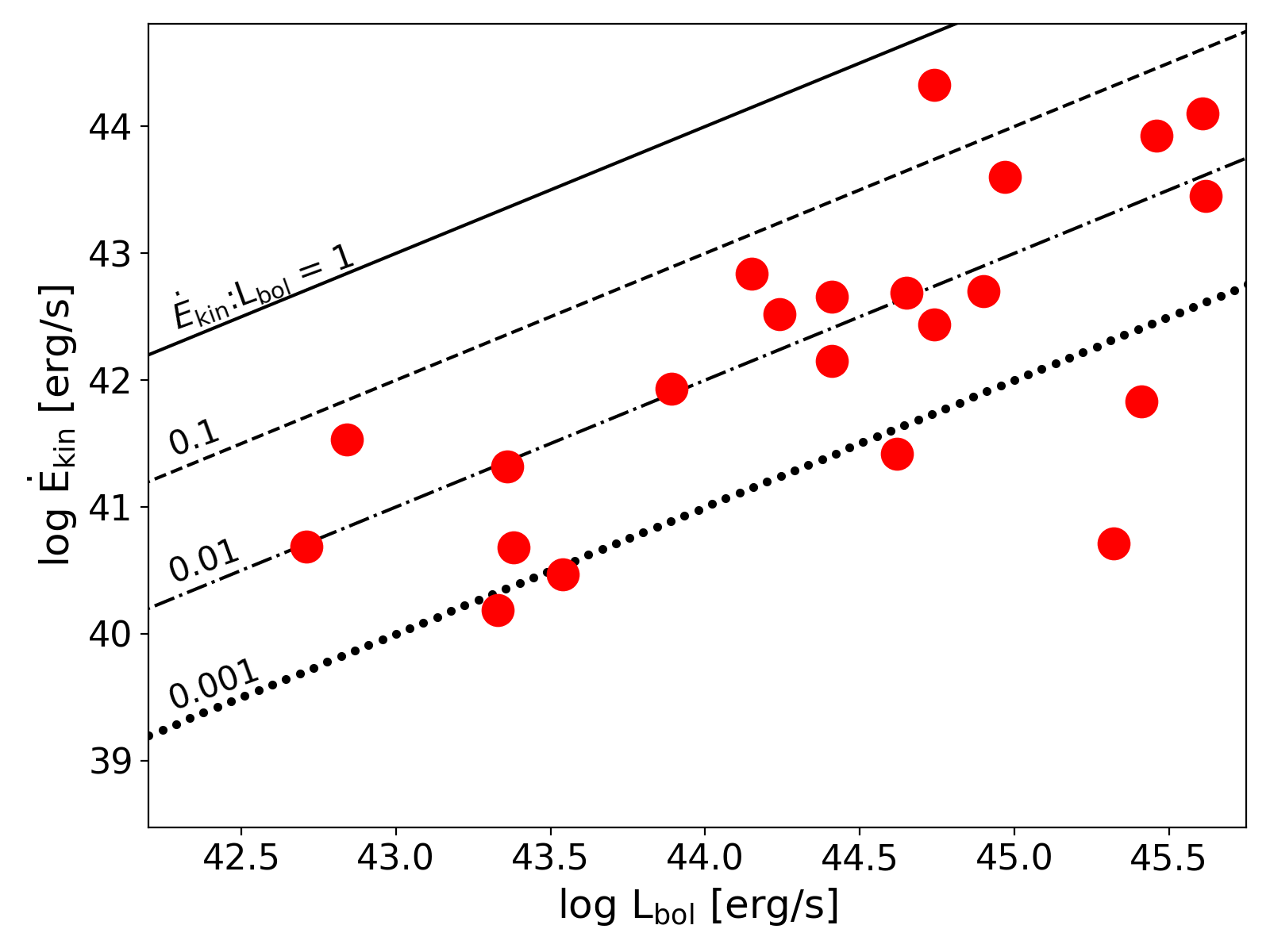}
\caption{The plot shows the outflow kinetic power of the targets presented in this paper (red circles) versus the bolometric luminosity of the AGN. The outflow kinetic power is derived from the resolved mass outflow rate map and the outflow velocity map. On average, the targets in this paper show 0.1\% coupling with the bolometric luminosity. \label{fig:Ekin_Lbol}}
\end{figure}

Lastly, we calculate the fraction of AGN bolometric luminosity that is coupled to the ISM via kinetic energy of the observed ionised outflows, $\dot{E}_{\rm kin}$ = 1/2$\cdot\dot{M}v^{2}$. Cosmological simulations invoking AGN feedback are able to reproduce observed AGN host galaxy correlations, such as the ones presented in this paper, by requiring that 5-10\% of the energy from the AGN is coupled with the ISM of the host galaxy \citep[e.g.,][]{fabian99, springel05, kurosawa09}. Observationally, determination of this coupling efficiency has been challenging due to the large systematic uncertainties in the derivation of this quantity, which we attempt to resolve using the IFS data presented in this paper. We derive $\dot{M}$ and $v$ values from the resolved maps resulting in a resolved kinetic energy map. Figure \ref{fig:Ekin_Lbol} shows the total kinetic energy obtained from the resolved map versus $L_{\rm bol}$. The different linear curves correspond to the coupling factors of 0.1\%, 1\%, 10\% and 100\%. We find a wide range in the coupling efficiency with ionised gas outflows from $<$0.1\% to $\sim$40\% (median $\sim$1\%), which seems to monotonically increase with the bolometric luminosity of the AGN. This is similar to trends previously found in the literature, although with more scatter \citep[e.g.,][]{carniani15}. The kinetic energy values reported in this paper are higher than the ones reported in the analysis of BASS sub-sample in \citet{rojas20}, which could be a combination of the fact that integrated spectra was used in \citet{rojas20}, resulting in time-averaged outflow properties, and that a much higher electron density value of 10$^{4.5}$ cm$^{-3}$ was used (both of which would approximately compensate for the observed differences between the two papers).
 
Calculation of the AGN luminosity coupling with outflow kinetic energy is further complicated with the fact that the outflows may not be in single gas phase, but may consist of ionised, molecular and neutral gas components \citep[e.g.,][]{rupke13, feruglio15, cicone18, husemann19, baron20, fluetsch20, herrera-camus20, perna20, treister20}. Similar mass outflow rate maps with different gas phases need to be traced to get a true picture of the extent of AGN influence on the ISM. While this paper characterises the properties of spatially resolved outflows, the impact that these outflows have on the host galaxy properties such as the overall molecular gas content and star formation rate will be addressed in future publications. 


\section{Summary and Conclusions} \label{sect7}

In this paper, we presented the \oiii$\lambda$5007 based ionised gas outflow properties, specifically the outflow velocity, mass outflow rates and kinetic energy, in 22 low redshift X-ray AGN host galaxies selected from the BASS survey. We have used optical IFS capabilities of MUSE instrument on board the VLT to understand how observational and analysis methods can influence these outflow properties and their correlations with the AGN properties, namely $L_{\rm bol}$, $M_{\rm BH}$, $\lambda_{\rm Edd}$ and $P_{\rm 1.4 GHz}$. Specifically, we used a 3\arcsec ~circular aperture and a $1.5\arcsec\times10\arcsec$ rectangular aperture, to mimic the fibre and slit observations that are frequently used in the literature. We modelled the emission lines using multi-Gaussian functions and the broad Gaussian function was used as a proxy for the outflowing component. From the spectra extracted from these apertures, we computed the \oiii ~velocity dispersion and time-averaged outflow rates and kinetic energies over the lifetime of the outflows. The electron density of the outflows are determined using the outflowing component of \sii$\lambda\lambda$6716, 6731 emission lines. We also derived instantaneous mass outflow rate in every pixel for all the galaxies, which incorporates variable outflow flux, density and velocity within the outflowing media. Finally, we investigated the presence of scaling relations between these outflow properties from multiple methods versus the AGN properties. The scaling relations are explored for the following range in the AGN properties: $L_{\rm bol} = 10^{42.6}$--$10^{45.6}$ erg s$^{-1}$; $M_{\rm BH} = 10^{6}$--$10^{8.9}$ M$_{\odot}$; $\lambda_{Edd} = $ 0.002--1.062 and $P_{\rm 1.4 GHz} =  10^{20.6}$--$10^{26}$ W Hz$^{-1}$. We enumerate here the main results from the analysis presented in this paper:

\begin{itemize}
\item[1.] Nearly 80\% of the targets show consistent \oiii ~$w_{80}$ values in the integrated spectra, irregardless of whether they are obtained from fibre or slit spectra. The differences in the 20\% of the galaxies result from whether the slit orientation is aligned with the outflow. The \oiii ~$w_{80}$ parameter shows a linear correlation with $L_{\rm bol}$  and a weak or no correlation with $M_{\rm BH}$, $\lambda_{\rm Edd}$ and $P_{\rm 1.4 GHz}$. The results of the scaling relations do not depend on whether the spectra is obtained from the fibre or slit aperture. 
\item[2.] A comparison between the $w_{80}$ maps and the BPT maps in each galaxy shows that the \oiii ~$w_{80}$ values have higher values ($\sim$120 km s$^{-1}$) on average in AGN ionised regions compared to regions ionised purely by star formation, suggesting an AGN impact on the ionised gas kinematics. Furthermore, the narrow Gaussian component reproduces the rotation profiles of the host galaxies and the kinematic maps show that the FWHM of the narrow component is $<$250 km s$^{-1}$ in most cases. The broad Gaussian map, on the other hand, shows a much wider distribution in its width (FWHM) with values reaching >600 km s$^{-1}$ and therefore, treated as tracing the outflowing component.
\item[3.] The summed local instantaneous mass outflow rate, obtained from the resolved maps is, on average, $\sim$2 orders of magnitude higher than the time-averaged mass outflow rate obtained from the integrated fibre or slit spectra. This difference is a due to a combination of higher outflow masses obtained via the larger area in the resolved maps and the respective equations used in the outflow rate computations of instantaneous and time-averaged values. The summed instantaneous mass outflow rates from the resolved maps is in the range 0.2--275 M$_{\odot}$ yr$^{-1}$, while the time-averaged outflow rate from the fibre aperture is in the range 0.001--38 M$_{\odot}$ yr$^{-1}$ and slit aperture 0.001--15 M$_{\odot}$ yr$^{-1}$. The time-averaged outflow rates are consistent with the values found in the literature using fibre or slit spectra.
\item[4.] Both the instantaneous and the time-averaged mass outflow rates from fibre and slit apertures show a linear correlation with $L_{\rm bol}$ (p-value for non-correlation $\leq$0.01). The correlations are strongest (smaller p-value) with time-averaged outflow rate values obtained from the fibre aperture. In the case of instantaneous outflow rates from resolved maps, smaller p-values are obtained for correlations with $L_{\rm bol}$ when the outflow rates are summed over the central 3\arcsec ~region, compared to the outflow rates summed over the entire FoV. These results suggest an efficient coupling between the AGN radiation and the ISM closer to the AGN location. We do find a weak or no correlation between the outflow rates obtained with the different methods and $M_{\rm BH}$, $\lambda_{\rm Edd}$ or $P_{\rm 1.4 GHz}$. This suggests that no single mechanism is dominant for driving the outflows in the low redshift sample presented in this paper.
\item[5.] Lastly, we find the median outflow coupling, i.e. the ratio between the outflow kinetic energy and the bolometric luminosity to be $\sim$1\%, although the entire range of coupling efficiency is between $<$0.1 and 40\%.
\end{itemize}

Although this paper presents the outflow properties in a single gas phase i.e. the ionised gas, many of these galaxies are known to have outflows existing in other gas phases such as the molecular gas phase, which may have a larger energy budget compared to the ionised gas. Therefore future work will present outflows from other gas phases using current and upcoming instruments/facilities such as ALMA, JVLA, PdBI, NIRSpec/JWST, MIRI/JWST and ERIS/VLT. Furthermore, the high velocity outflows with \oiii ~width >600 km s$^{-1}$ are found closest to the AGN and are at times not resolved with the currently available MUSE data. Therefore, a future targeted follow-up at high resolution using AO instruments such as the Narrow Field Mode of MUSE would be key in characterising the outflow properties close to the AGN. Instruments on board the Extremely Large Telescope such as HARMONI \citep[e.g.,][]{thatte10} will also play a key role in pushing such studies to even higher redshifts (where AGN with higher bolometric luminosity are detected) with increased sensitivity and spatial resolution. 

\section*{Acknowledgements}

We thank the referee for their useful suggestions. ET and FEB acknowledge support from ANID grants CATA-Basal AFB-170002 and FB210003 and FONDECYT Regular grant 1190818. ET acknowledges support from ANID Anillo ACT172033 and Millennium Nucleus NCN19\_058 (TITANs). FEB acknowledges support from FONDECYT regular grant 1200495, Millennium Science Initiative ICN12\_009. ARL acknowledges aupport from FONDECYT Postdoctorado project No. 3210157. MK acknowledges support from NASA through ADAP award NNH16CT03C. KO acknowledges support from the National Research Foundation of Korea (NRF-2020R1C1C1005462). CR acknowledges support from the Fondecyt Iniciacion grant 11190831 and ANID BASAL project FB210003. BT acknowledges support from the Israel Science Foundation (grant number 1849/19) and from the European Research Council (ERC) under the European Union's Horizon 2020 research and innovation program (grant agreement number 950533). DK, ES, NDM and NG acknowledge the support from ESO for their studentship programme and Science Support Discretionary Funds. NDM and NG are grateful for the hospitality by ESO-Chile. FR acknowledges support from PRIN MIUR 2017 project “Black Hole winds and the Baryon Life Cycle of Galaxies: the stone-guest at the galaxy evolution supper”, contract 2017PH3WAT. RR thanks Conselho Nacional de Desenvolvimento Cient\'{i}fico e Tecnol\'ogico ( CNPq, Proj. 311223/2020-6,  304927/2017-1 and 400352/2016-8), Funda\c{c}\~ao de amparo 'a pesquisa do Rio Grande do Sul (FAPERGS, Proj. 16/2551-0000251-7 and 19/1750-2), Coordena\c{c}\~ao de Aperfei\c{c}oamento de Pessoal de N\'{i}vel Superior (CAPES, Proj. 0001). Based on observations made with ESO telescopes at the La Silla Paranal Observatory under programmes 60.A-9100(K), 60.A-9339(A), 094.B-298, 094.B-0321, 094.B-0345, 095.B-0015, 095.B-0482, 095.B-0532, 095.B-0934, 096.D-0263, 096.B-0309, 097.B-0080, 097.D-0408, 099.B-0137, 0100.B-0116. 


\section*{Data Availability}
All the raw data analysed in this paper are publicly available in the ESO archive. The respective programme IDs from which these data are obtained are mentioned in the footnote in Sect. \ref{sect2}.

\bibliographystyle{mnras}
\bibliography{reference}

\newpage
\appendix

\section{Appendix} \label{sect:appendix}
In the following section, we show the plots for the rest of the sample presented in this paper.

\begin{figure*}
\centering
\includegraphics[scale=0.5]{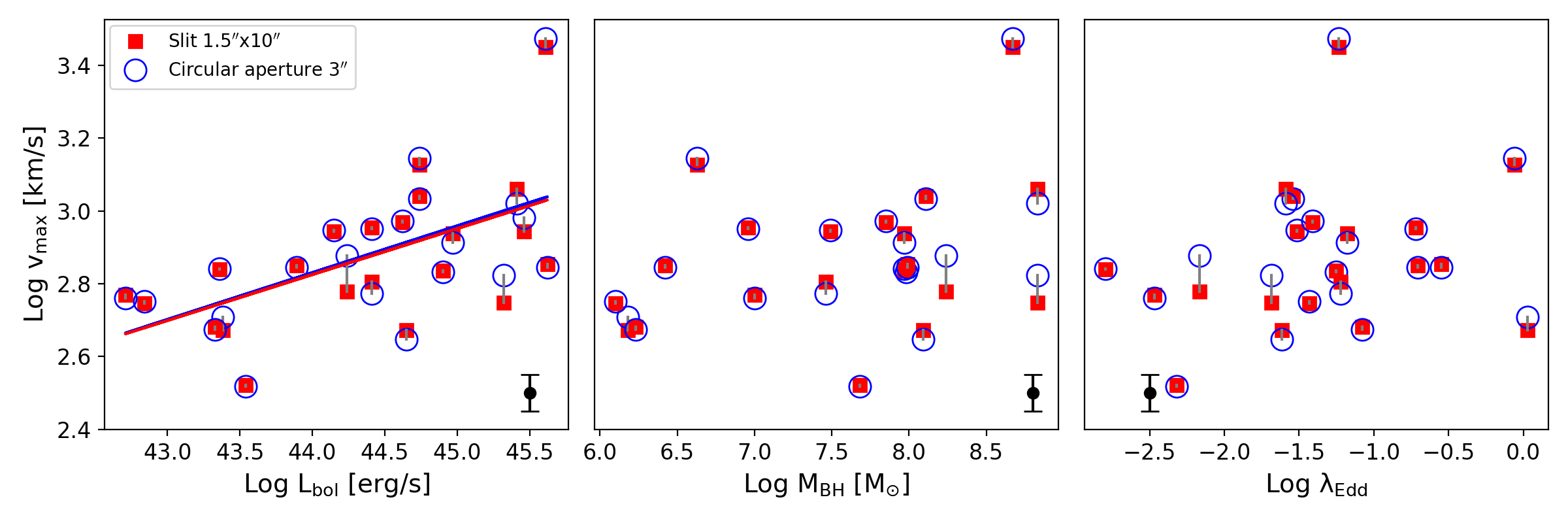}
\caption{These plots show the relation between the parametric velocity, $v_{\rm max}$, defined as $\Delta\lambda$ + 2$\sigma_{\rm broad}$ (the former term refers to the difference in the centroid of narrow and broad Gaussian components) and the AGN properties, L$_{\rm bol}$ (left panel), M$_{\rm BH}$ (middle panel) and $\lambda_{\rm Edd}$ (right panel). The parametric velocity values are obtained from spectra extracted from a fibre and slit aperture. The colour coding in these plots is the same as in Fig. \ref{fig:w80_integrated}. All galaxies show consistent $v_{\rm max}$ values and are within the measurement errors ($\sim$150 km s$^{-1}$). The blue and red curves show the best-fit linear relations for the fibre and slit spectra respectively. Similar to the case of $w_{80}$ scaling relations, the strongest correlation is observed with $L_{\rm bol}$, while weak or no correlations are observed with $M_{\rm BH}$ and $\lambda_{\rm Edd}$. The black symbol with the error bar in the lower right (or lower left) region of the plots shows the uncertainty in the $v_{max}$ values.}
\label{fig:vmax_scalings}
\end{figure*}

\begin{figure*}
\centering
\includegraphics[scale=0.24]{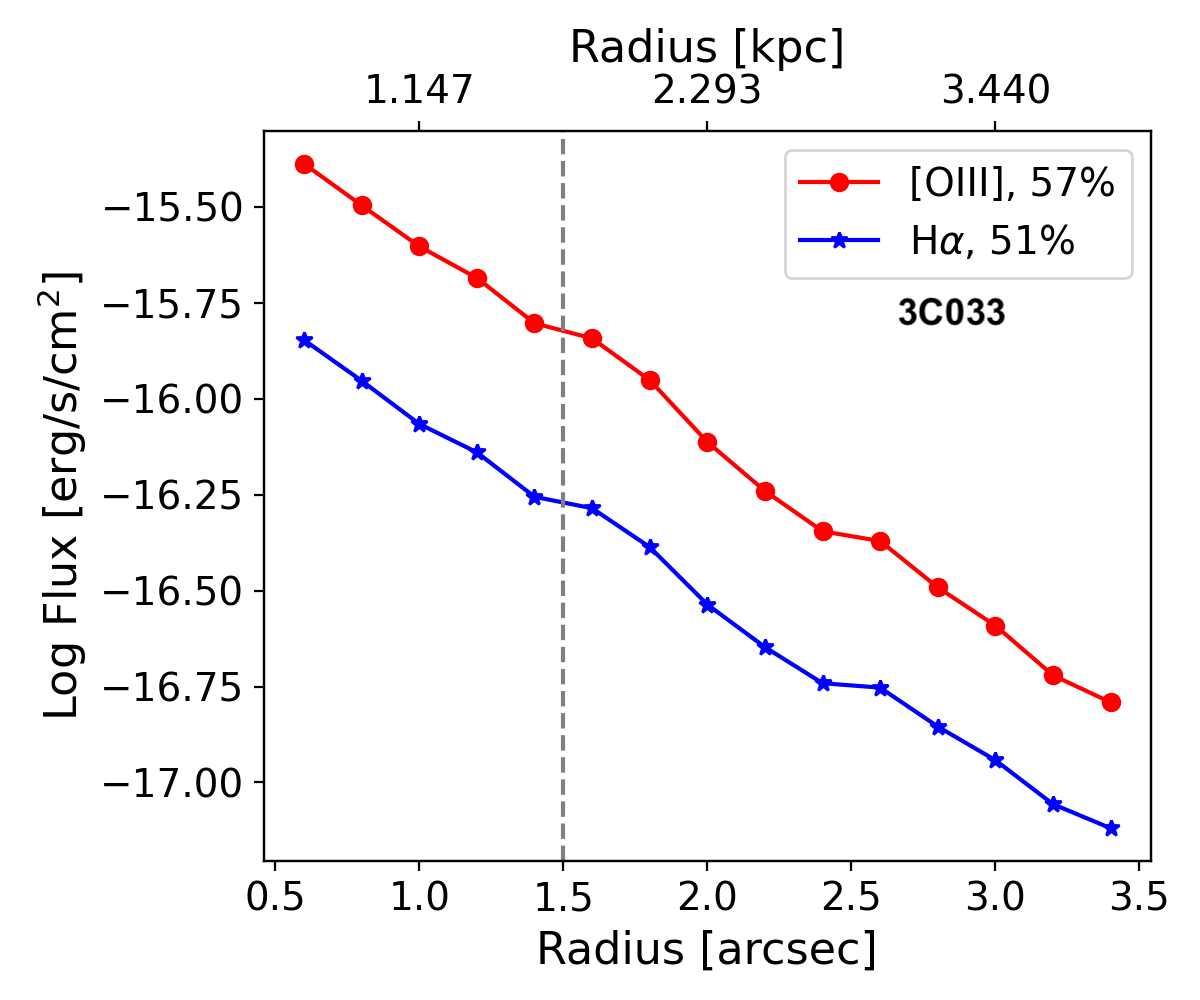}
\includegraphics[scale=0.24]{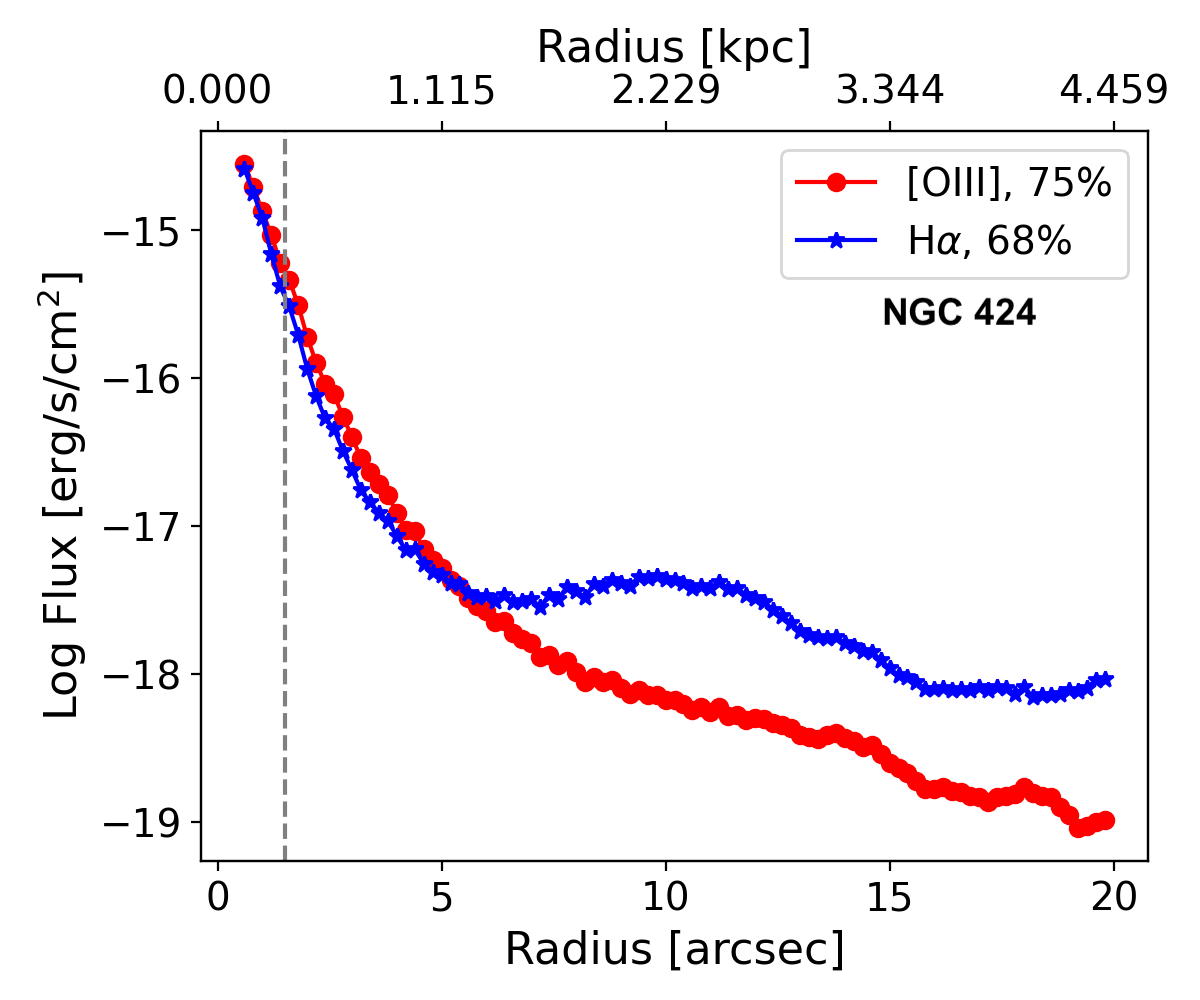}
\includegraphics[scale=0.24]{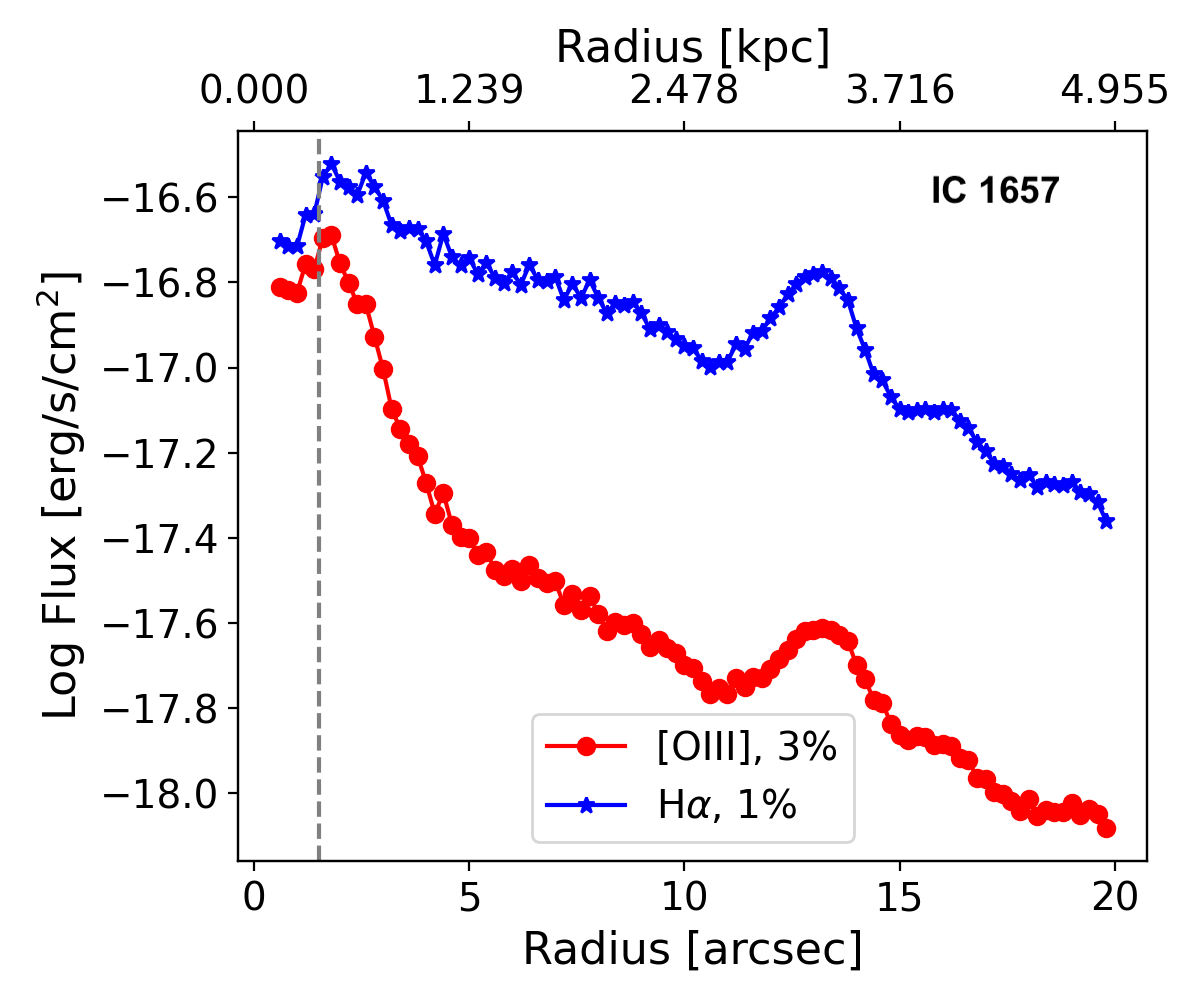}
\includegraphics[scale=0.24]{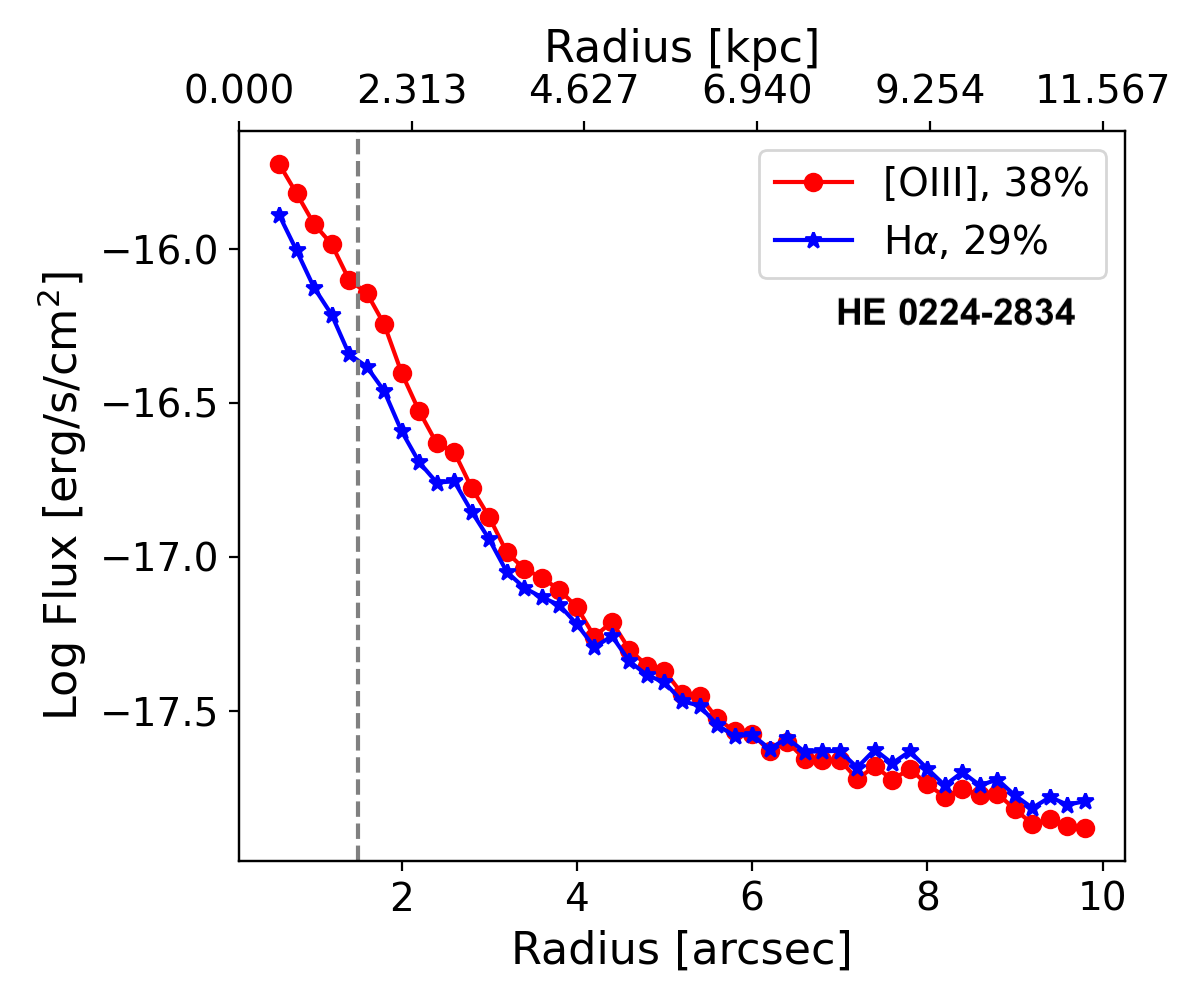}\\
\includegraphics[scale=0.24]{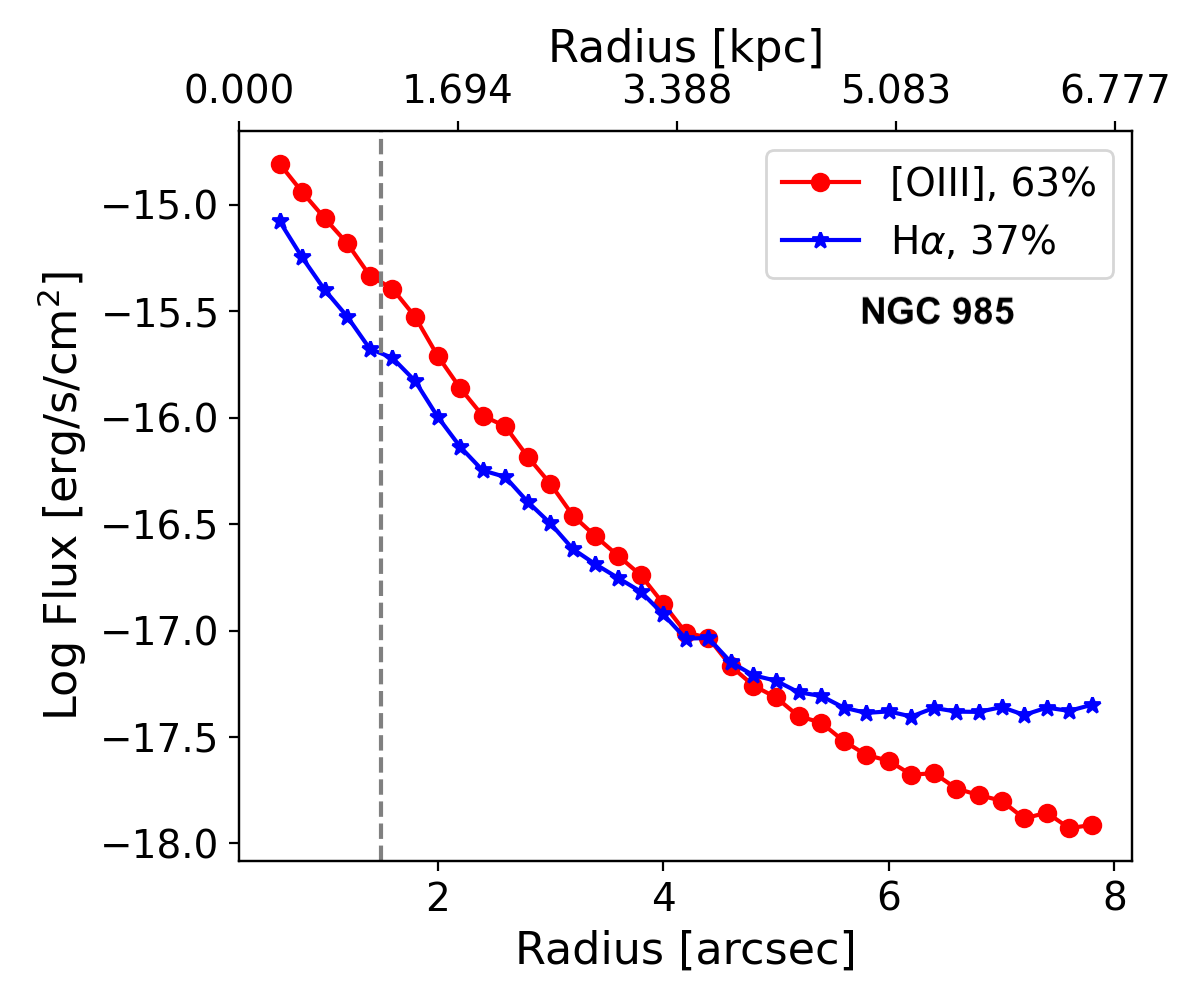}
\includegraphics[scale=0.24]{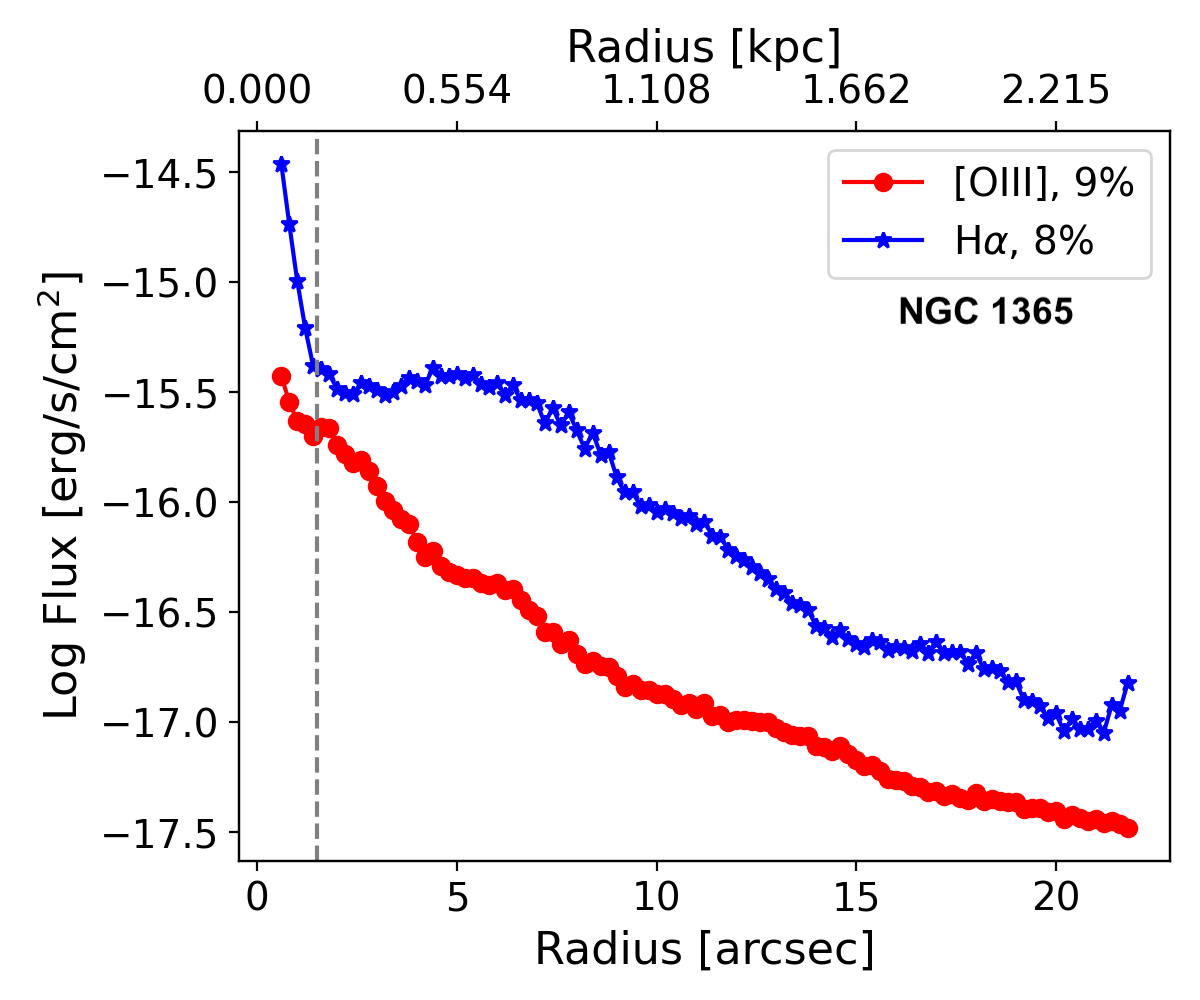}
\includegraphics[scale=0.24]{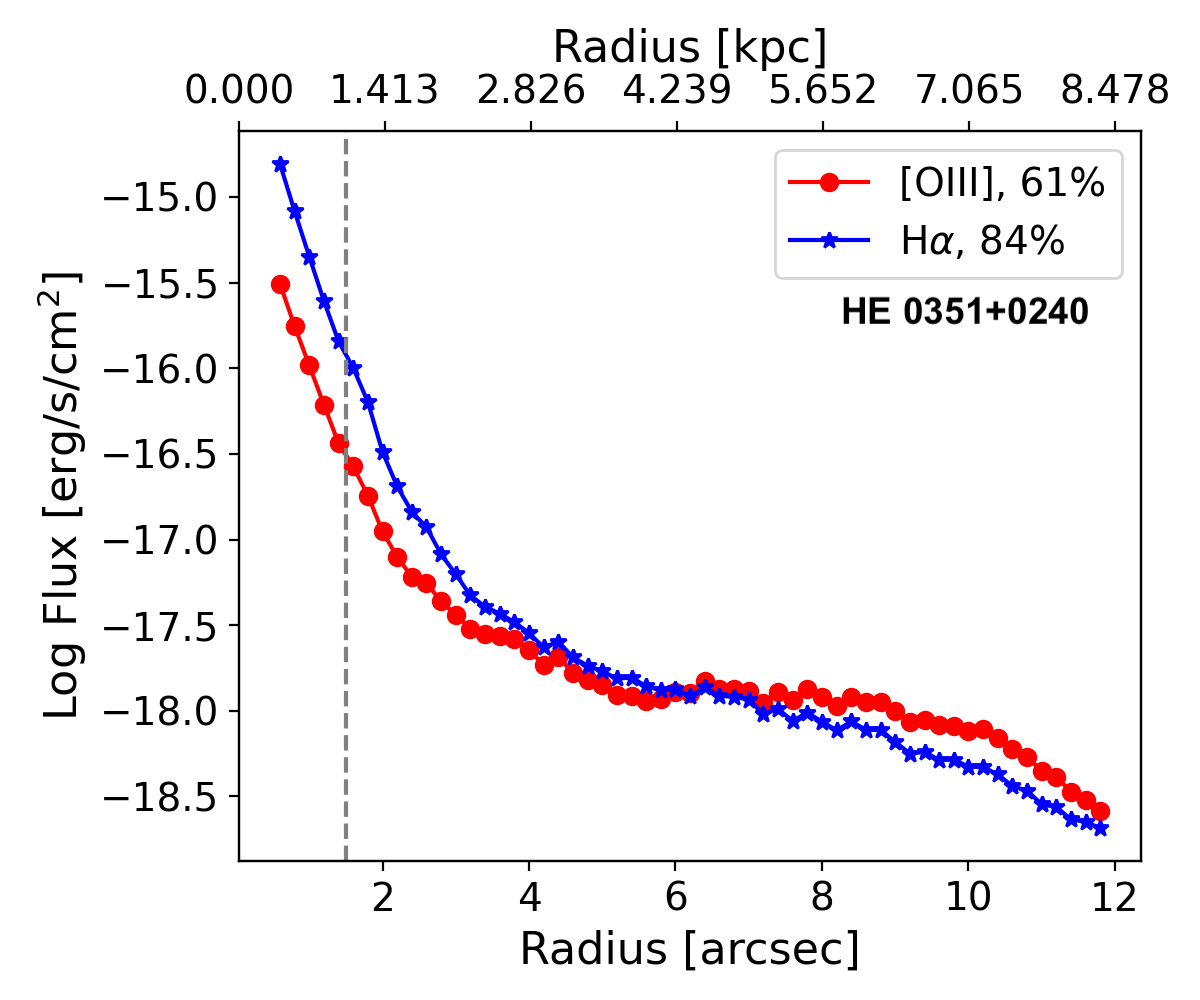}
\includegraphics[scale=0.24]{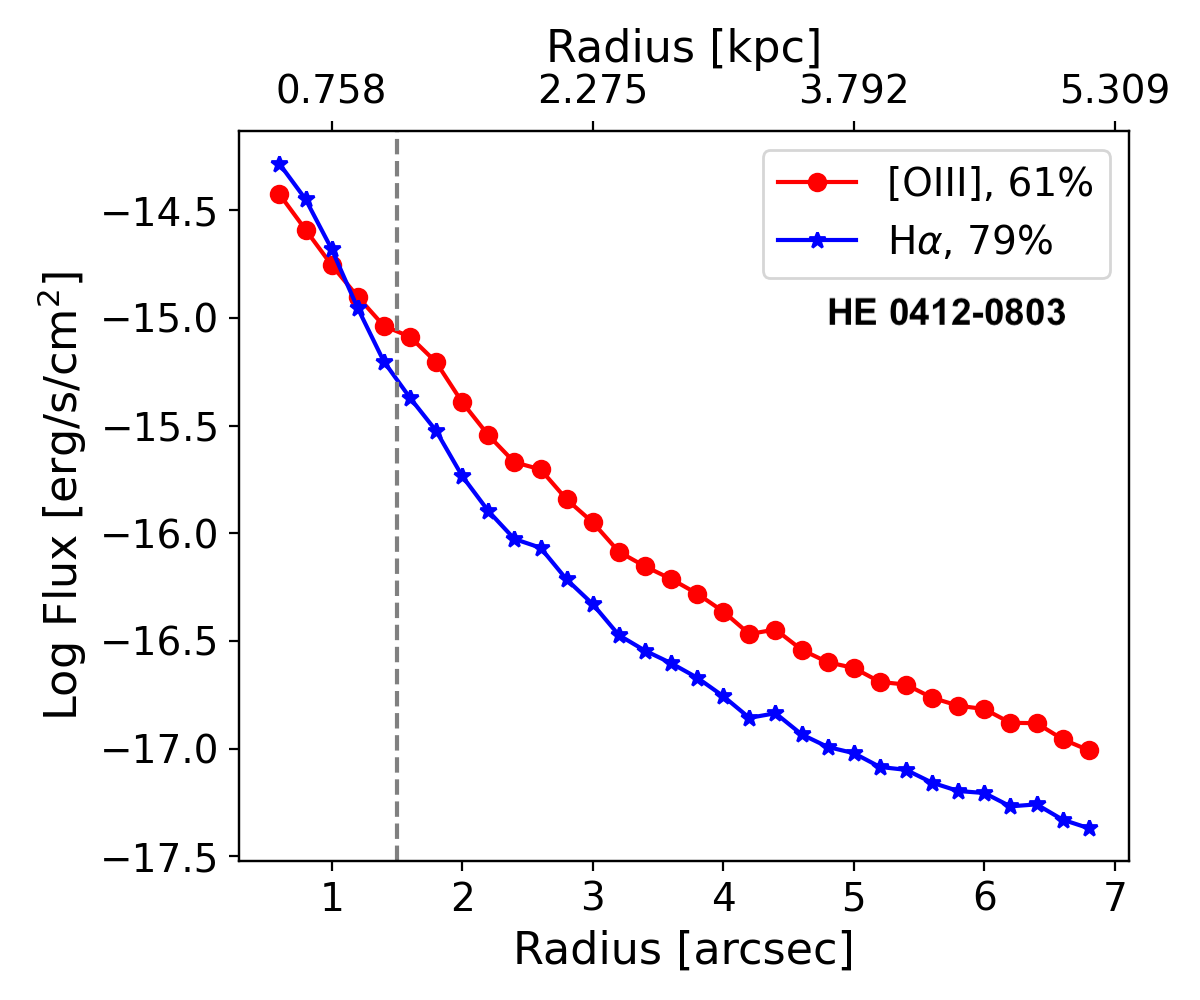}\\
\includegraphics[scale=0.24]{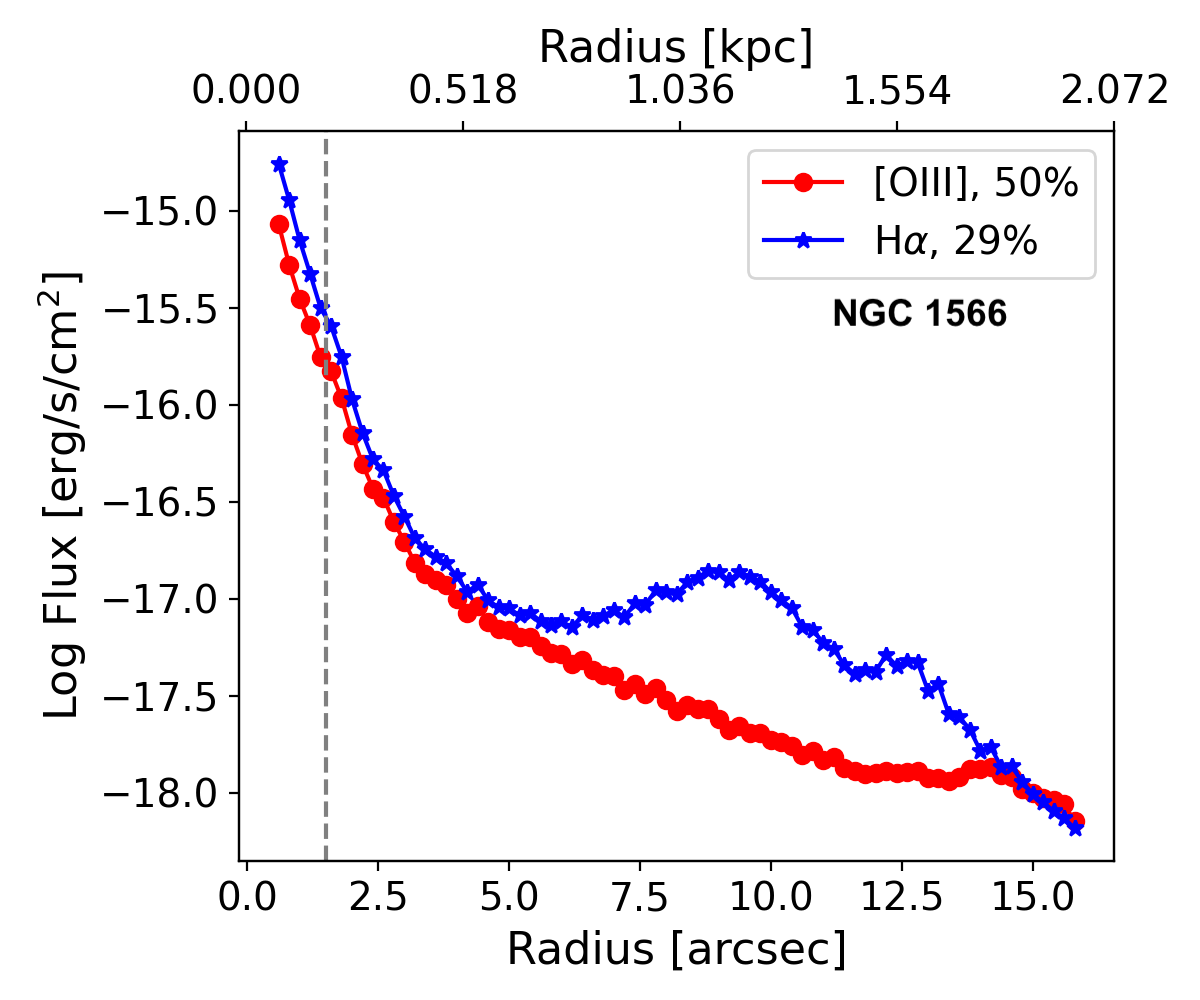}
\includegraphics[scale=0.24]{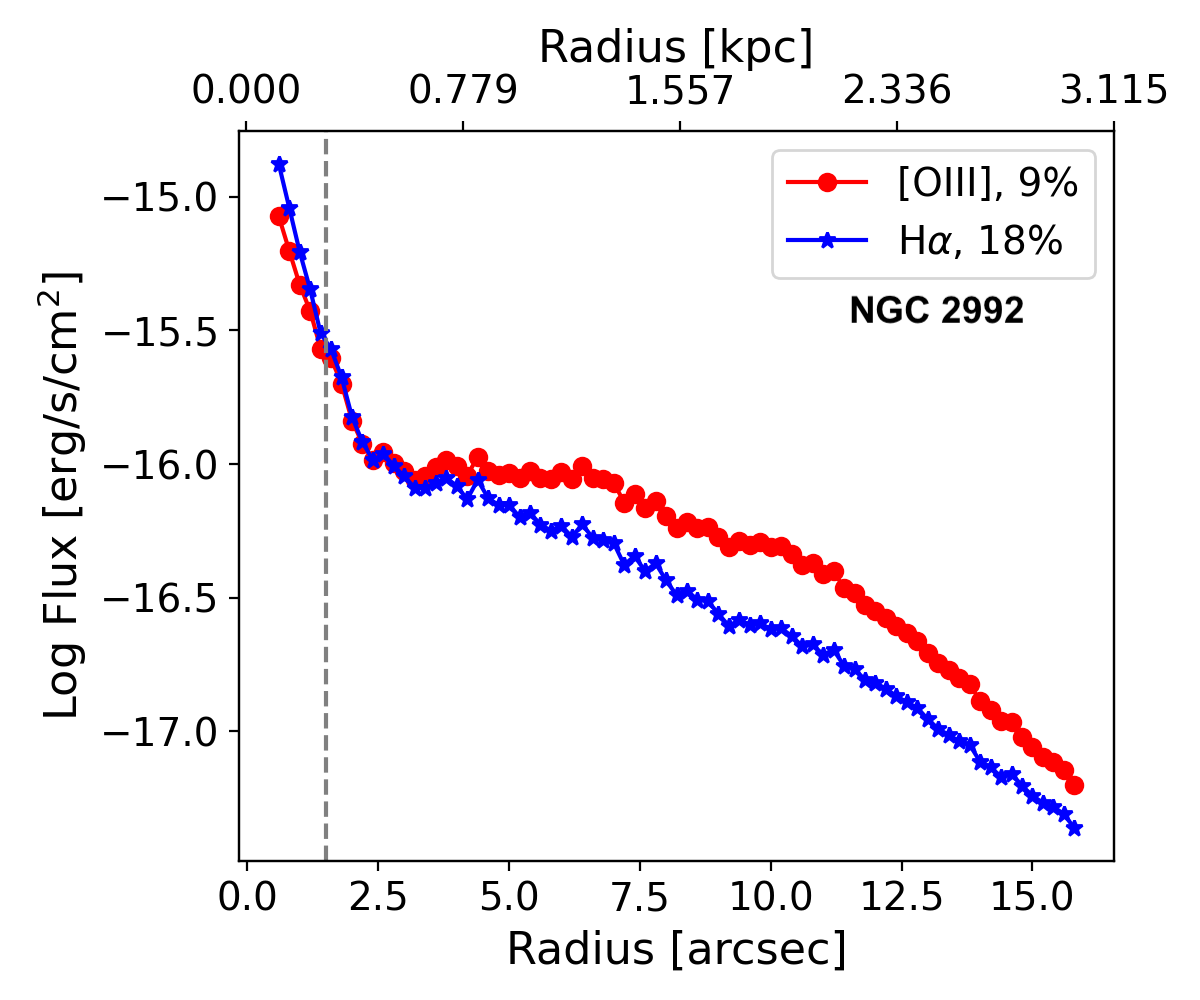}
\includegraphics[scale=0.24]{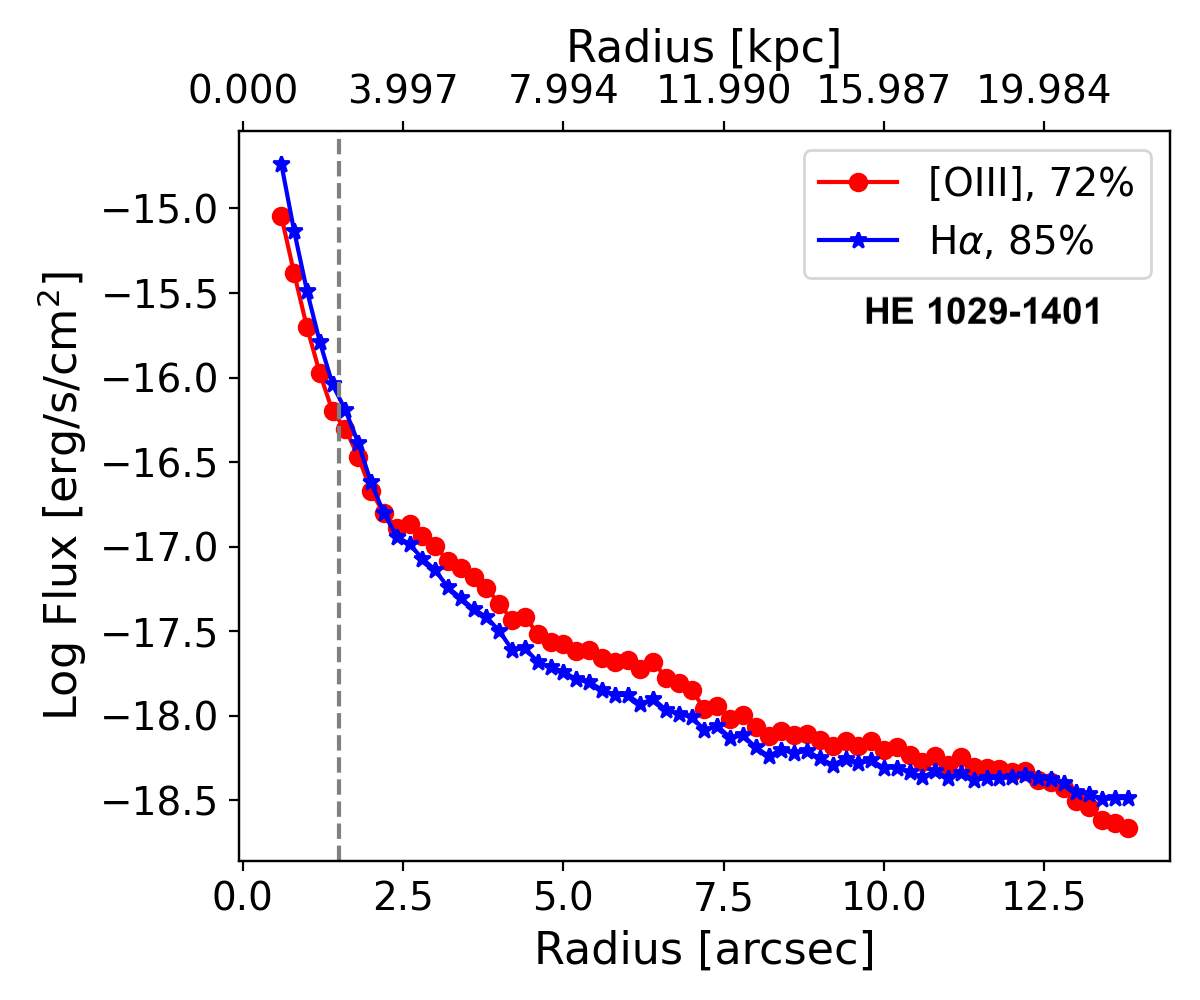}
\includegraphics[scale=0.24]{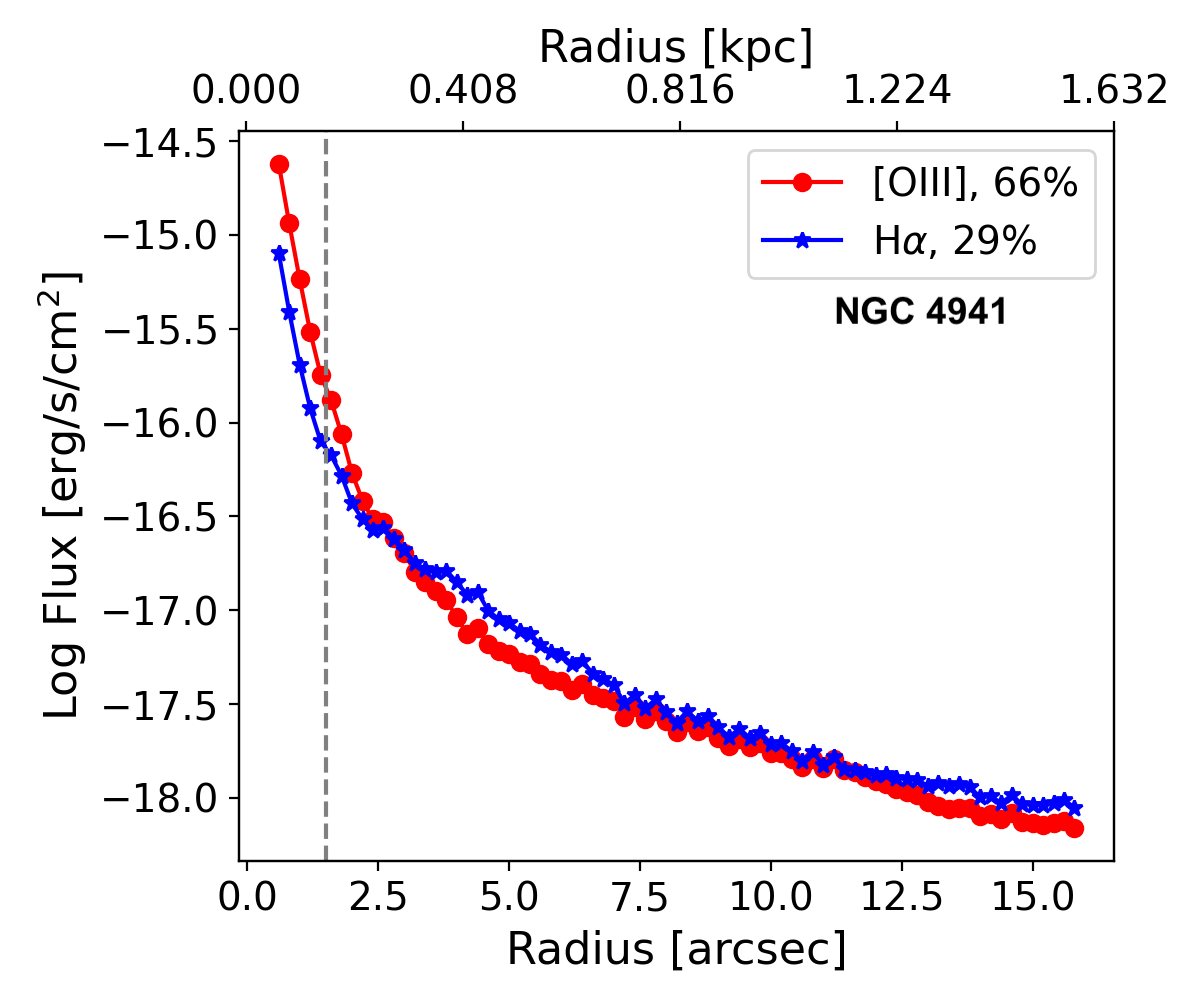}\\
\includegraphics[scale=0.24]{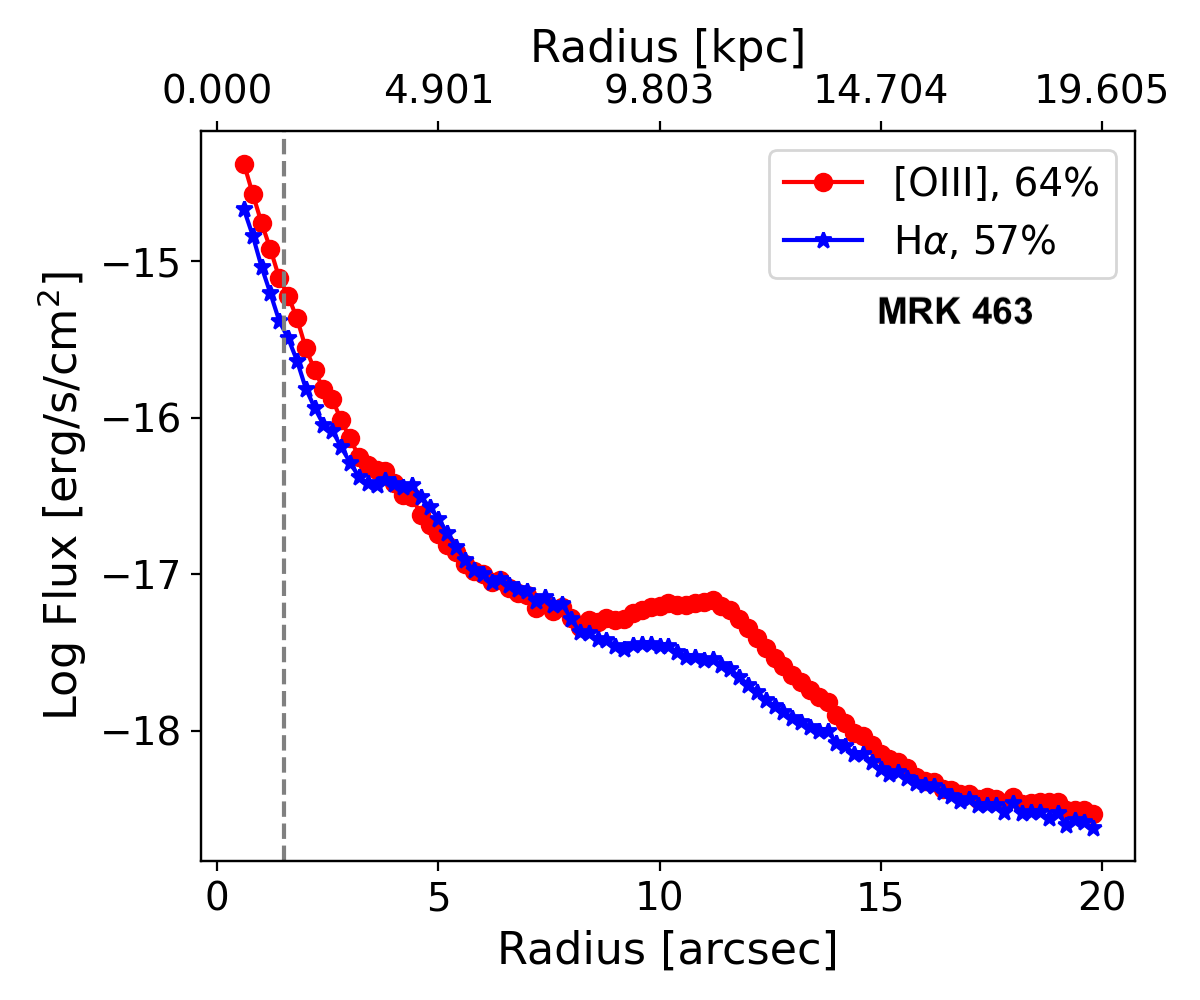}
\includegraphics[scale=0.24]{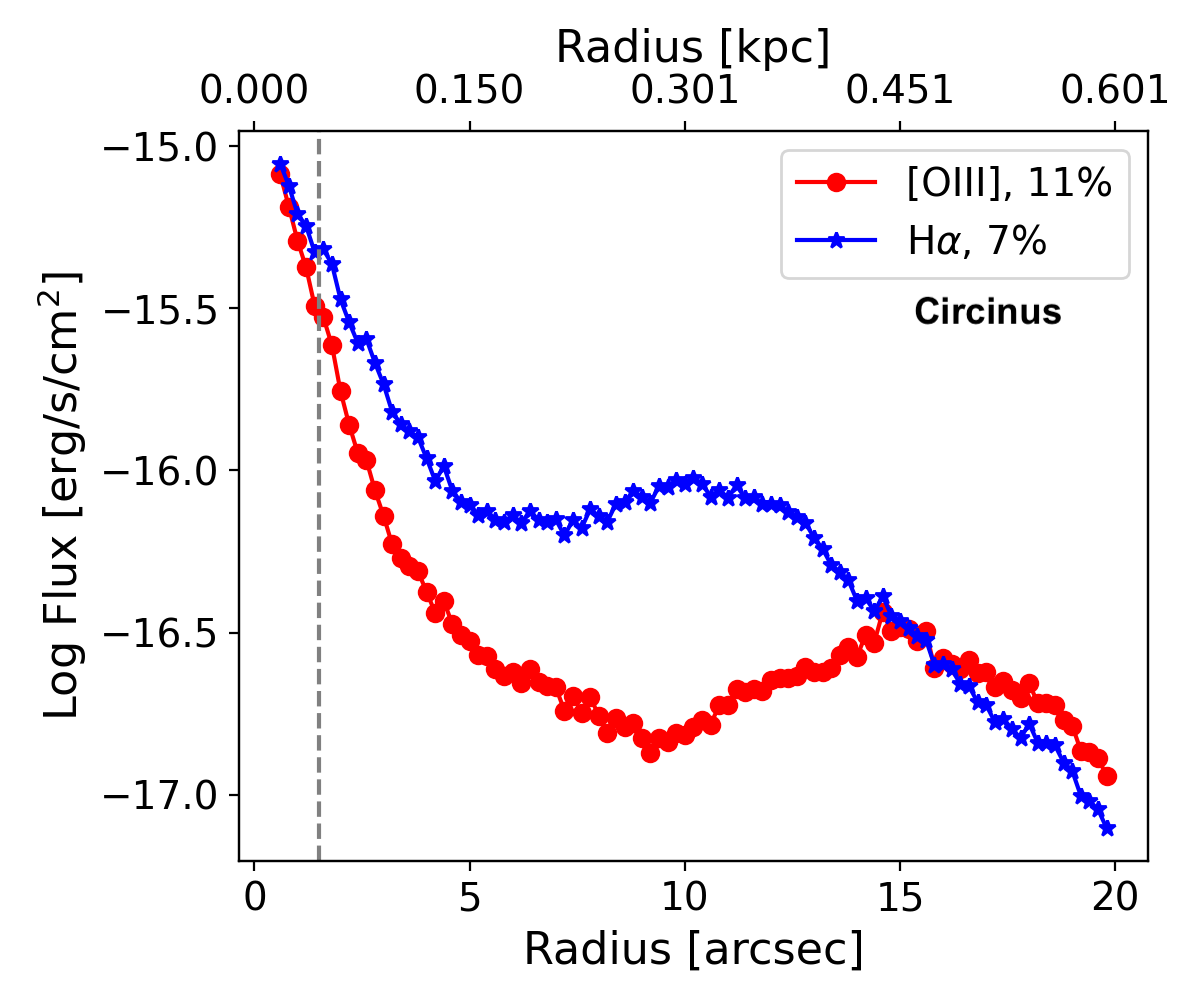}
\includegraphics[scale=0.24]{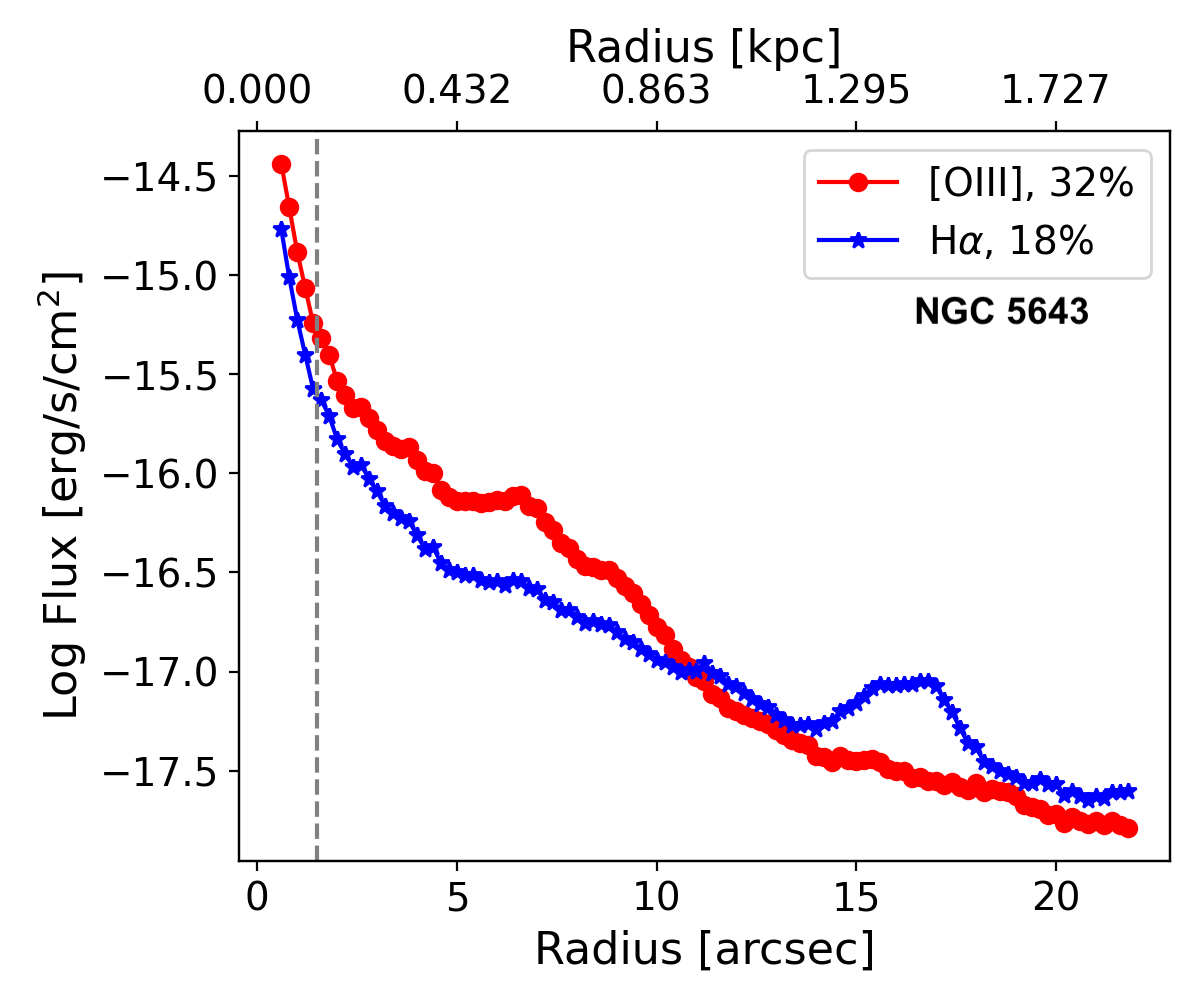}
\includegraphics[scale=0.24]{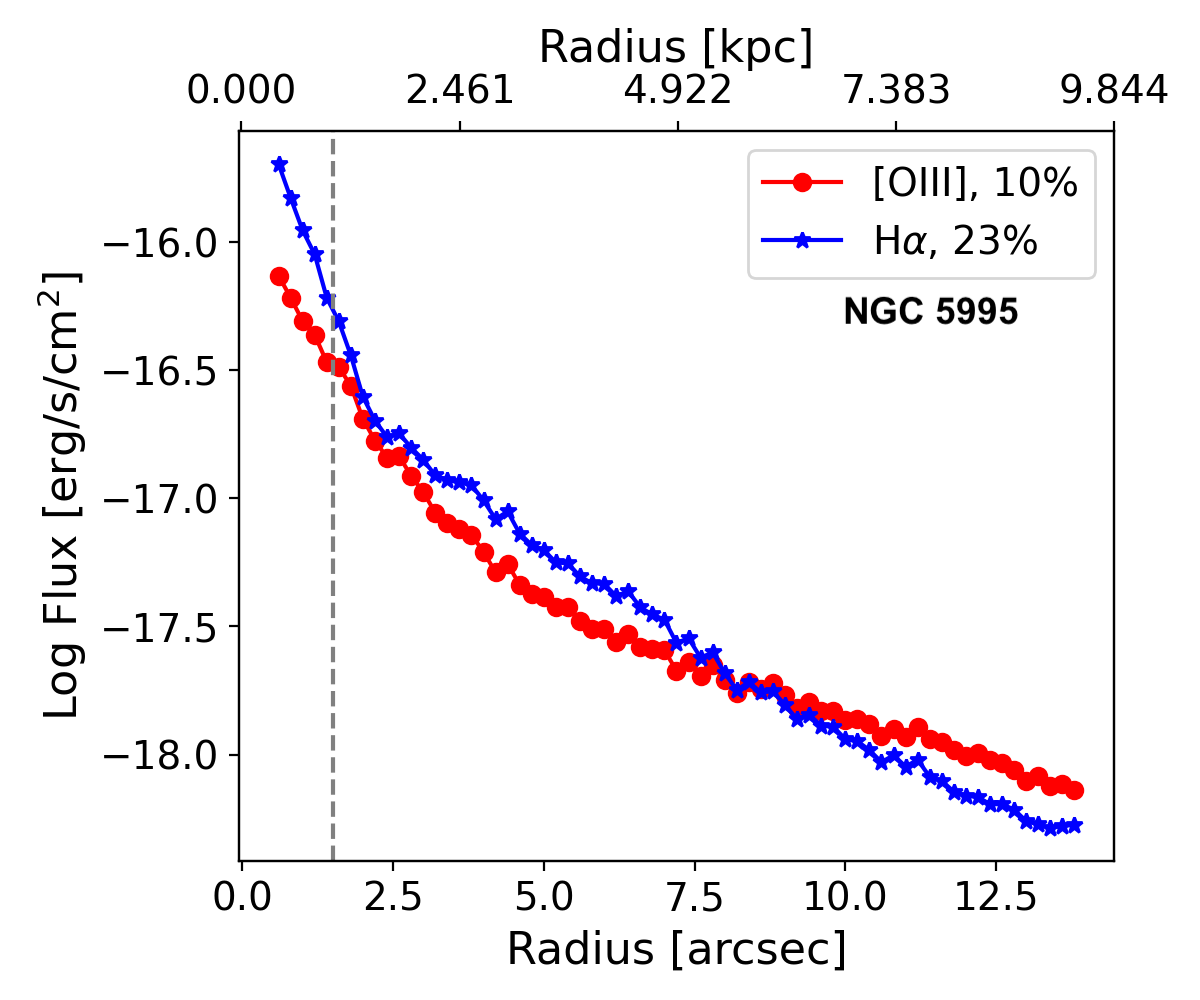}\\
\includegraphics[scale=0.24]{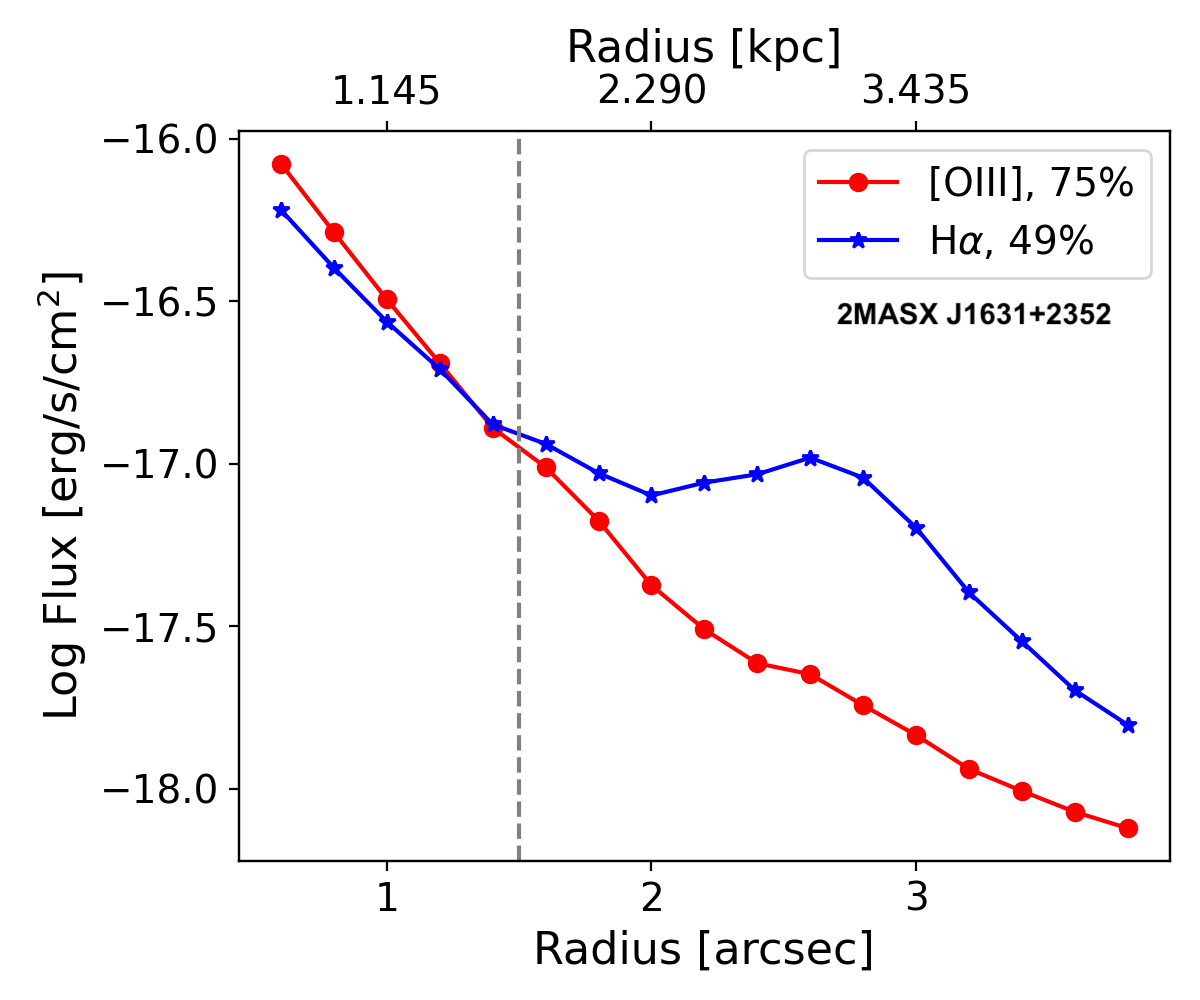}
\includegraphics[scale=0.24]{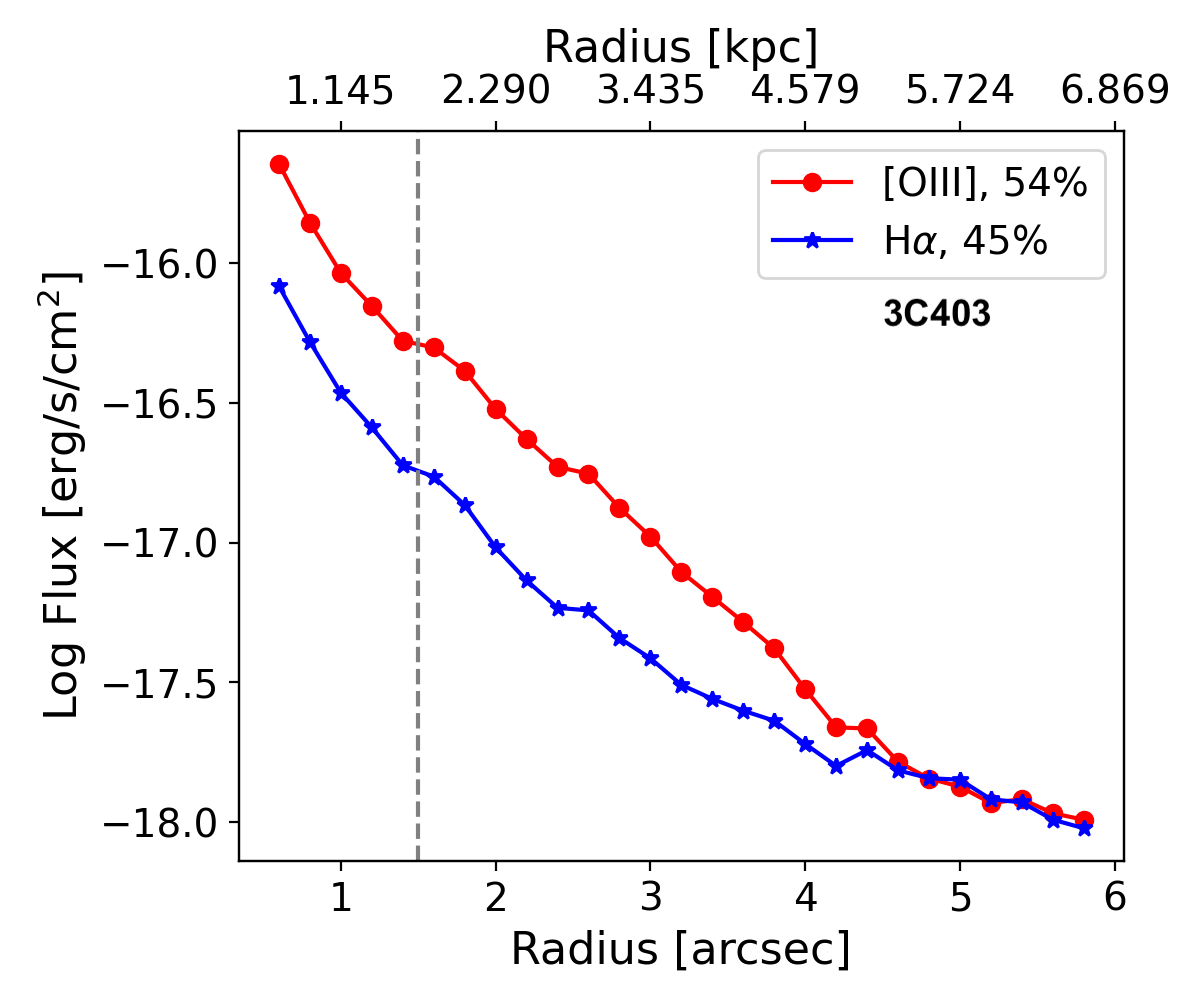}
\includegraphics[scale=0.24]{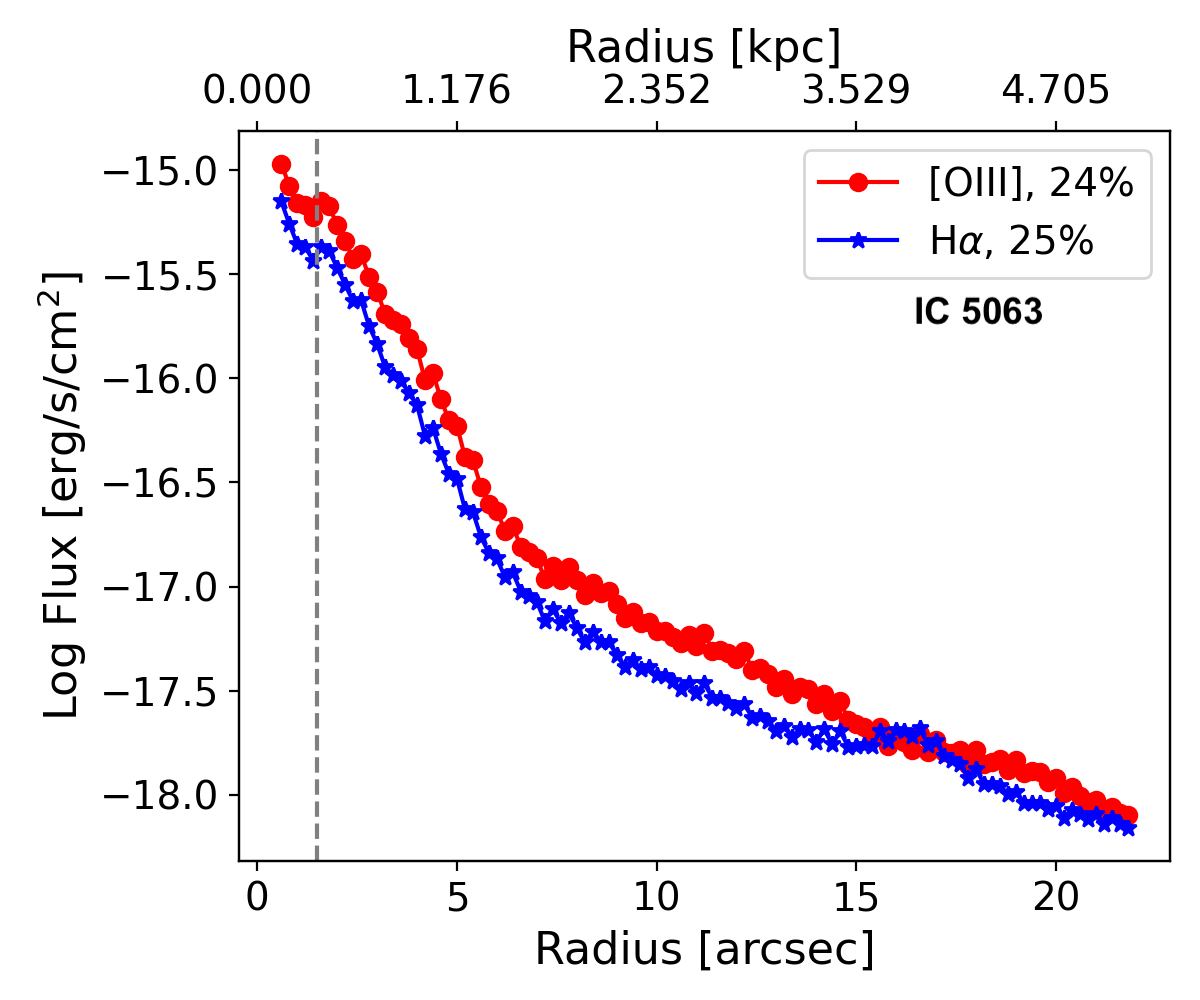}
\includegraphics[scale=0.24]{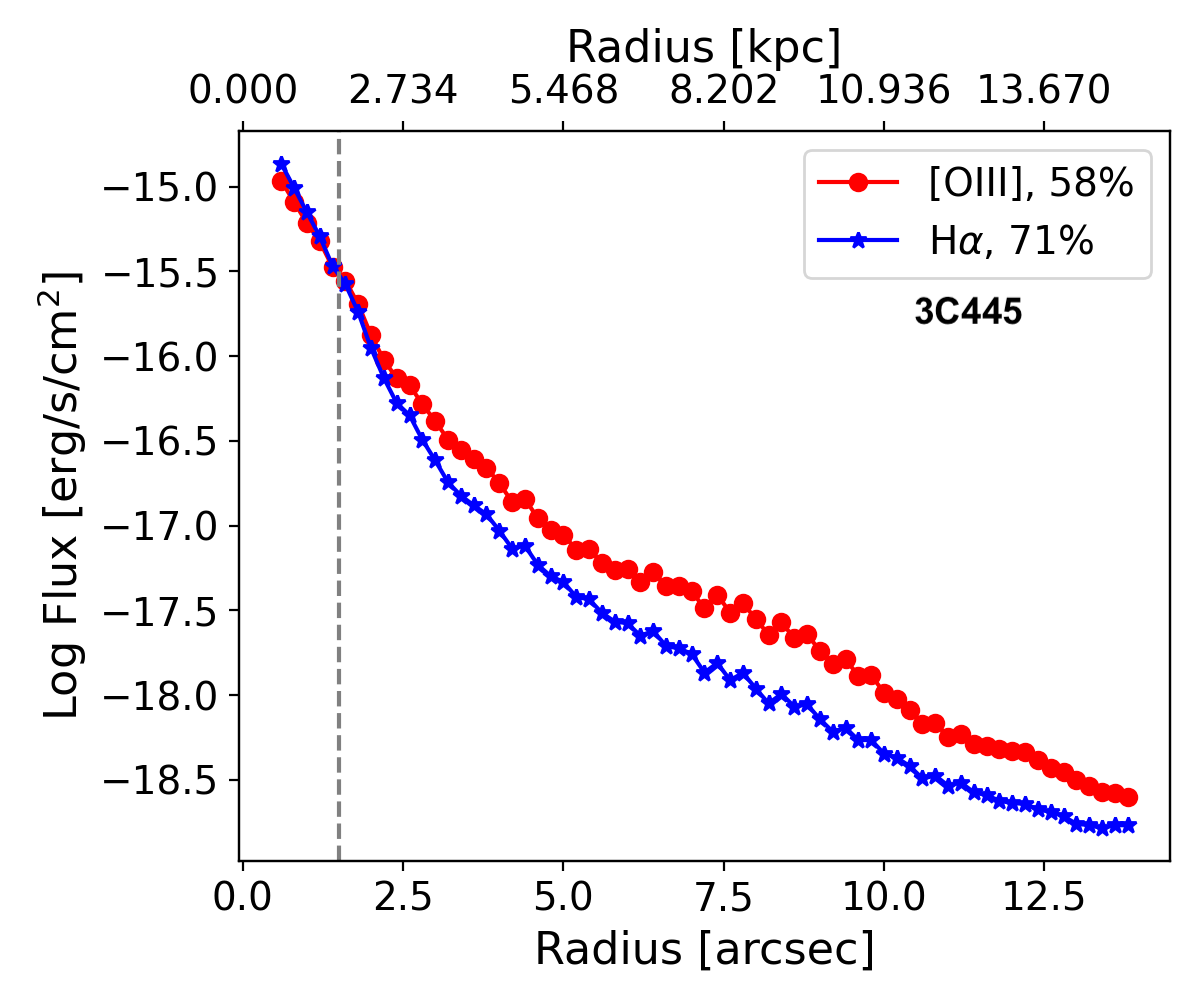}\\
\includegraphics[scale=0.24]{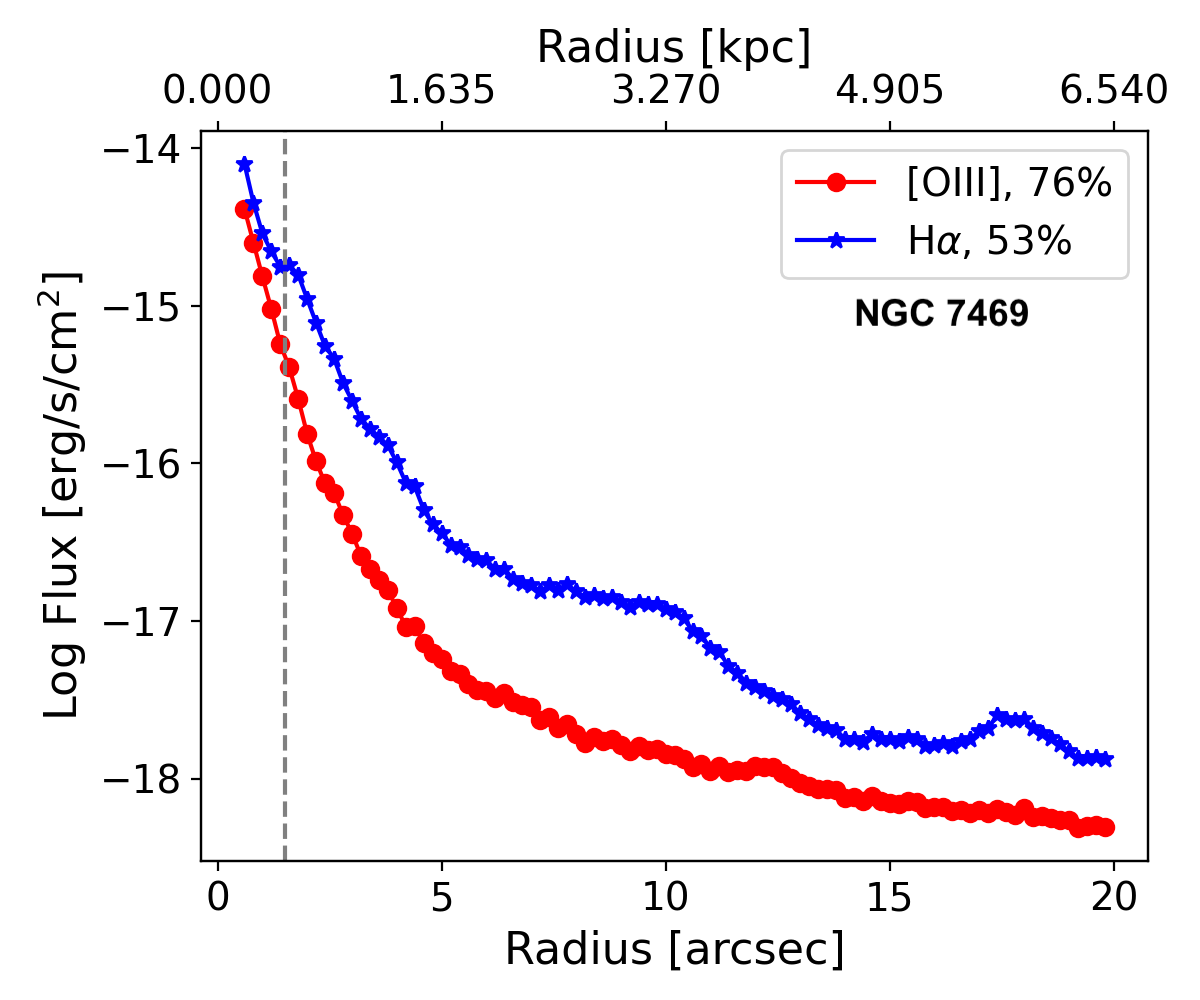}
\includegraphics[scale=0.24]{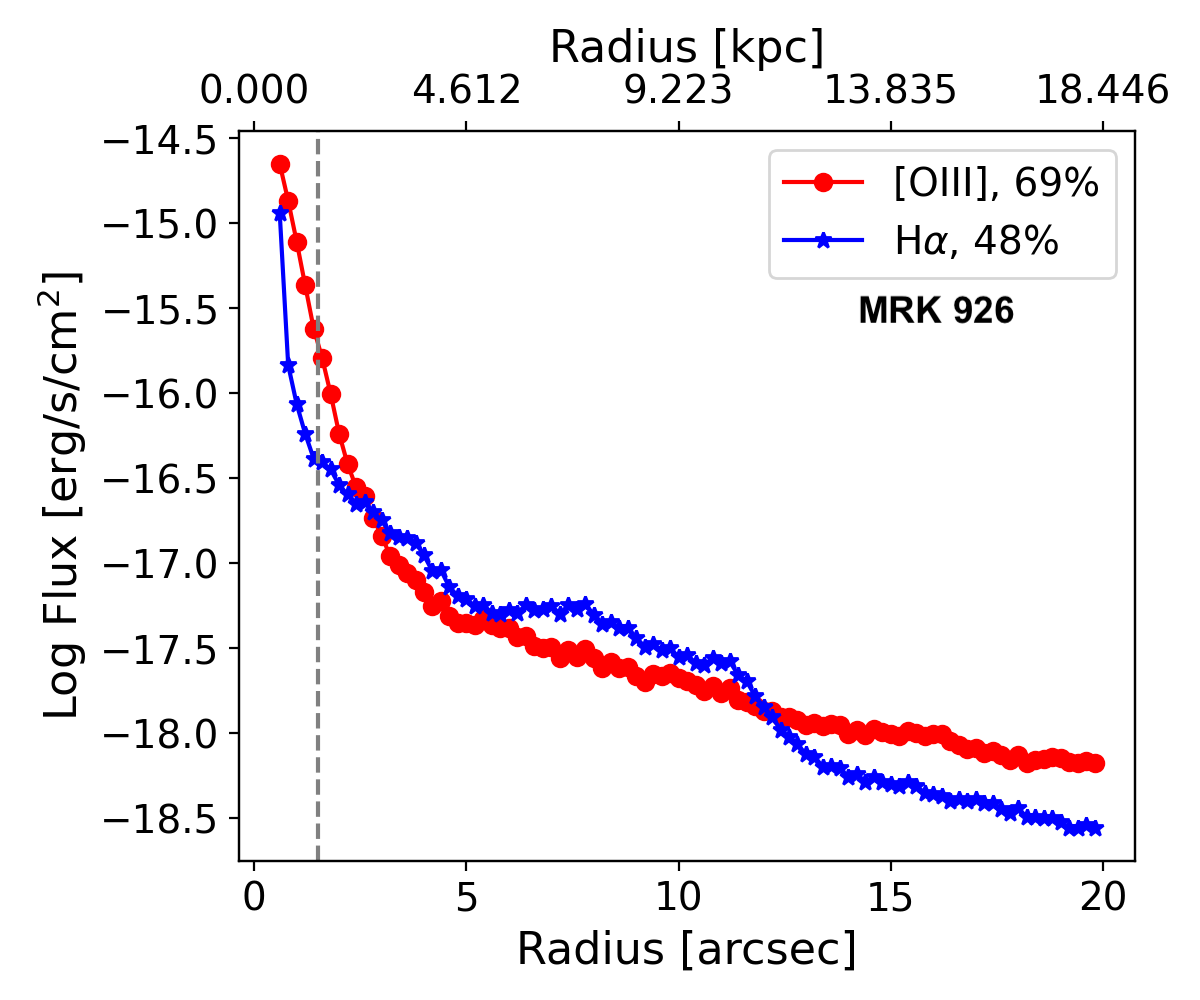}
\caption{The plots show the radial flux gradients of \oiii ~(red) and H$\alpha$ (blue) emission lines for the targets presented in this paper. The flux gradients are obtained from increasing circular shells in steps of 0.2\arcsec ~(the spatial sampling MUSE) centered on the AGN. The vertical dashed line marks the radius of the 3\arcsec ~fibre used in this paper for integrated fibre measurements. The percentages next to the legend represent the fraction of the total flux within the 3\arcsec ~radius. In most cases, the flux uniformly decreases with increasing distance from the AGN, except in targets like IC 1657, NGC 1365 and NGC 1566, where the flux gradients show a bump, often associated with one of the spiral arms or HII regions.}
\label{fig:all_flux_grads}
\end{figure*}

\begin{figure*}
\centering
\includegraphics[scale=0.5]{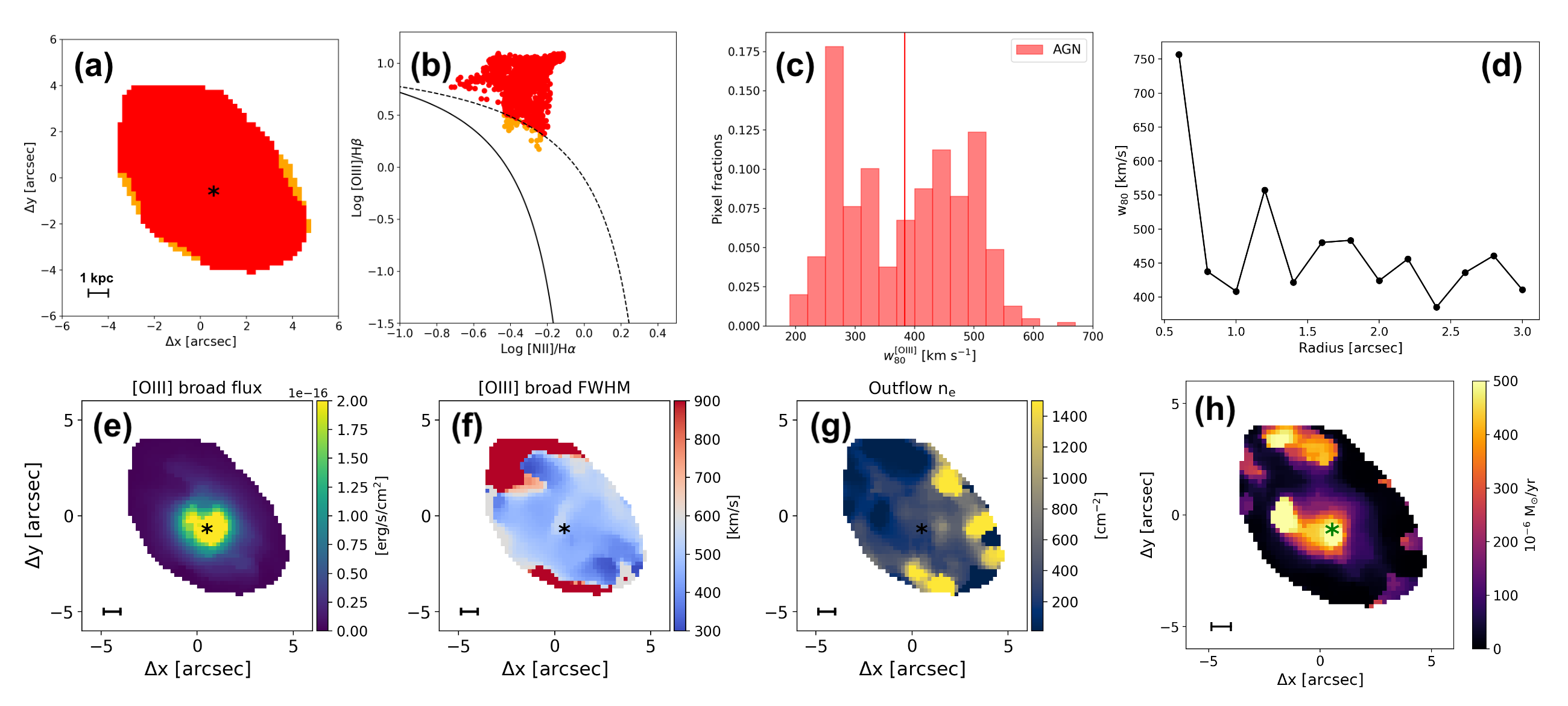}
\caption{The figure shows individual plots of 3C033. Panels (a) and (b) show the spatially resolved BPT maps. The red colour shows the region ionised by the AGN, while orange shows composite ionisation and blue (not detected in this galaxy) shows the star forming ionised regions. Panel (c) shows the \oiii ~$w_{80}$ distribution obtained from all the pixels from panel (a). In the case of ionisation by star formation, a histogram with blue colour is also shown, similar to Fig. \ref{fig:resolved_BPT_NGC7469}. Panel (d) shows the $w_{80}$ radial gradient which shows a peak at the centre in this galaxy before dropping off to lower values at larger radii. Panels (e), (f) and (g) show the \oiii ~broad component parameter maps (flux, velocity (FWHM) and electron density from \sii ~doublet respectively), which is used as an indicator of the outflowing component. Panel (h) shows the mass outflow rate map, similar to the one presented in Fig. \ref{fig:outflow_rate_map_NGC7469}. The black star in all the maps shows the AGN location and the horizontal black bar represents 1 kpc in physical scale.}
\label{fig:plots_3C033}
\end{figure*}

\begin{figure*}
\centering
\includegraphics[scale=0.5]{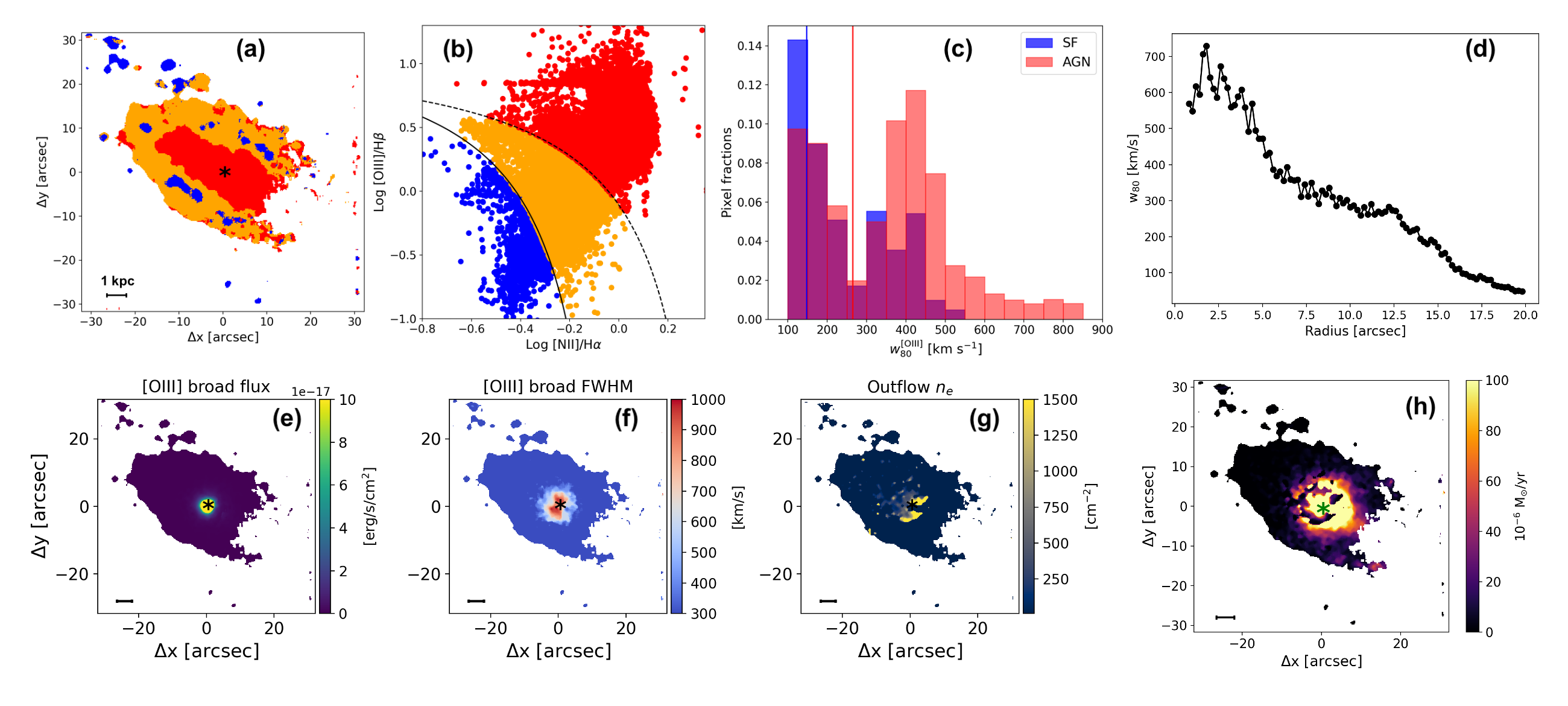}
\caption{Same as Fig. \ref{fig:plots_3C033}, for NGC 424}
\label{fig:plots_NGC424}
\end{figure*}

\begin{figure*}
\centering
\includegraphics[scale=0.5]{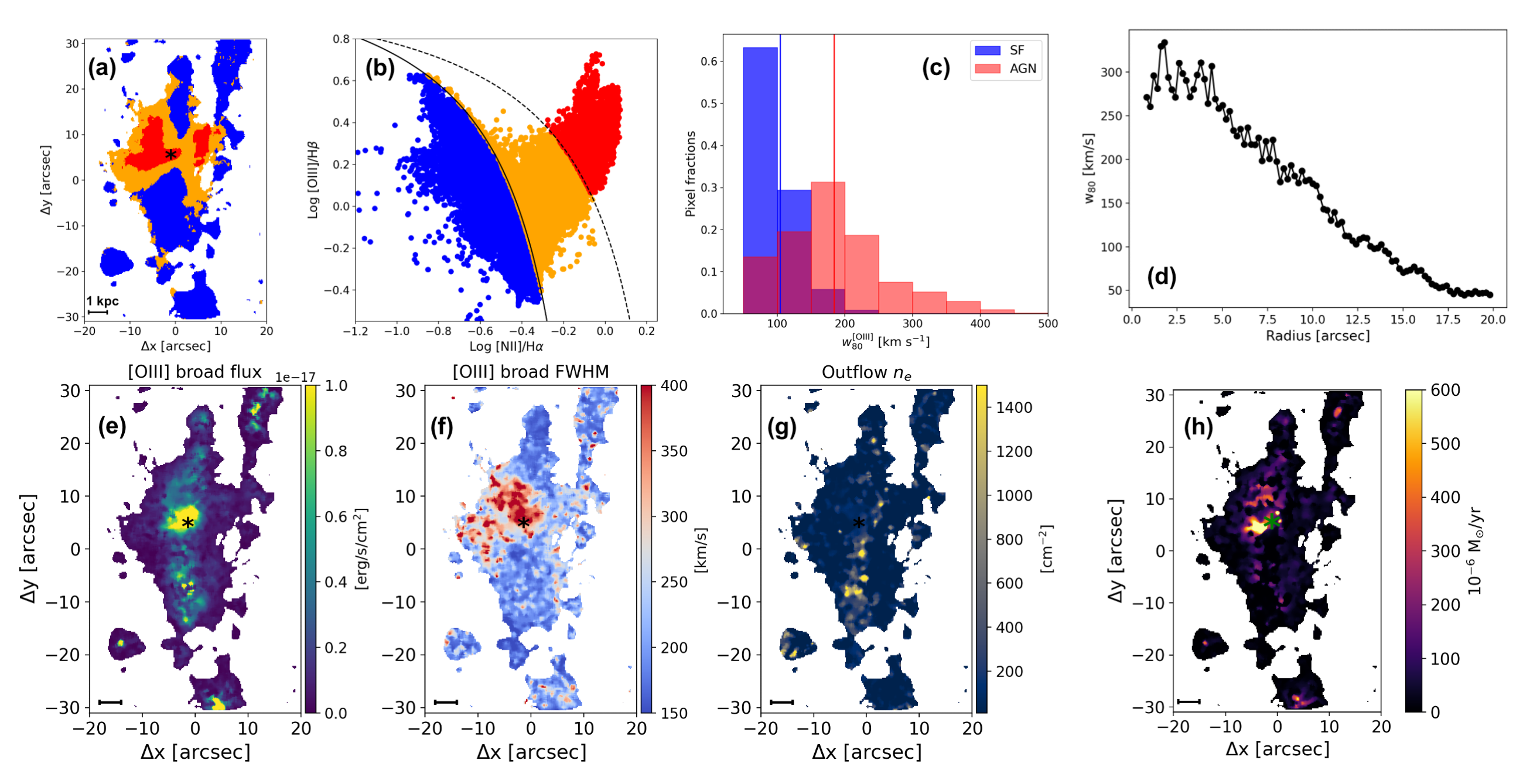}
\caption{Same as Fig. \ref{fig:plots_3C033}, for IC 1657}
\label{fig:plots_IC1657}
\end{figure*}

\begin{figure*}
\centering
\includegraphics[scale=0.5]{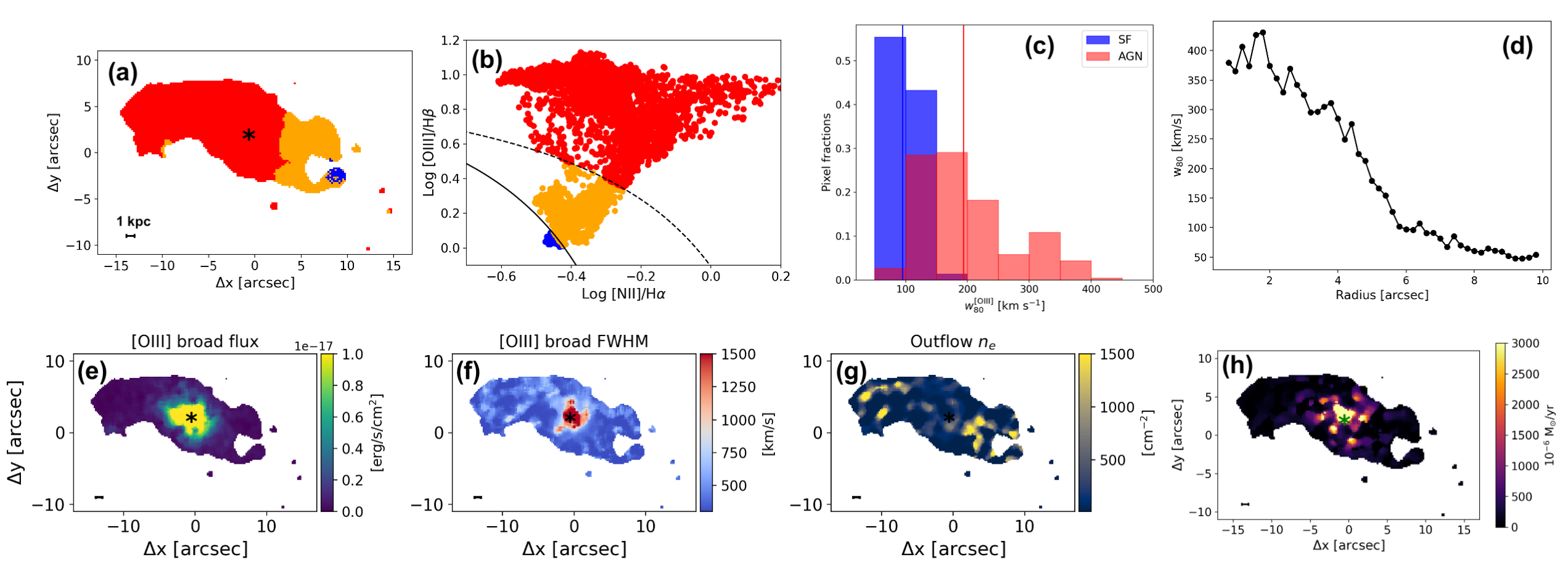}
\caption{Same as Fig. \ref{fig:plots_3C033}, for HE 0224-2834}
\label{fig:plots_HE0224_2834}
\end{figure*}

\begin{figure*}
\centering
\includegraphics[scale=0.5]{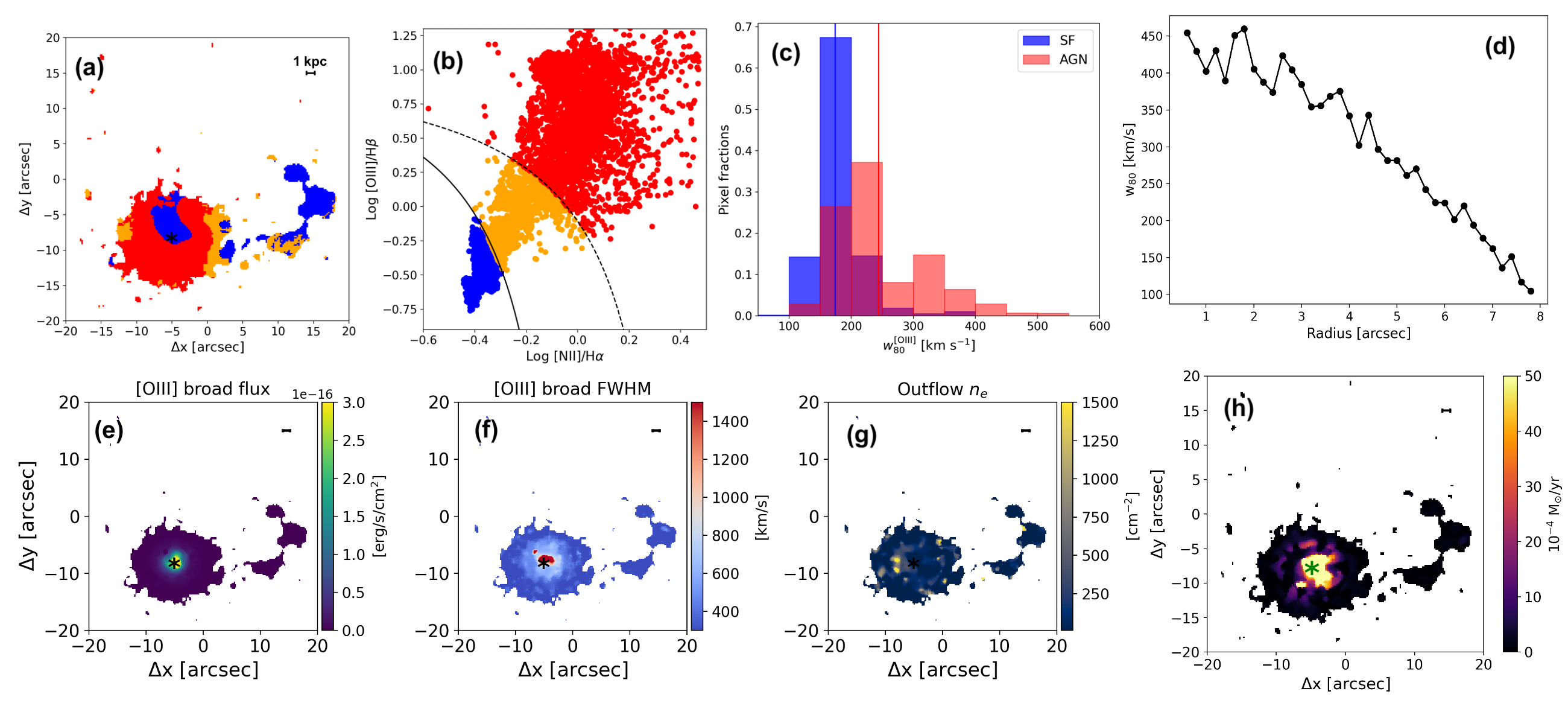}
\caption{Same as Fig. \ref{fig:plots_3C033}, for NGC 985}
\label{fig:plots_HE0232_0900}
\end{figure*}

\begin{figure*}
\centering
\includegraphics[scale=0.5]{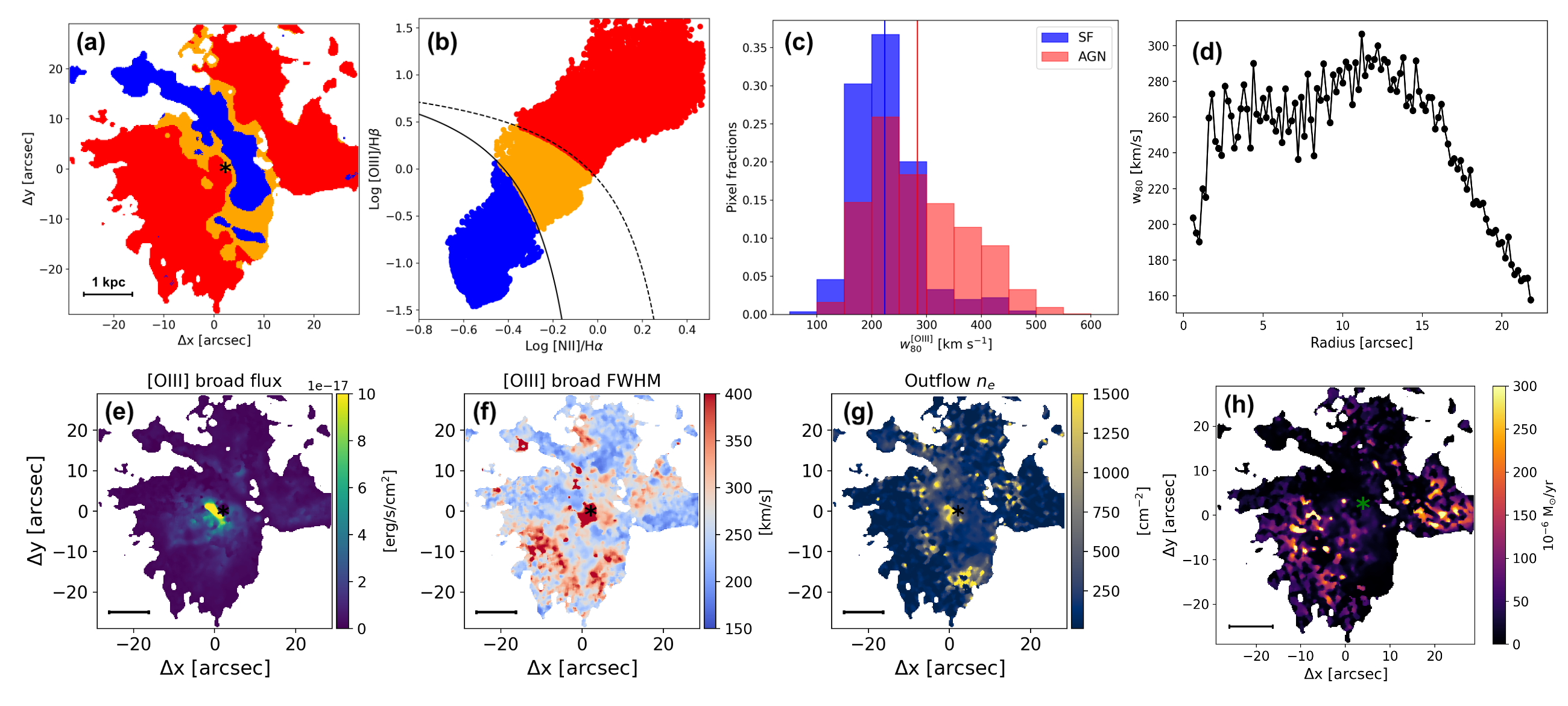}
\caption{Same as Fig. \ref{fig:plots_3C033}, for NGC 1365}
\label{fig:plots_NGC1365}
\end{figure*}

\begin{figure*}
\centering
\includegraphics[scale=0.5]{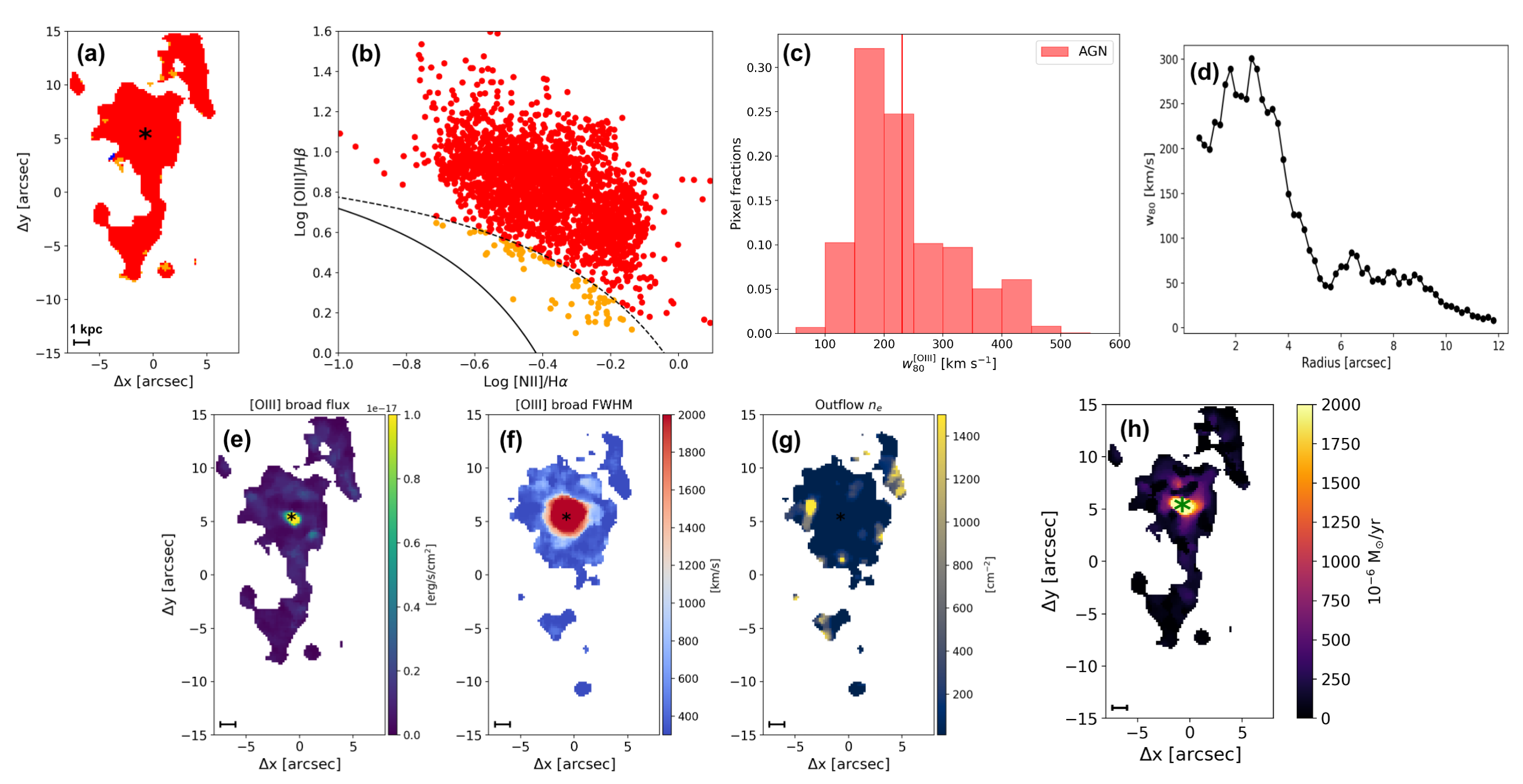}
\caption{Same as Fig. \ref{fig:plots_3C033}, for HE 0351+0240}
\label{fig:plots_HE0351_0240}
\end{figure*}

\begin{figure*}
\centering
\includegraphics[scale=0.5]{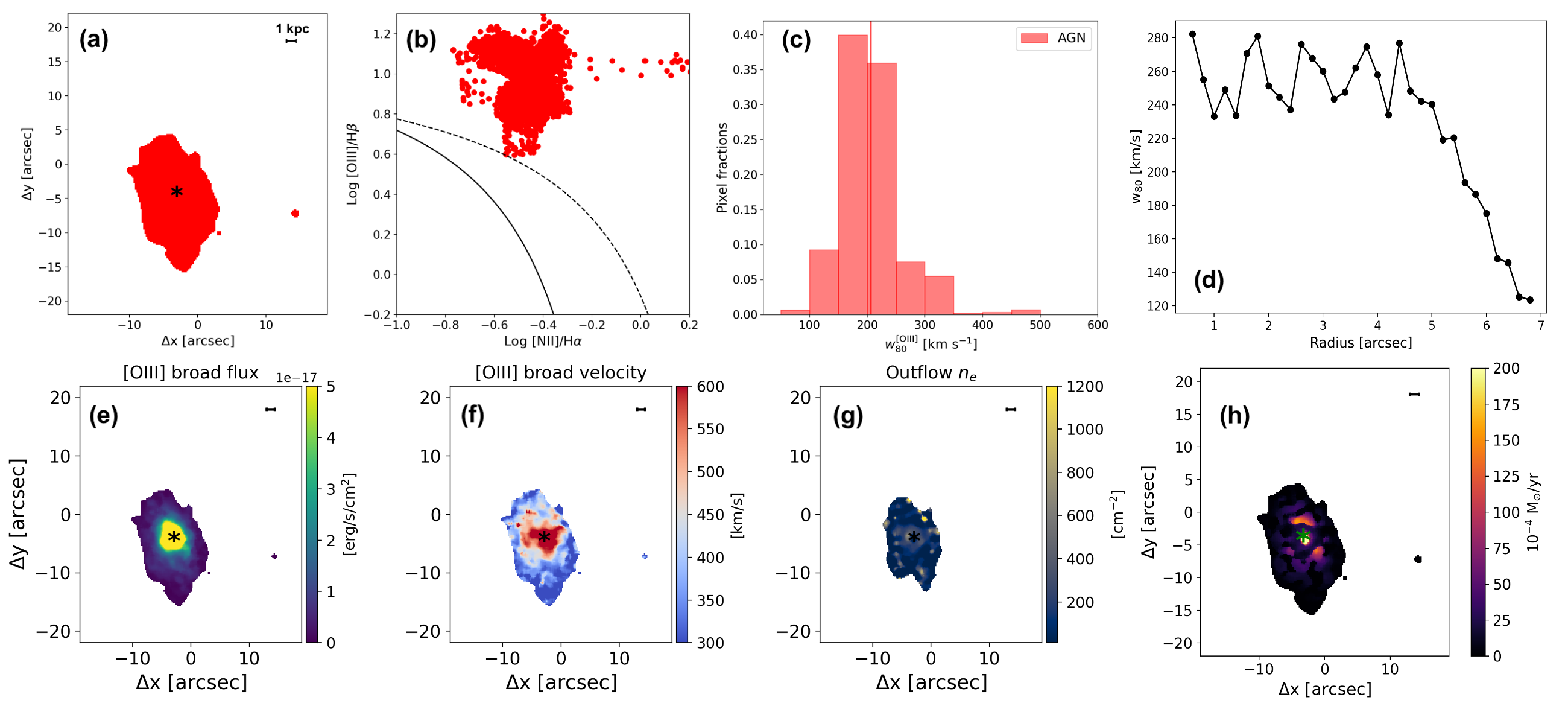}
\caption{Same as Fig. \ref{fig:plots_3C033}, for HE 0412-0803}
\label{fig:plots_HE0412_0803}
\end{figure*}

\begin{figure*}
\centering
\includegraphics[scale=0.5]{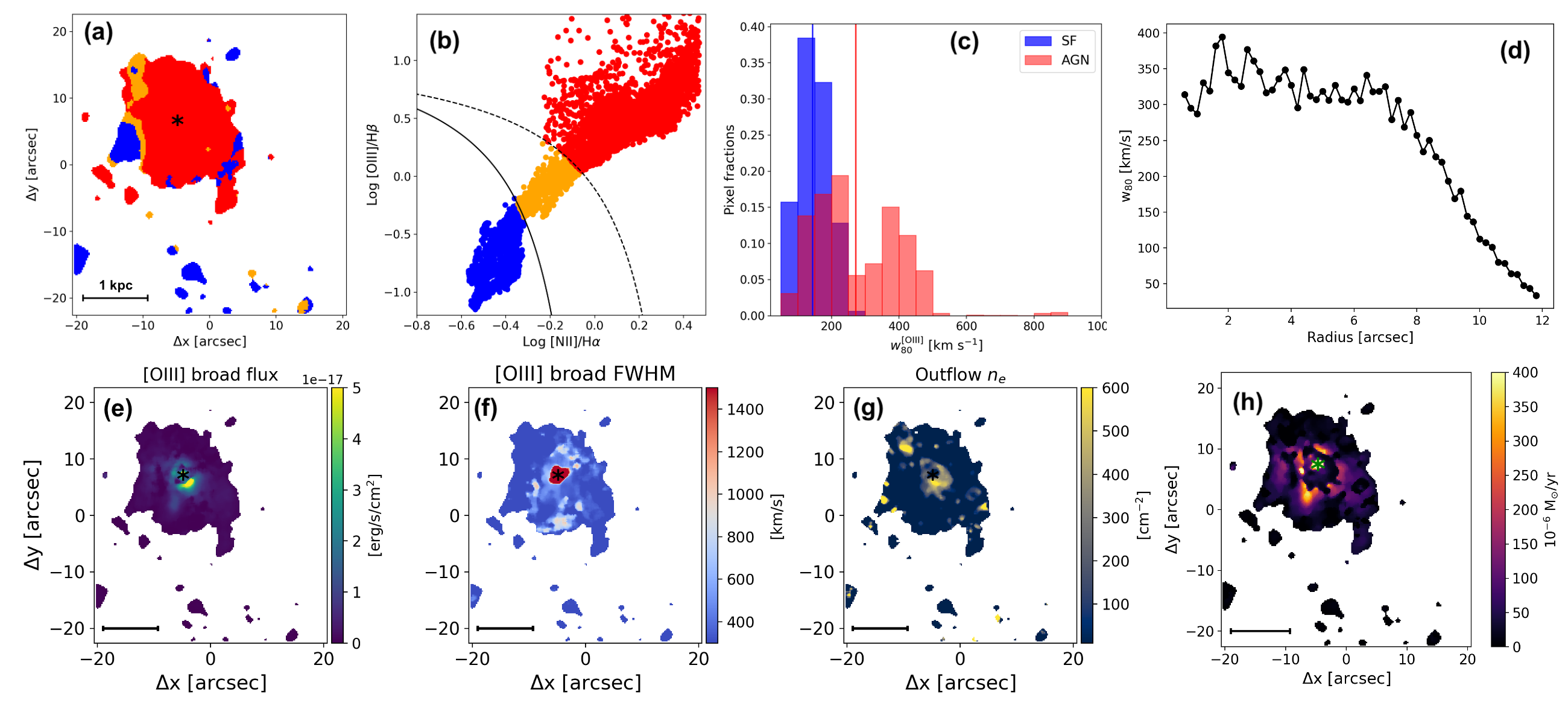}
\caption{Same as Fig. \ref{fig:plots_3C033}, for NGC 1566}
\label{fig:plots_NGC1566}
\end{figure*}

\begin{figure*}
\centering
\includegraphics[scale=0.5]{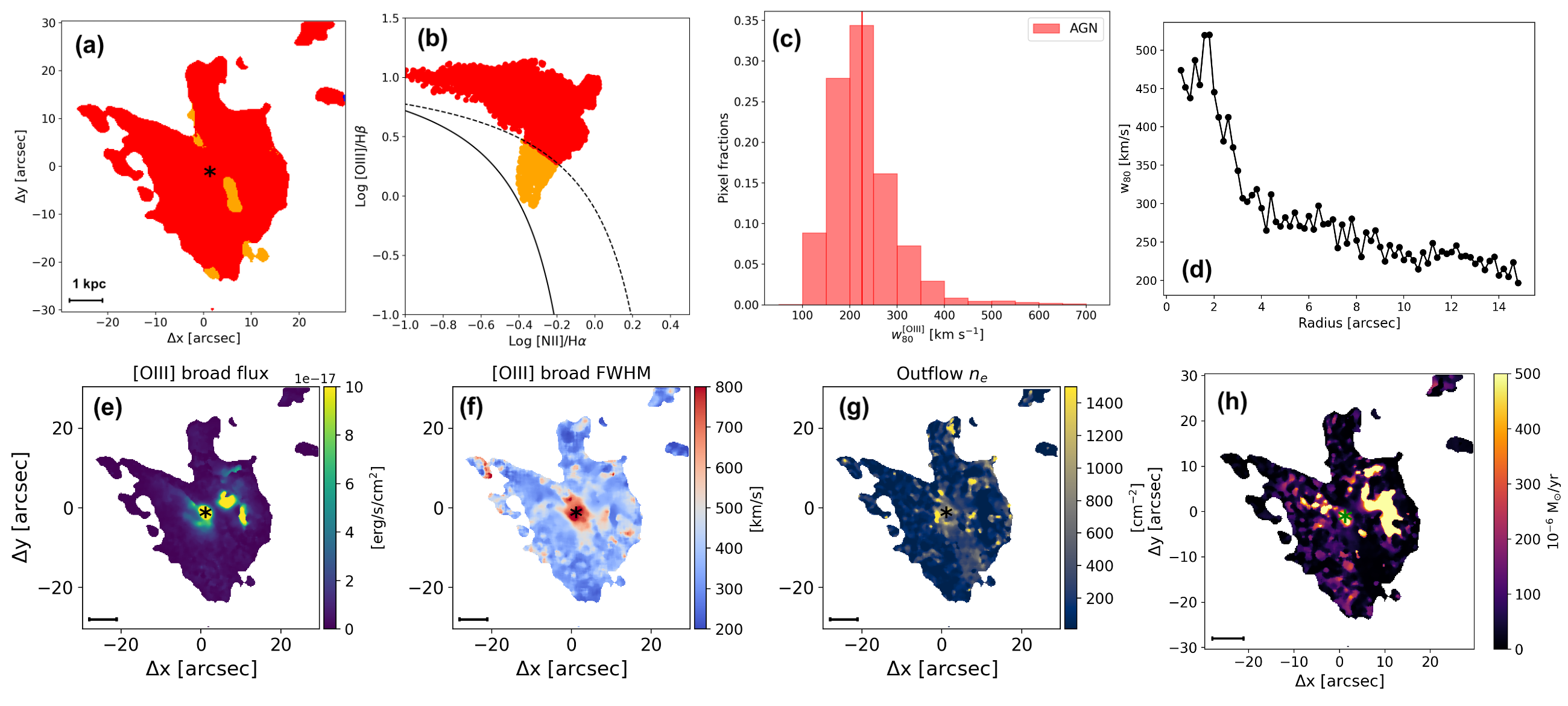}
\caption{Same as Fig. \ref{fig:plots_3C033}, for NGC 2992}
\label{fig:plots_NGC2992}
\end{figure*}

\begin{figure*}
\centering
\includegraphics[scale=0.5]{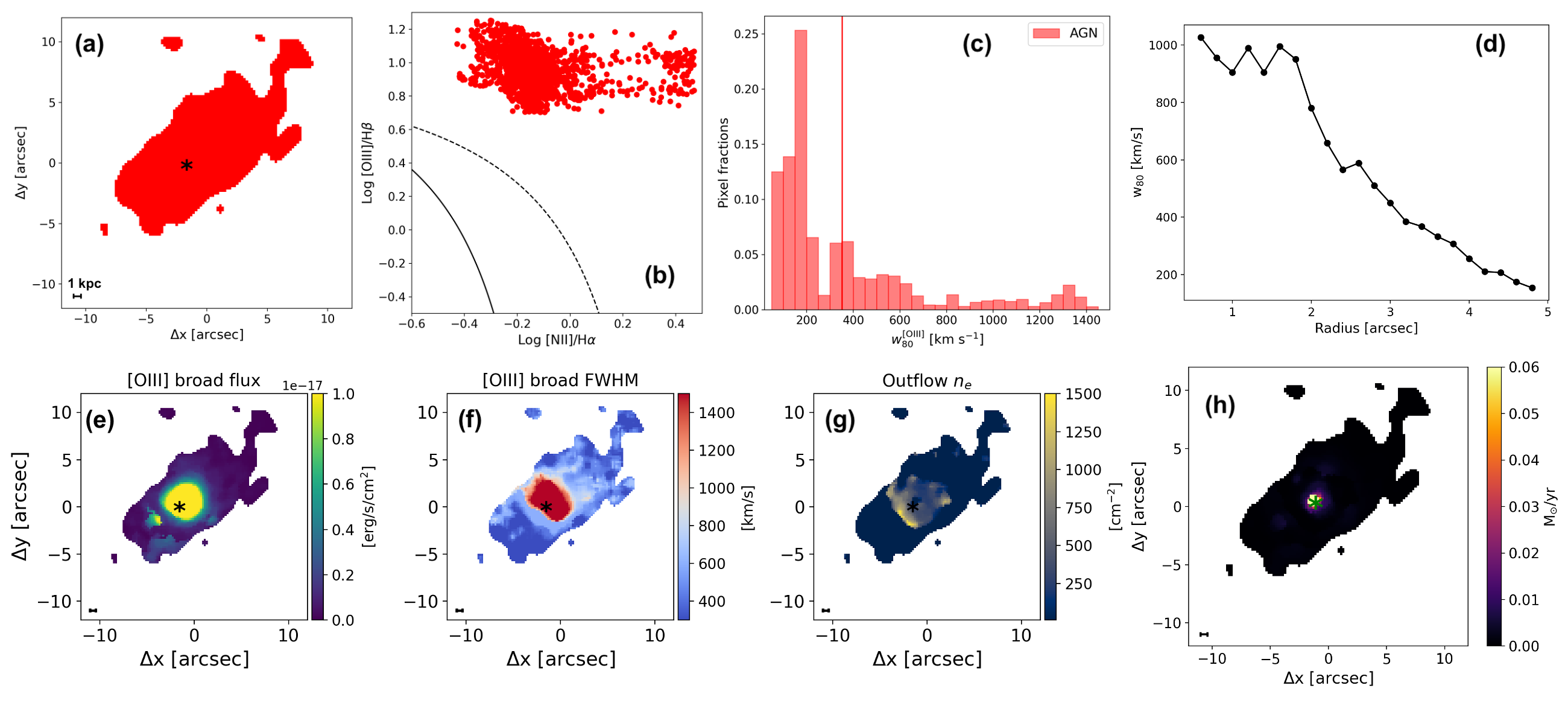}
\caption{Same as Fig. \ref{fig:plots_3C033}, for HE 1029-1401}
\label{fig:plots_HE1029_1401}
\end{figure*}

\begin{figure*}
\centering
\includegraphics[scale=0.5]{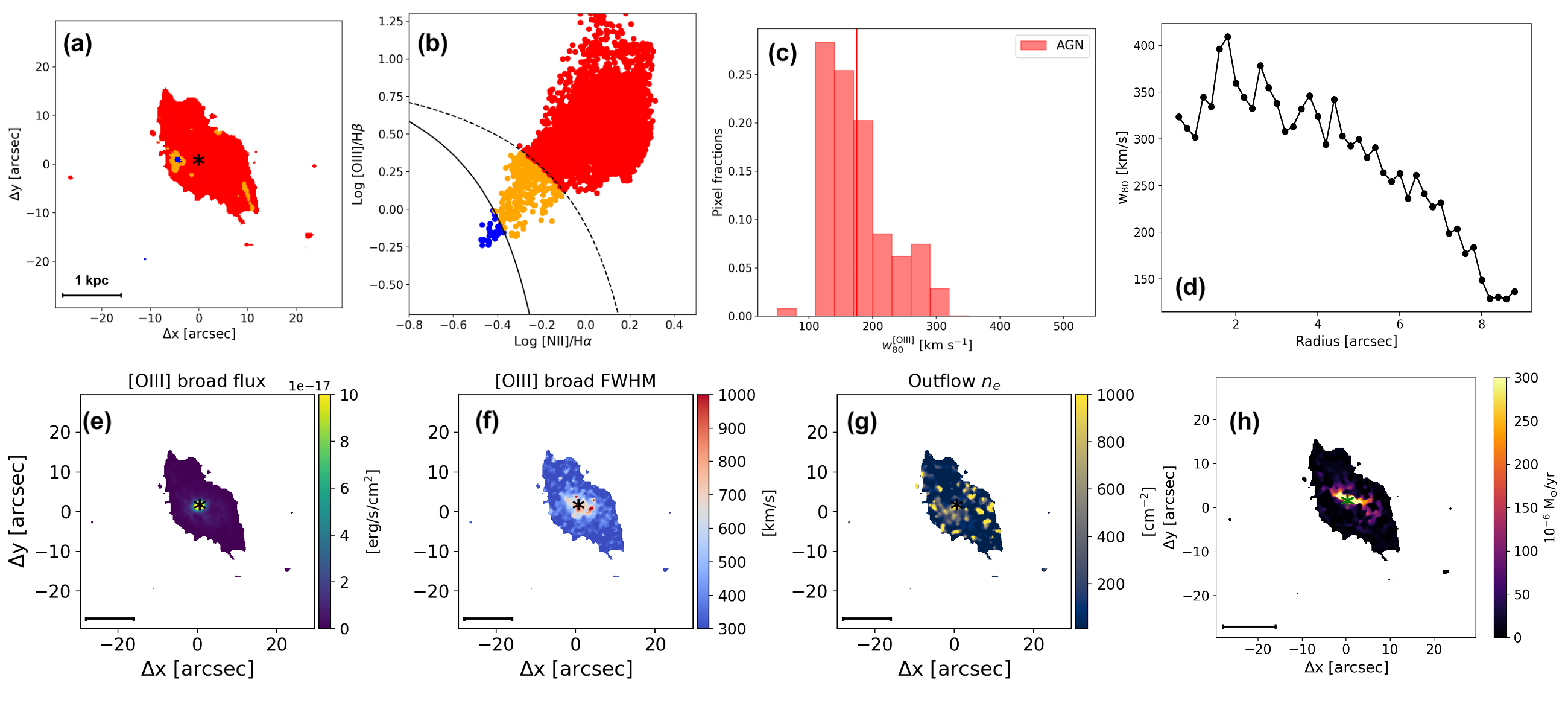}
\caption{Same as Fig. \ref{fig:plots_3C033}, for NGC 4941}
\label{fig:plots_NGC4941}
\end{figure*}

\begin{figure*}
\centering
\includegraphics[scale=0.5]{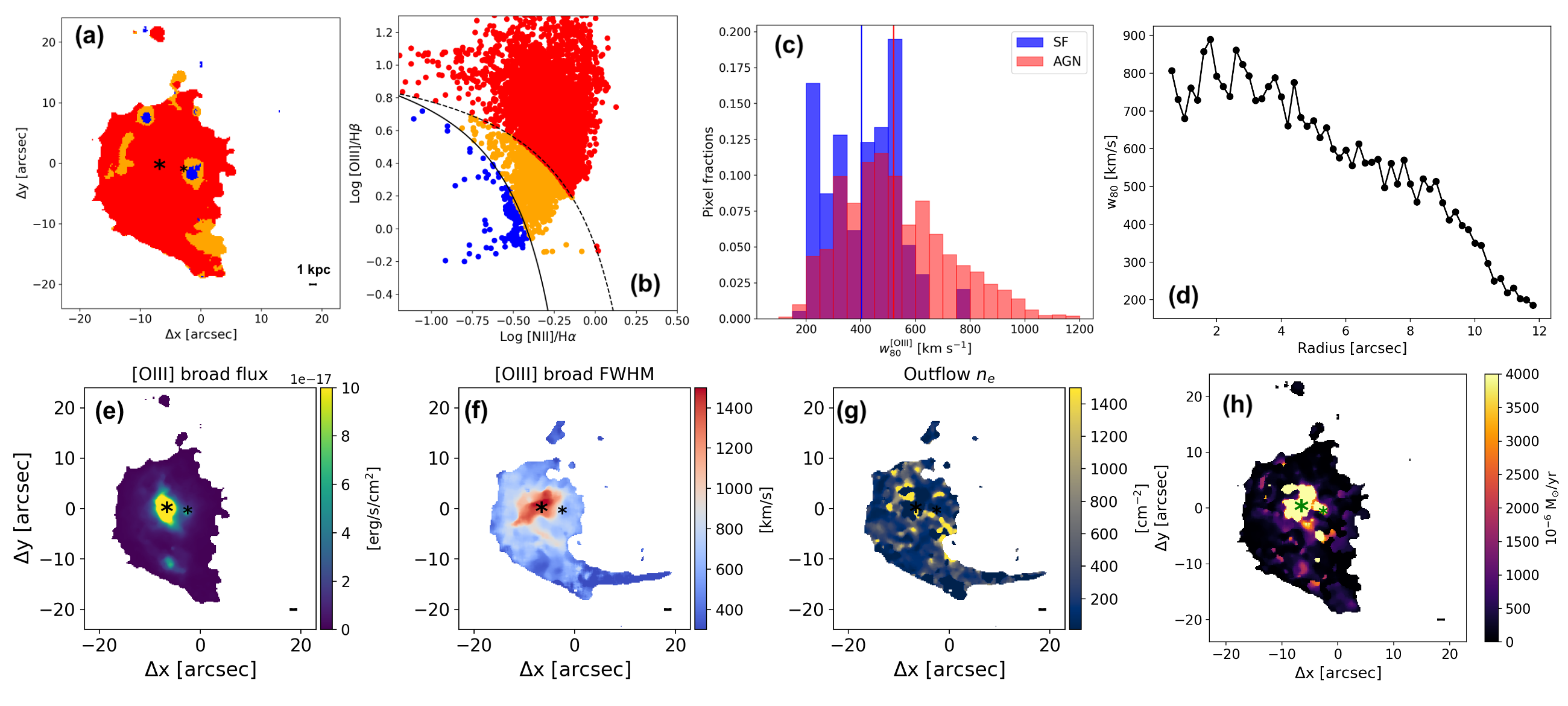}
\caption{Same as Fig. \ref{fig:plots_3C033}, for Mrk 463}
\label{fig:plots_MRK463}
\end{figure*}

\begin{figure*}
\centering
\includegraphics[scale=0.5]{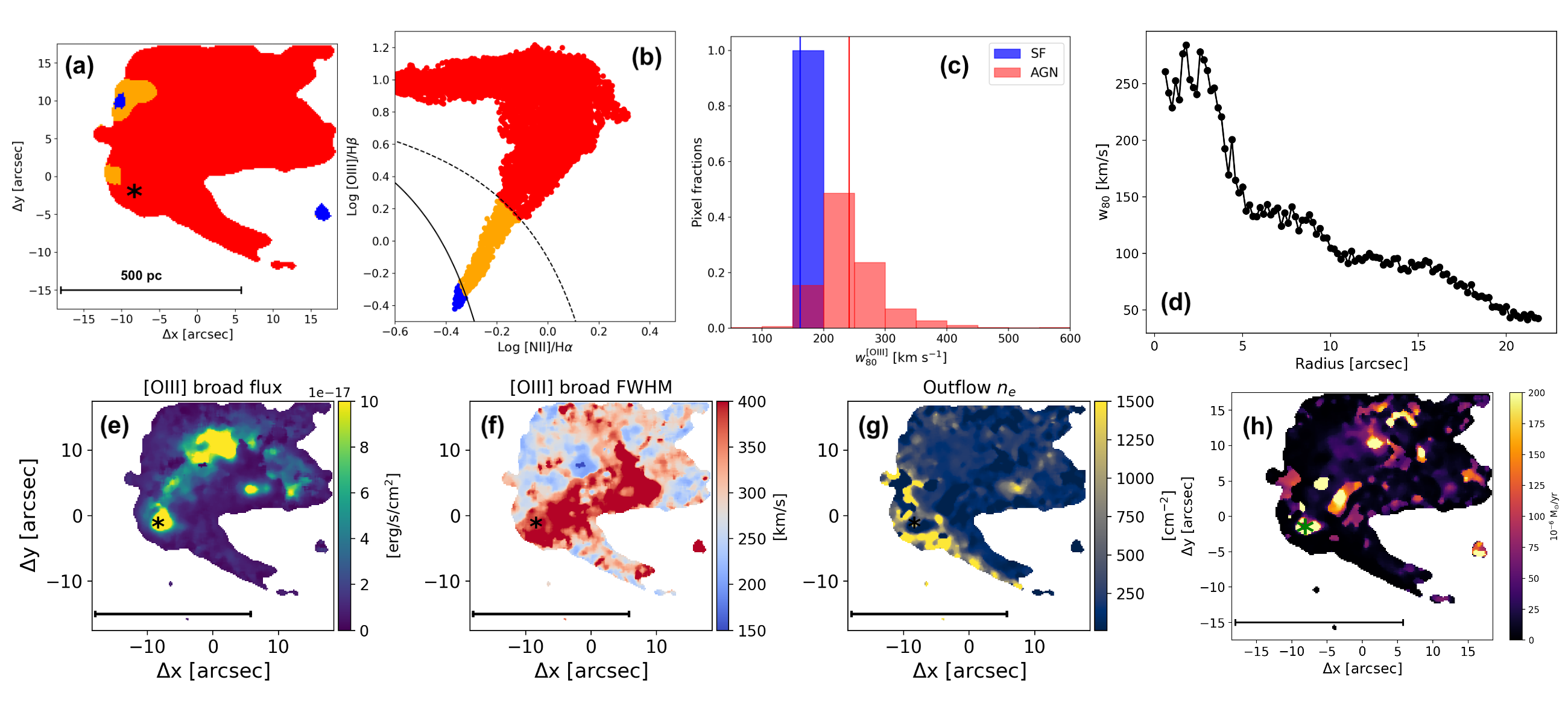}
\caption{Same as Fig. \ref{fig:plots_3C033}, for Circinus}
\label{fig:plots_circinus}
\end{figure*}

\begin{figure*}
\centering
\includegraphics[scale=0.5]{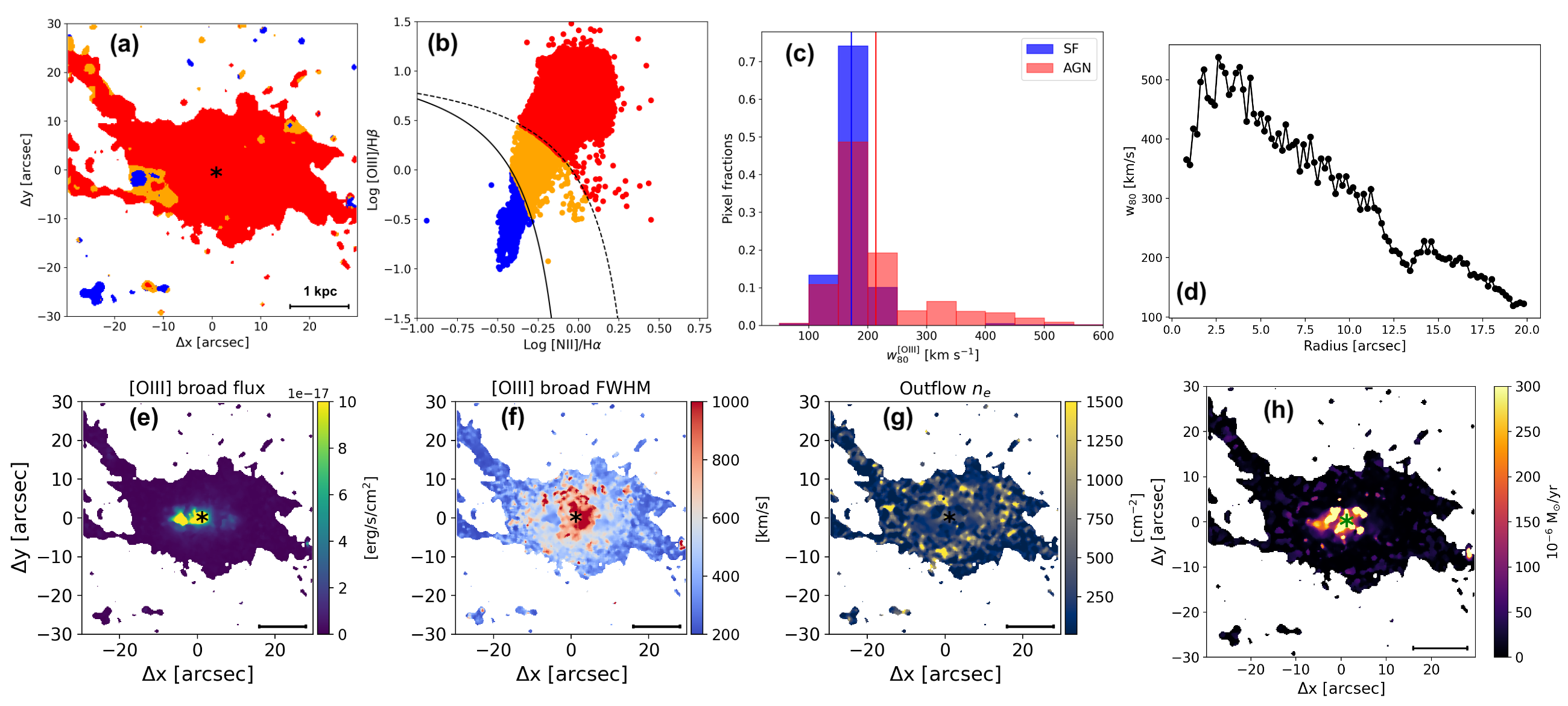}
\caption{Same as Fig. \ref{fig:plots_3C033}, for NGC 5643}
\label{fig:plots_NGC5643}
\end{figure*}

\begin{figure*}
\centering
\includegraphics[scale=0.5]{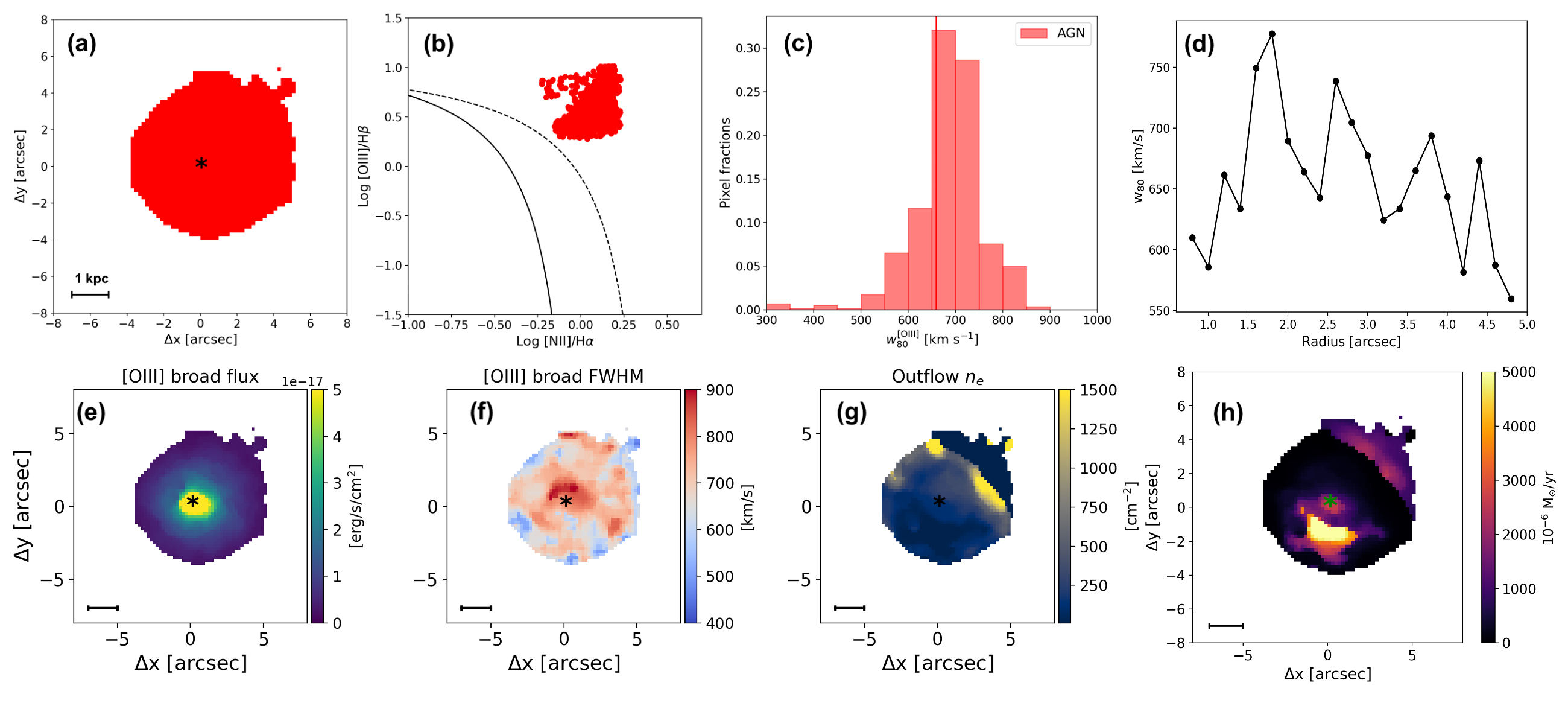}
\caption{Same as Fig. \ref{fig:plots_3C033}, for NGC 5995}
\label{fig:plots_NGC5995}
\end{figure*}

\begin{figure*}
\centering
\includegraphics[scale=0.5]{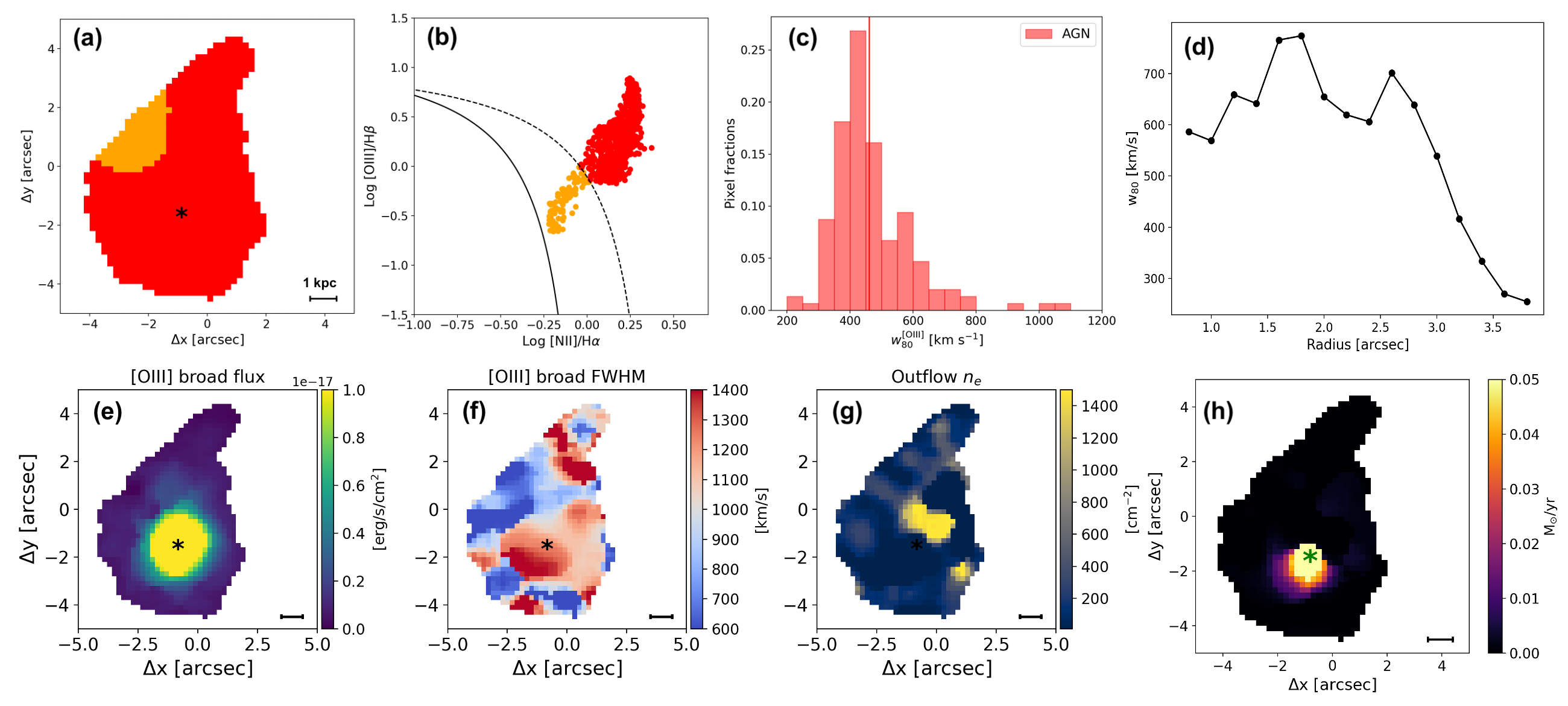}
\caption{Same as Fig. \ref{fig:plots_3C033}, for 2MASX J1631+2352}
\label{fig:plots_2MASX}
\end{figure*}

\begin{figure*}
\centering
\includegraphics[scale=0.5]{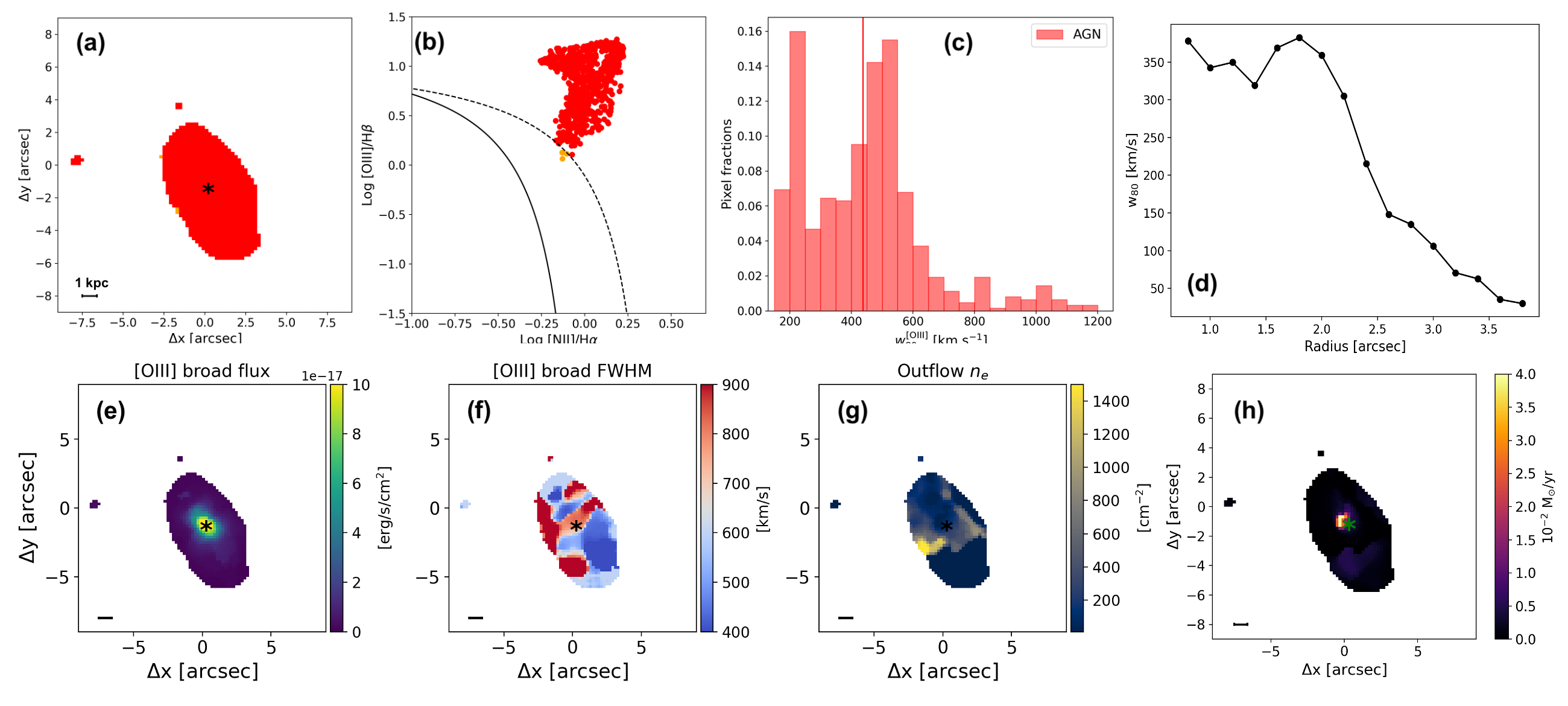}
\caption{ASame as Fig. \ref{fig:plots_3C033}, for 3C403}
\label{fig:plots_3C403}
\end{figure*}

\begin{figure*}
\centering
\includegraphics[scale=0.5]{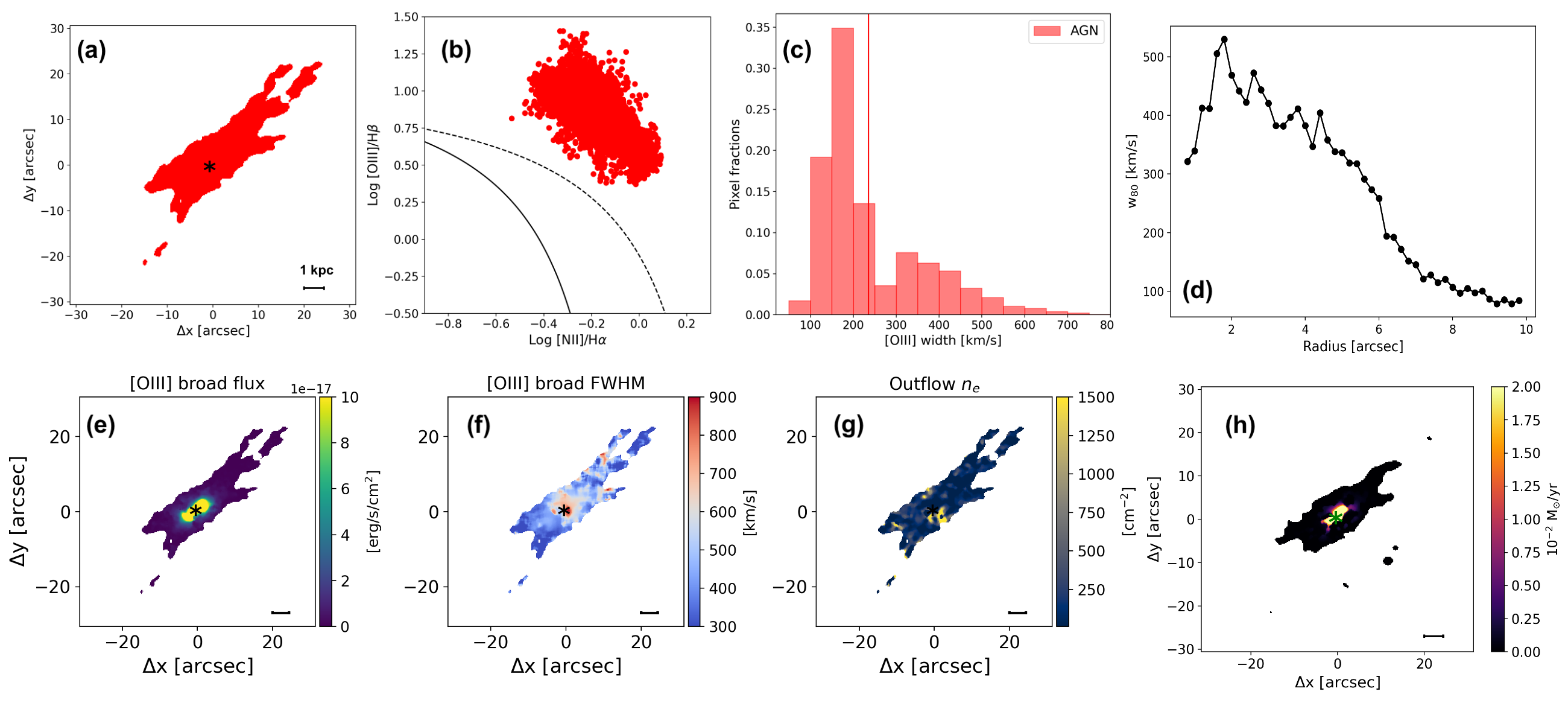}
\caption{Same as Fig. \ref{fig:plots_3C033}, for IC 5063}
\label{fig:plots_IC5063}
\end{figure*}

\begin{figure*}
\centering
\includegraphics[scale=0.5]{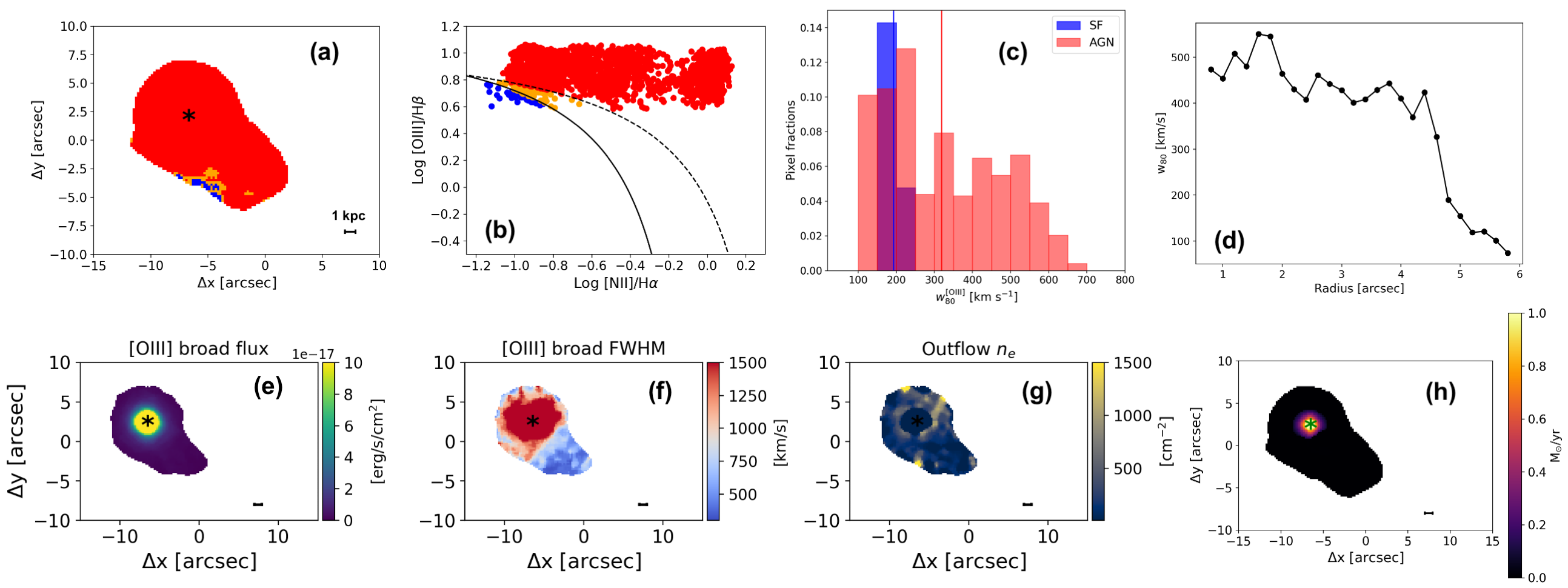}
\caption{Same as Fig. \ref{fig:plots_3C033}, for 3C445}
\label{fig:plots_3C445}
\end{figure*}

\begin{figure*}
\centering
\includegraphics[scale=0.5]{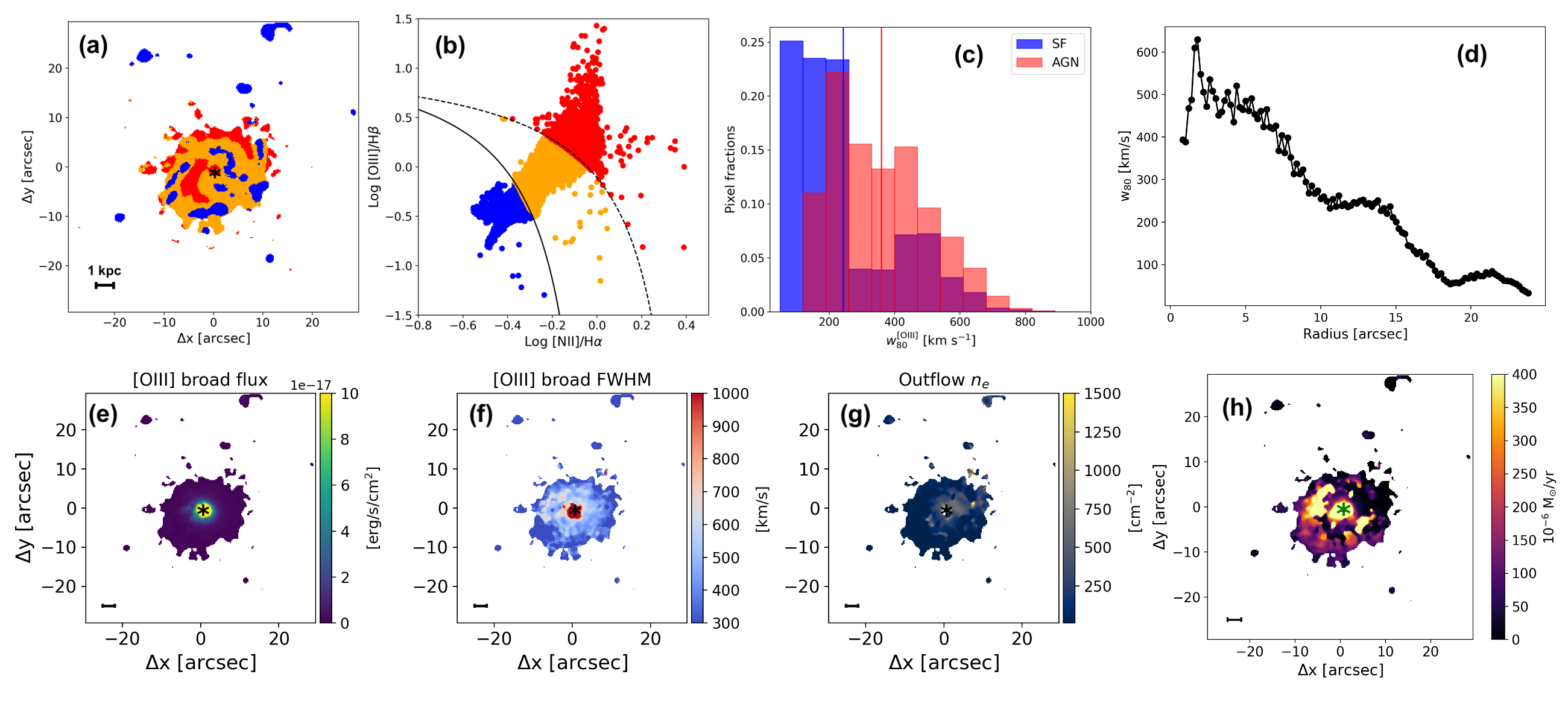}
\caption{Same as Fig. \ref{fig:plots_3C033}, for NGC 7469}
\label{fig:plots_NGC7469}
\end{figure*}

\begin{figure*}
\centering
\includegraphics[scale=0.5]{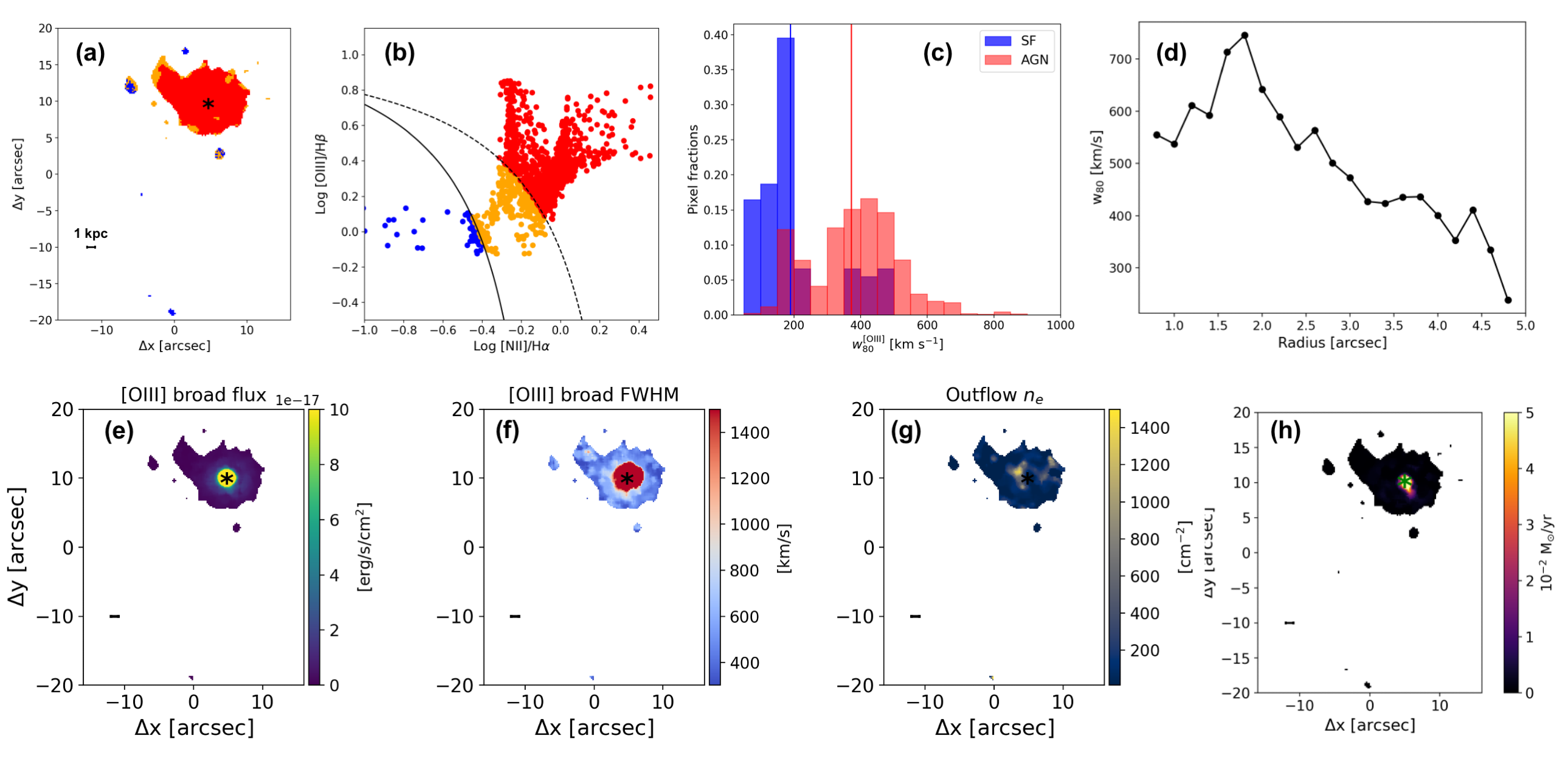}
\caption{Same as Fig. \ref{fig:plots_3C033}, for Mrk 926}
\label{fig:plots_MRK926}
\end{figure*}

\bsp	
\label{lastpage}
\end{document}